\documentclass[12pt,twoside,reqno]{article}
\usepackage{amsmath}
\usepackage{amsthm}
\usepackage{latexsym}
\usepackage{epsf}
\usepackage{graphicx, color}
\usepackage{ifpdf}  
\usepackage{epstopdf}

\begin{document}

\theoremstyle{remark}
\newtheorem{remark}{Remark}[section]
\numberwithin{equation}{section}

\parskip 4pt
\abovedisplayskip 7pt
\belowdisplayskip 7pt

\parindent=12pt

\newcommand{\bn}{{\bf n}}
\newcommand{\bnabla}{{\boldsymbol{\nabla}}}
\newcommand{\bu}{{\bf u}}
\newcommand{\bU}{{\bf U}}
\newcommand{\bPhi}{{\bf \Phi}}
\newcommand{\bPsi}{{\bf \Psi}}
\newcommand{\bv}{{\bf v}}
\newcommand{\bw}{{\bf w}}
\newcommand{\bx}{{\bf x}}
\newcommand{\bV}{{\bf V}}
\newcommand{\bz}{{\bf 0}}
\newcommand{\cth}{{\mathcal{T}_h}}
\newcommand{\ct}{{\mathcal{T}}}
\newcommand{\f}{{\bf f}}
\newcommand{\g}{{\bf g}}
\newcommand{\gx}{{\Gamma_{-} (0, \Delta t)}}
\newcommand{\into}{{\displaystyle{\int_{\Omega}}}}
\newcommand{\intG}{{\displaystyle{\int_{\Gamma}}}}
\newcommand{\oo}{{\overline{\Omega}}}
\newcommand{\ox}{{\Omega \times (0, \Delta t)}}
\newcommand{\oxot}{{\Omega \times (0,T)}}
\newcommand{\R}{{I\!\!R}}

\newpage
\thispagestyle{empty}
\noindent{\Large\bf A numerical study of the transition to oscillatory flow in 3D lid-driven cubic cavity flows}
\vskip 2ex
\normalsize
\noindent{Shang-Huan Chiu,  Tsorng-Whay Pan, Jiwen He, Aixia Guo,  and Roland Glowinski}
\vskip 1ex
\noindent{Department of Mathematics, University of Houston, Houston, Texas, USA}

\vskip 5ex

\noindent{\bf Abstract:}
In this article, three dimensional (3D) lid-driven cubic cavity flows have been studied numerically 
for various values of Reynolds number ($Re$).  The numerical solution of the Navier-Stokes equations 
modeling incompressible viscous fluid flow in a cubic cavity is obtained via a methodology 
combining a first order accurate operator-splitting, $L^2$-projection Stokes solver,
a wave-like equation treatment of the advection and finite element methods.  The numerical results 
obtained for Re$=$400, 1000, and 3200 show a good agreement with available numerical and experimental 
results in literature.  Simulation results predict that the critical Re$_{cr}$ for the transition from 
steady flow to oscillatory (a Hopf bifurcation) is somewhere in [1870, 1875] for the mesh size $h=1/96$.
Via studying the flow field distortion of fluid flow at Re before and after Re$_{cr}$, the occurrence of 
the first pair of Taylor-G\"ortler-like vortices is connected to the flow field distortion at the transition 
from steady flow to oscillatory flow in 3D lid-driven cubic cavity flows for Re $< 2000$.

\vskip 1ex
\noindent{{\bf Classification:} AMS: 65M60}

\section{Introduction.}

\normalsize
Lid-driven cavity flow is a classical flow problem that has attracted attention due to its 
flow configuration relevant to many industrial applications, such as coating  and melt-spinning 
processes pointed out in \cite{Aidun1991}, and its importance to the basic study of fluid mechanics, 
including boundary layers, eddies, secondary flows, complex three-dimensional patterns, various 
instabilities and transition, chaotic, and turbulence, as discussed in a review paper by Shankar 
and Deshpande in \cite{Shankar2000}. Also it is due to its geometrical simplicity and unambiguous boundary 
conditions which facilitate experimental calibrations and numerical computations, thus providing 
an idea benchmark problem for validating numerical results and 
comparing results obtained from experiments and computations.

It is known that,  depending on the methodologies, boundary conditions and mesh sizes used in the simulation, 
the critical Reynolds number ($Re_{cr}$) for the occurrence of the transition from a steady flow 
to an oscillatory flow (a Hopf bifurcation)  in two-dimensional square lid-driven cavity 
flow varies between  8000 and 10000 (e.g., see \cite{Shen1991}, \cite{Goyon1996}, \cite{Auteri2002}, 
\cite {Sahin2003}, \cite {Bruneau2006} and \cite  {Pan2008}). The oscillatory instability in  cubic 
lid-driven cavity flows has been studied recently in \cite{Feldman2010}, \cite{Liberzon2011} and 
\cite{Anupindi2014}. Numerically, Feldman and Gelfgat \cite{Feldman2010} obtained that the critical Reynolds number for 
the occurrence of such Hopf bifurcation is at $Re_{cr}=1914$.  Anupindi {\it et al.} \cite{Anupindi2014}
reported that their critical value is $Re_{cr}=2300$ (but it was obtained with regularized boundary conditions).  
Both are more precise than a much earlier result predicted by Iwatsu {et al.}, who gave the range between 
2000 and 3000 for the critical Reynolds number in \cite{Iwatsu1989}.  Experimentally, Liberzon {\it et al.}
\cite{Liberzon2011} reported that the critical Reynolds number is  in the range between 1700 and 1970,
which is slightly lower than Re$=2000$, at which Iwatsu {\it et al.}  \cite{Iwatsu1990}  obtained a pair of 
Taylor-G\"ortler-like (TGL) vortices in a cubic lid-driven cavity flow. Giannetti {\it et al.} obtained that 
cubic lid-driven cavity flow becomes unstable for  $Re$ just above 2000 via three-dimensional global linear 
stability analysis in  \cite{Giannetti2009}. Those aforementioned experimental and computational results 
suggest that there is a connection between oscillatory flows and TGL vortices.

In this article, we have studied numerically the transition from steady flow to oscillatory flow and  
occurrence of the TGL vortices in order to find out the relation between them if there is any.
We have applied  a first order accurate operator-splitting scheme, Lie's scheme \cite{chorin1978}, to  
the numerical solution of the Navier-Stokes equations, which is a continuation of the works discussed in 
\cite{pan2000} and \cite{pan2008}. The resulting methodology is easy to implement and quite modular since,
for each time step,  there is a sequence of three simpler subproblems needed to be solved.
For the first subproblem we have used a $L^2$-projection  Stokes solver \`a la Uzawa to force the incompressibility condition. 
To solve the advection problem as the second subproblem,  we have applied a wave-like equation method \cite{glowinski2003}.
The third one is a diffusion problem which can be solved easily. The numerical results obtained for Re$=$400, 1000 and 
3200 show a good agreement with available numerical and experimental  results in literature. Our
simulation results predict that the critical Re$_{cr}$ for the transition from steady flow to oscillatory 
(a Hopf bifurcation) is somewhere in (1870, 1875) (resp., (1860, 1865)) for $h=1/96$ (resp., $h=1/60$).
For the connection between the occurrence of TGL vortices and the transition from steady flow to oscillatory flow, 
we have investigated the flow field at $Re$ close to $Re_{cr}$. The difference of flow fields at different instants of time
shows  a pair of vortices reminiscent of the GTL ones, but with much smaller magnitude  for Re slightly smaller than
Re$_{cr}$. For Re slightly larger than Re$_{cr}$, this flow field distortion is  visible and then invisible  periodically 
as a pair of TGL vortices. When increasing Re even higher (but < 2000), two tertiary vortices next to the corner vortices 
(one from each corner) start interacting with this pair periodically and the number of TGL pairs varies between one and two.
All these pairs are symmetric. Thus our computational results suggest that the first pair of symmetric TGL vortices for Re < 2000 
is actually the first  oscillating mode flow, which exists all the way through the transition.
The outline of this paper is as follows:  We first introduce the formulation of the problem  and then 
numerical method briefly in Section 2. 
In Section 3, the numerical results obtained for the classical lid-driven cavity flow problem at Re$=$400, 1000,
and 3200 are compared with numerical and experimental results for validation purpose. Then the critical Reynolds number
for the transition from steady flow to oscillatory flow has been investigated in Section 4. Finally, the connection
between  oscillatory flow and TGL vortices has been studied in Section 5.

\section{Formulation of the problem.}
\begin{figure} [ht]
\begin{center}
\leavevmode
\includegraphics[width=2.5in]{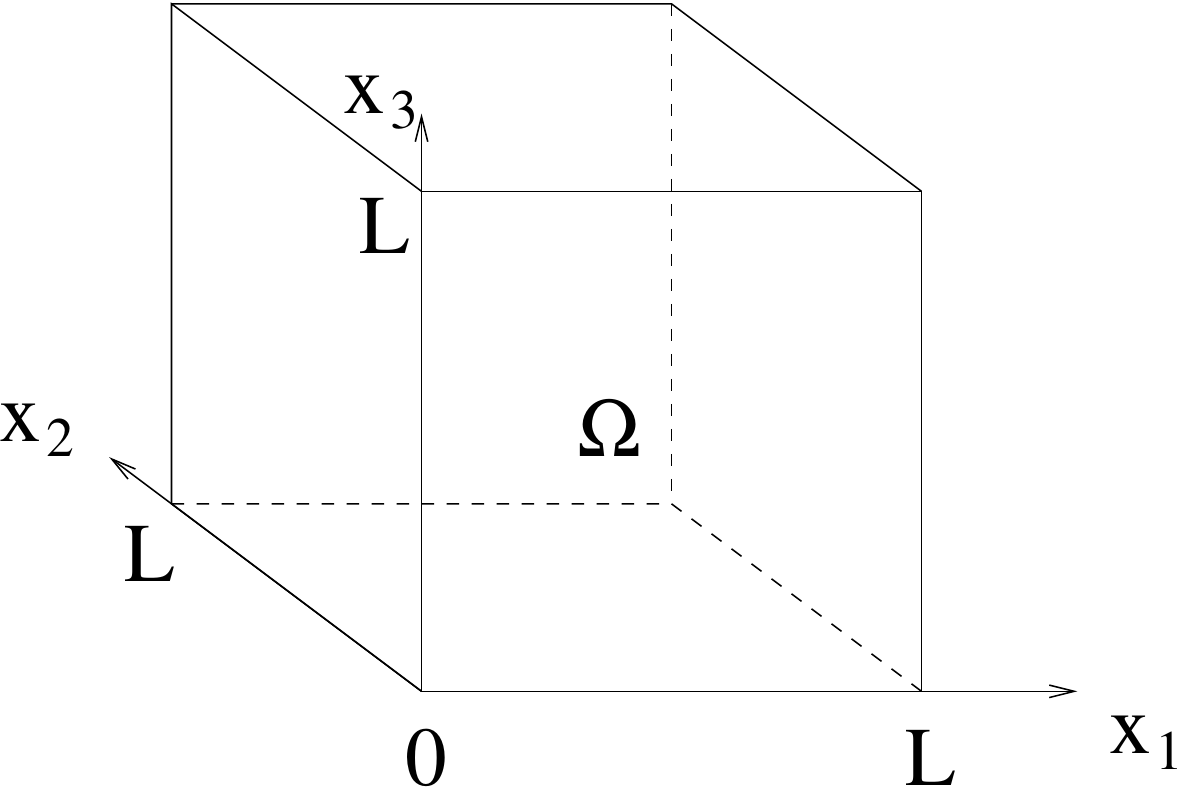} 
\end{center}
\caption{The computation domain $\Omega$ and coordinate system.}\label{fig1}
\end{figure}
The governing equations for modeling incompressible viscous fluid flow in $\Omega \subset \R^3$ (see Fig. \ref{fig1}) for $T>0$ are the Navier-Stokes equations, namely
\begin{eqnarray}
&\dfrac{\partial \bu}{\partial t} - \nu \Delta \bu + (\bu \cdot
\bnabla) \bu + \bnabla p = {\bf f} \ in \ \oxot,  \label{eqn:1.1}\\
&\bnabla \cdot \bu = 0 \ in \ \oxot,\label{eqn:1.2}\\
&\bu(0) = \bu_0, \ with \ \bnabla \cdot \bu_0 = 0,\label{eqn:1.3}\\
&\bu = \bu_B(\bx) \ on \ \partial\oxot \ with \ \displaystyle\int_{\partial\Omega} \bu_B 
\cdot \bn\, d\Gamma = 0 \ on \ (0,T),\label{eqn:1.4}
\end{eqnarray}
where $\bu$ and $p$ are the flow velocity and pressure, 
respectively, $\nu (=1/Re)$ is a viscosity coefficient, ${\bf f}$ is the body force, 
$\bu_B(\bx)$ is the  boundary condition,
and $\bn$ is the unit outward normal vector at $\Gamma=\partial \Omega$.  We denote by $v(t)$ the 
function $\bx \to v(\bx,t)$, $\bx$ being the generic point of $\R^3$.

The numerical solution of problem (\ref{eqn:1.1})-(\ref{eqn:1.4}) has generated a most abundant 
literature. Following Chorin \cite{chorin1967, chorin1968} and Temam  \cite{temam1969a, temam1969b}, 
most ``modern'' Navier-Stokes solvers are based on operator splitting algorithms 
(see, e.g., refs. \cite{marchuk1990}, \cite{turek1996}, \cite{marion1998} (Chapter 3) 
and \cite{glowinski2003} (Chapters 2 and 7)) in order to force the incompressibility 
condition via either $H^1$-projection or  $L^2$-projection Stokes solver method.
Among those methods for the numerical solution of (\ref{eqn:1.1})-(\ref{eqn:1.4}),
we have chosen  the   Lie scheme (see, e.g., see \cite{glowinski2003} and  
\cite{OS2016} for a general discussion of that scheme). It is first order accurate in time; but
its low order accuracy in time is compensated by its modular,  easy implementation, good stability, 
and robustness properties. 
To speed up the numerical solution of the cubic lid-driven cavity flow problem, we have time-discretized 
the related problem (\ref{eqn:1.1})-(\ref{eqn:1.4}), using a three stage Lie scheme scheme, 
namely: (i) using a $L^2$-projection  Stokes solver \`a la Uzawa to force the incompressibility 
condition, (ii) an advection step, and 
(iii) a diffusion step. The resulting methodology reads as follows: 
\begin{equation}
\bu^0=\bu_{0}. \label{eqn:1.14}
\end{equation}
\noindent{For $n \ge 0$,  $\bu^n \to \{\bu^{n+1/3},p^{n+1}\} \to \bu^{n+2/3} \to  \bu^{n+1}$ 
via the solution of:}
\begin{eqnarray}
\hskip -10pt \begin{cases}
&\rho \dfrac{\bu^{n+1/3}-\bu^{n}}{\triangle t}   + \bnabla p^{n+1} = {\bf 0}
\ \text{in} \ \Omega, \\
&\bnabla \cdot \bu^{n+1/3} = 0 \   \text{in} \ \Omega,\\
&\bu^{n+1/3}\cdot \bn=0 \  \text{on} \ \Gamma,  
\end{cases}\label{eqn:1.15}
\end{eqnarray}
\begin{eqnarray}
&&\begin{cases}
&\dfrac{\partial \bw}{\partial t} + (\bu^{n+1/3} \cdot \bnabla) \bw   = {\bf 0} \ 
\text{in} \ \Omega \times (t^n,t^{n+1}),\\
&\bw(t^n) = \bu^{n+1/3},\\
&\bw(t) = \bu_B(\bx) \ \text{on} \  \Gamma_{-}^{n+1}\times (t^n,t^{n+1}),
\end{cases}  \label{eqn:1.16} \\
&&\bu^{n+2/3}=\bw(t^{n+1}),  \label{eqn:1.17} \
\end{eqnarray}
\begin{eqnarray}
\hskip 30pt \begin{cases}
&\rho \dfrac{\bu^{n+1}-\bu^{n+2/3}}{\triangle t} - \mu \bnabla^2 \bu^{n+1} = {\bf f}^{n+1} 
\ \text{in} \ \Omega, \\
&\bu^{n+1} = \bu_B(\bx) \  \text{on} \ \Gamma.
\end{cases}\label{eqn:1.18}
\end{eqnarray}
Two simplifications take place for the lid-driven cavity flow problem considered here: 
namely, ${\bf f}= {\bf 0}$ and $\Gamma_{-}^{n+1} = \{\bx|\bx \in \Gamma, \bu_B(\bx)\cdot \bn(\bx) < 0\}=\emptyset$.

For the space discretization, we have used, as  in   \cite{glowinski2003} (Chapter 5) 
and \cite{bristeau1987}, a $P_1$-$iso$-$P_2$ (resp., $P_1$) finite element  approximation for
the velocity field (resp., pressure) defined on uniform ‘‘tetrahedral’’  meshes $\cth$ (resp., $\ct_{2h}$).
The problem (\ref{eqn:1.15}) is reminiscent of those encountered when applying Chorin's projection 
method (\cite{chorin1967}).  Three subproblems in (\ref{eqn:1.15})-(\ref{eqn:1.18}) are very classical problems 
and each one  can be solved by many different available methods. This is  the key point of the operator 
splitting methods.  The saddle point problem (\ref{eqn:1.15}) has been solved
by a Uzawa/preconditioned conjugate gradient algorithm as discussed in \cite{glowinski2003} (Section 21).  
Problem (\ref{eqn:1.16})-(\ref{eqn:1.17}) 
has been solved by   a wave-like equation method (e.g., see \cite{dean1997} and
\cite{glowinski2003} (Section 31)) which does not introduce numerical dissipation,
unlike the  commonly used {\it upwinding} schemes.  The problem (\ref{eqn:1.18}) is a 
classical discrete elliptic problem which can be solved easily.

\section{Numerical Results}

For the  lid-driven cavity flow problem in a cube, considered here, we took 
$\Omega = (0,1)^3$  as  computational domain and defined the Dirichlet data  
$\bu_B$ by
\begin{equation}
\bu_B(\bx) = \begin{cases} (1,0,0)^T \ \text{on} \ \{\bx \ | \ \bx = (x_1,x_2, 1)^T, 0 < x_1, x_2 < 1\},\\
 {\bf 0} \ \text{elsewhere \ on} \ \Gamma.
\end{cases}\label{eqn:3.1}
\end{equation}
We considered that the steady state has been reached  when 
the  change between two consecutive time step in the simulation, 
$\|\bu^n_h - \bu^{n-1}_h\|_{\infty} /{\triangle t}$, 
is less than $10^{-4}$,  and then took $\bu_h^n$  as the steady state solution.

\begin{figure} [tp]
\begin{center}
\leavevmode
\includegraphics[width=2.25in]{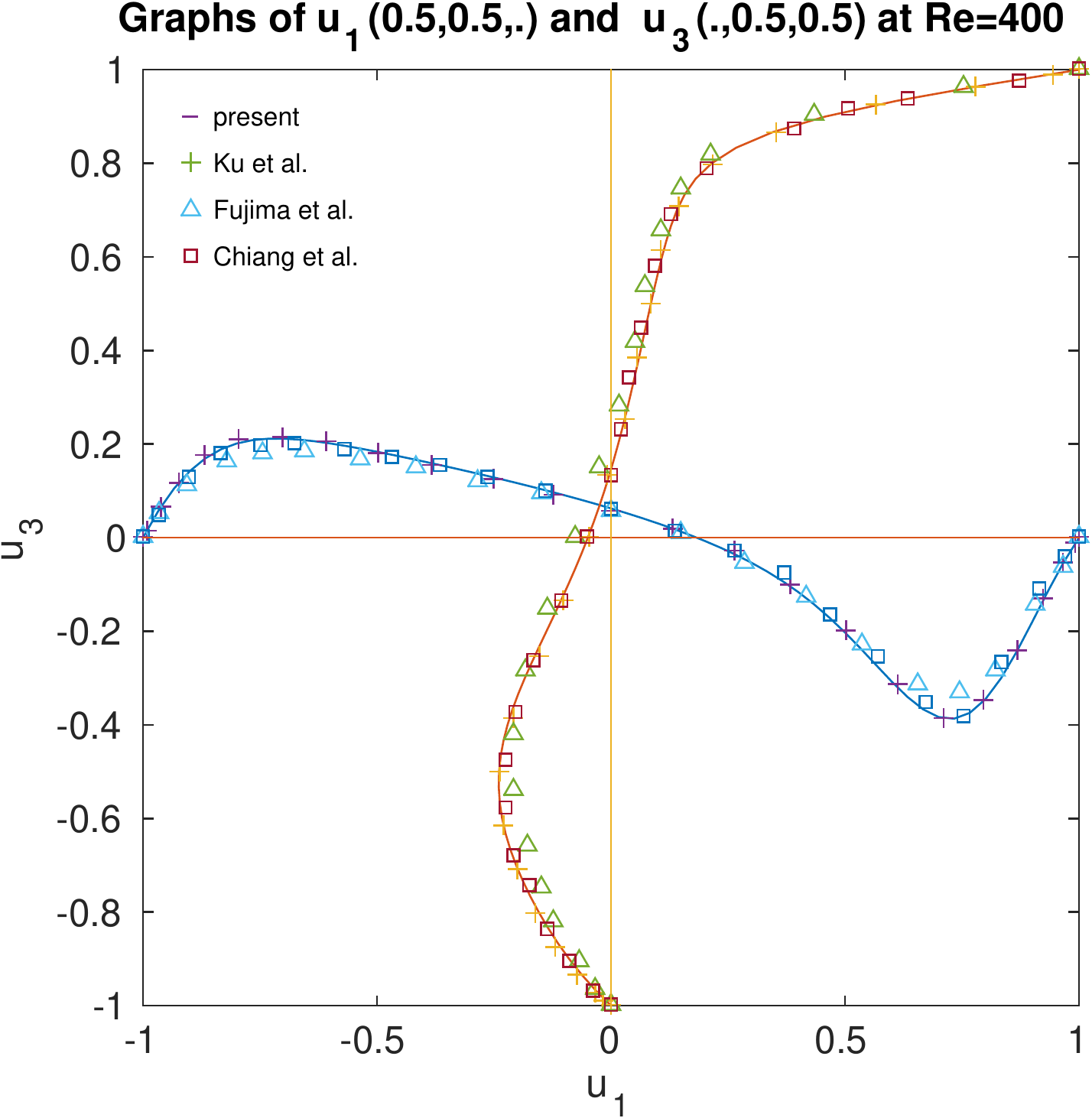} 
\includegraphics[width=2.25in]{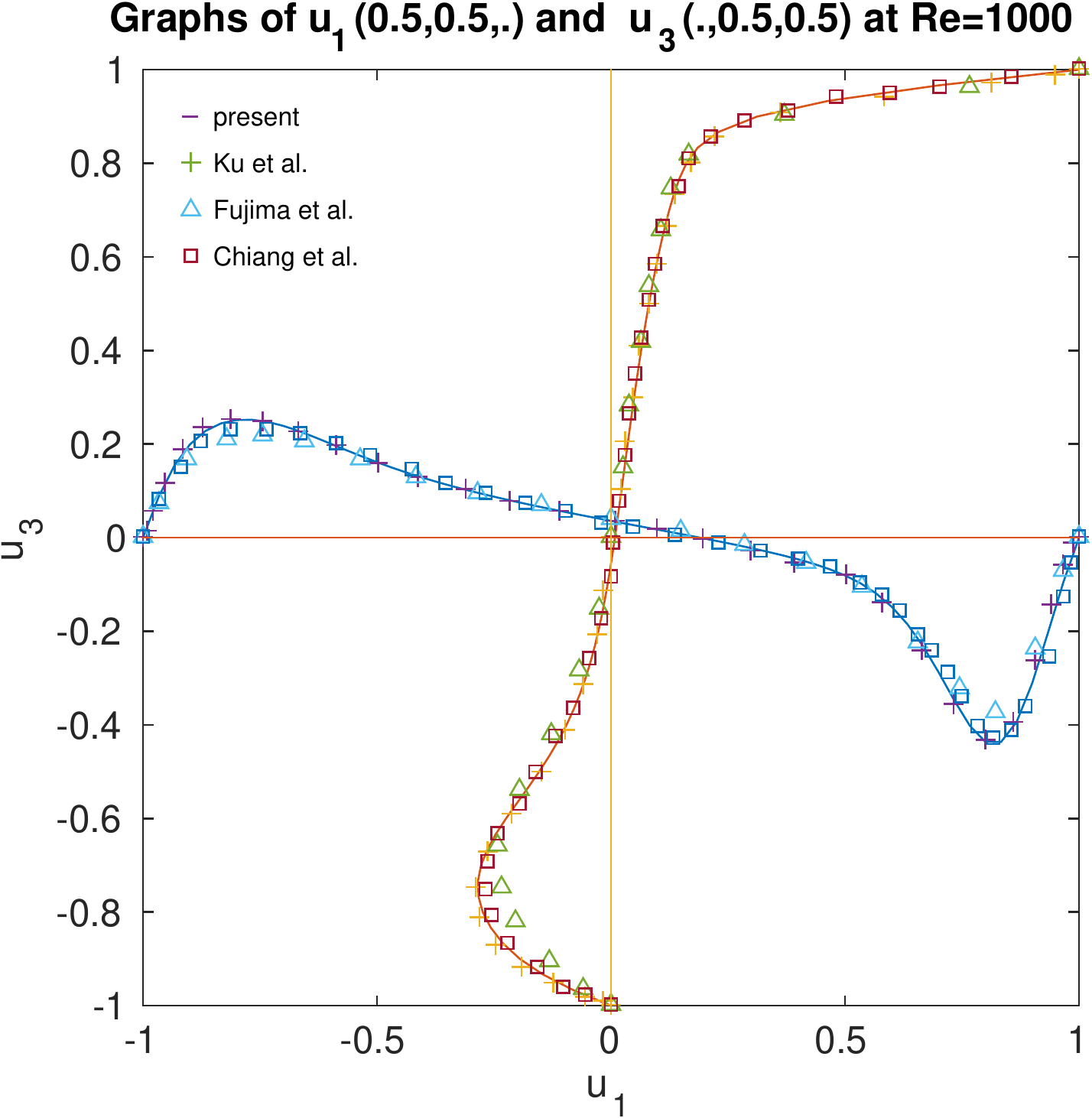} \\
\includegraphics[width=2.25in]{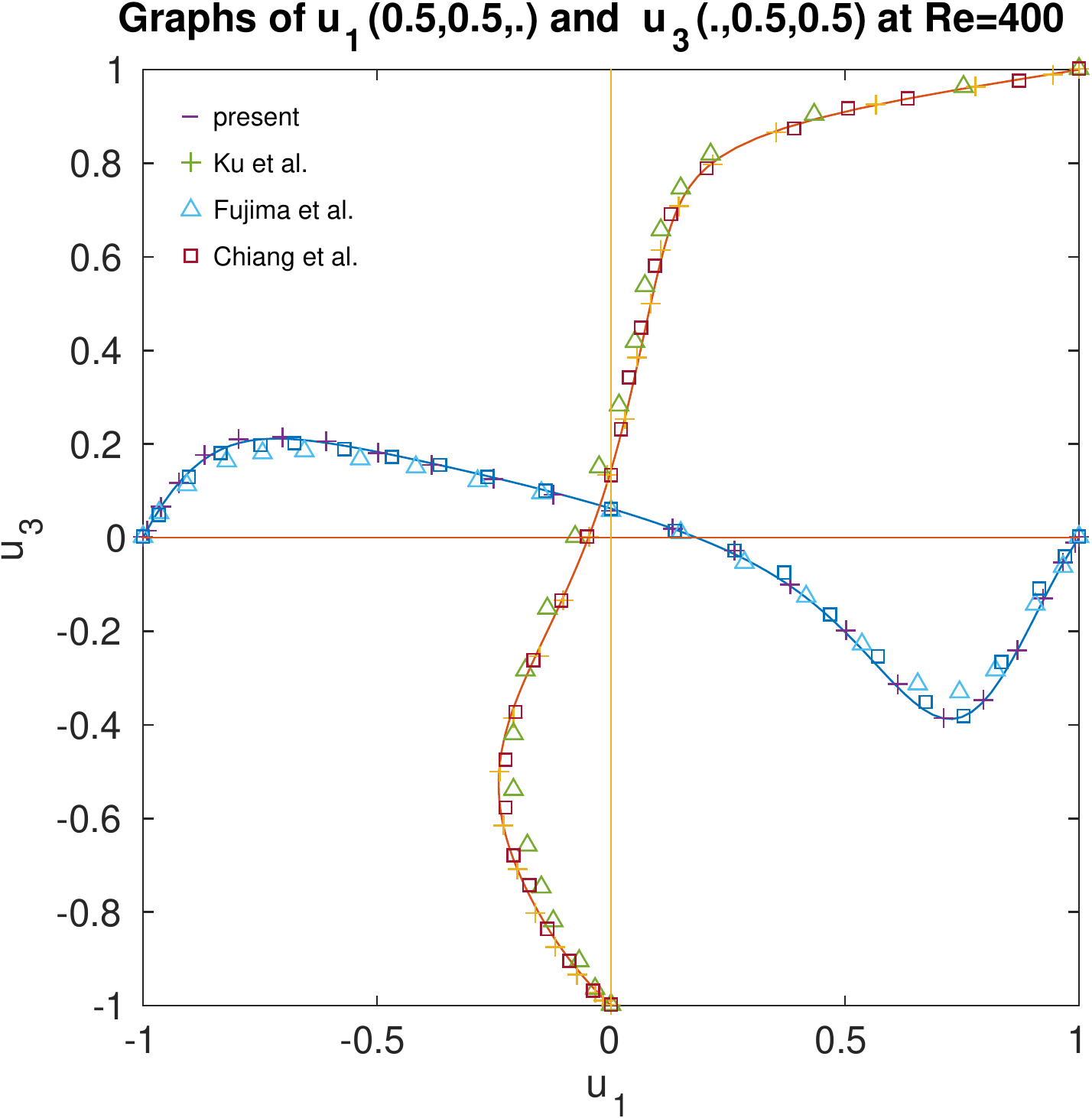} 
\includegraphics[width=2.25in]{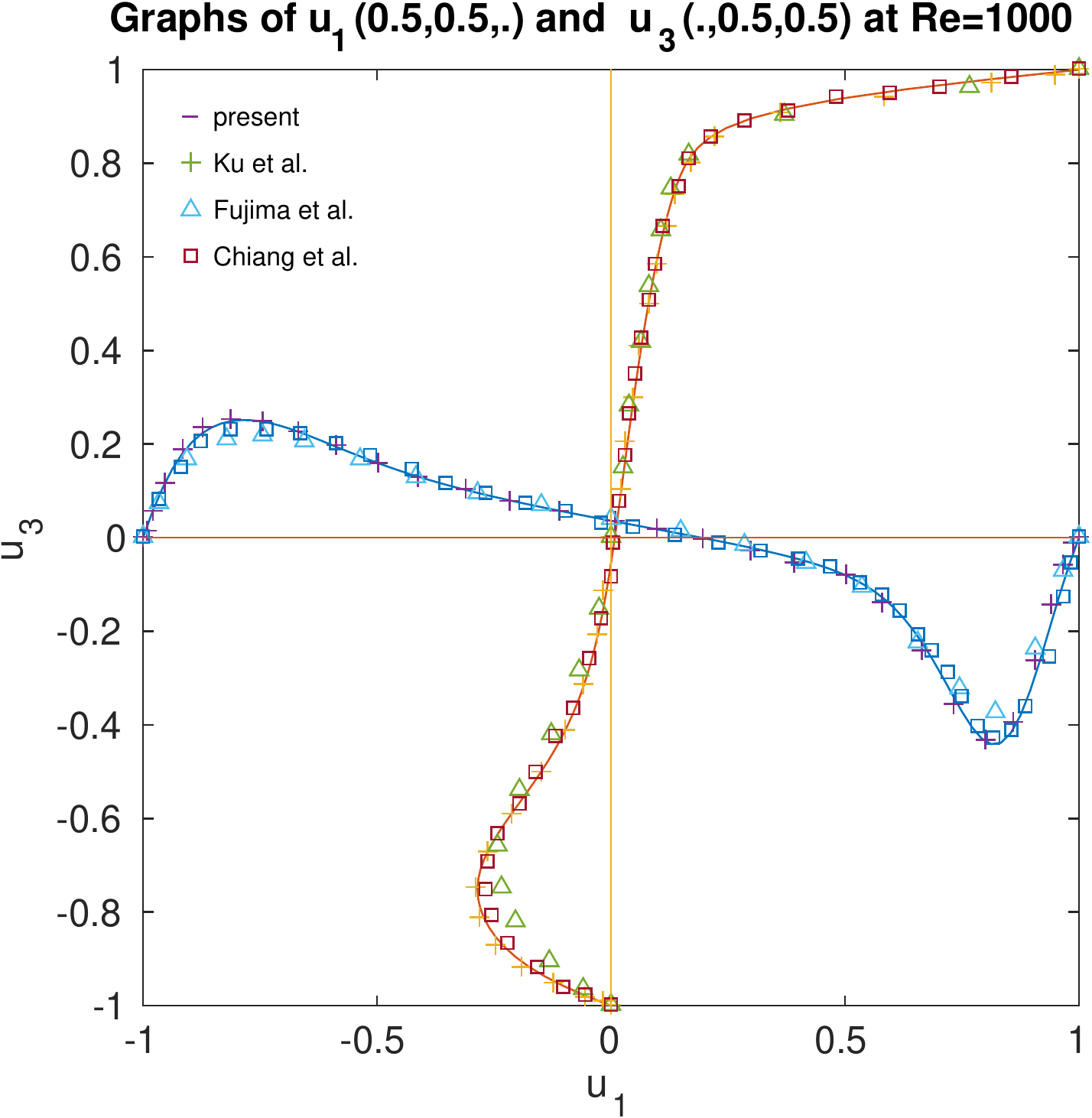} 
\end{center}
\caption{Comparisons of the numerical results obtained for $h=1/60$ (top) and  1/96 (bottom) at Re=400 (left) and 1000 (right).}\label{fig2}
\begin{center}
\leavevmode
\includegraphics[width=2.25in]{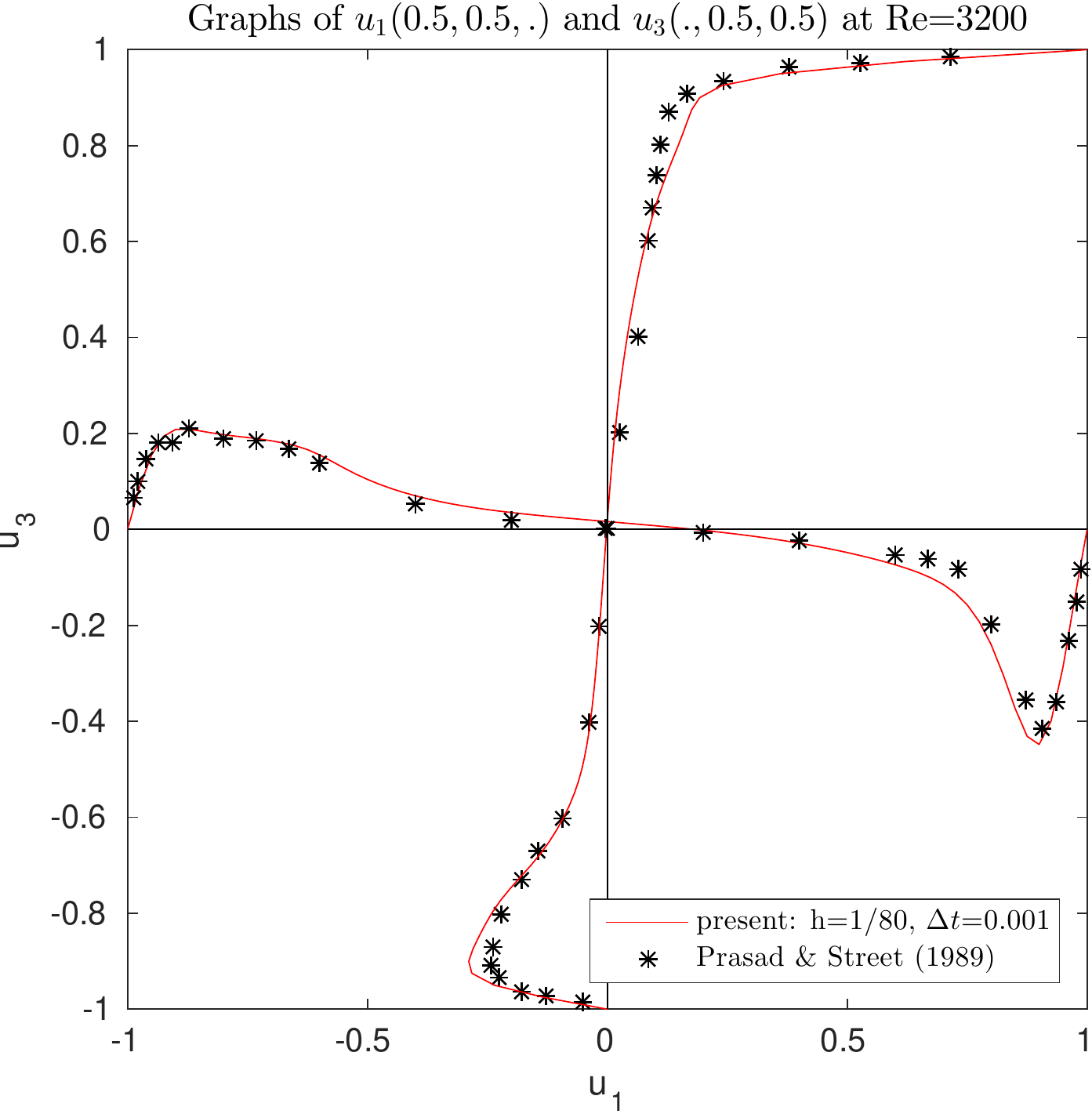}  \hskip 2pt
\includegraphics[width=2.25in]{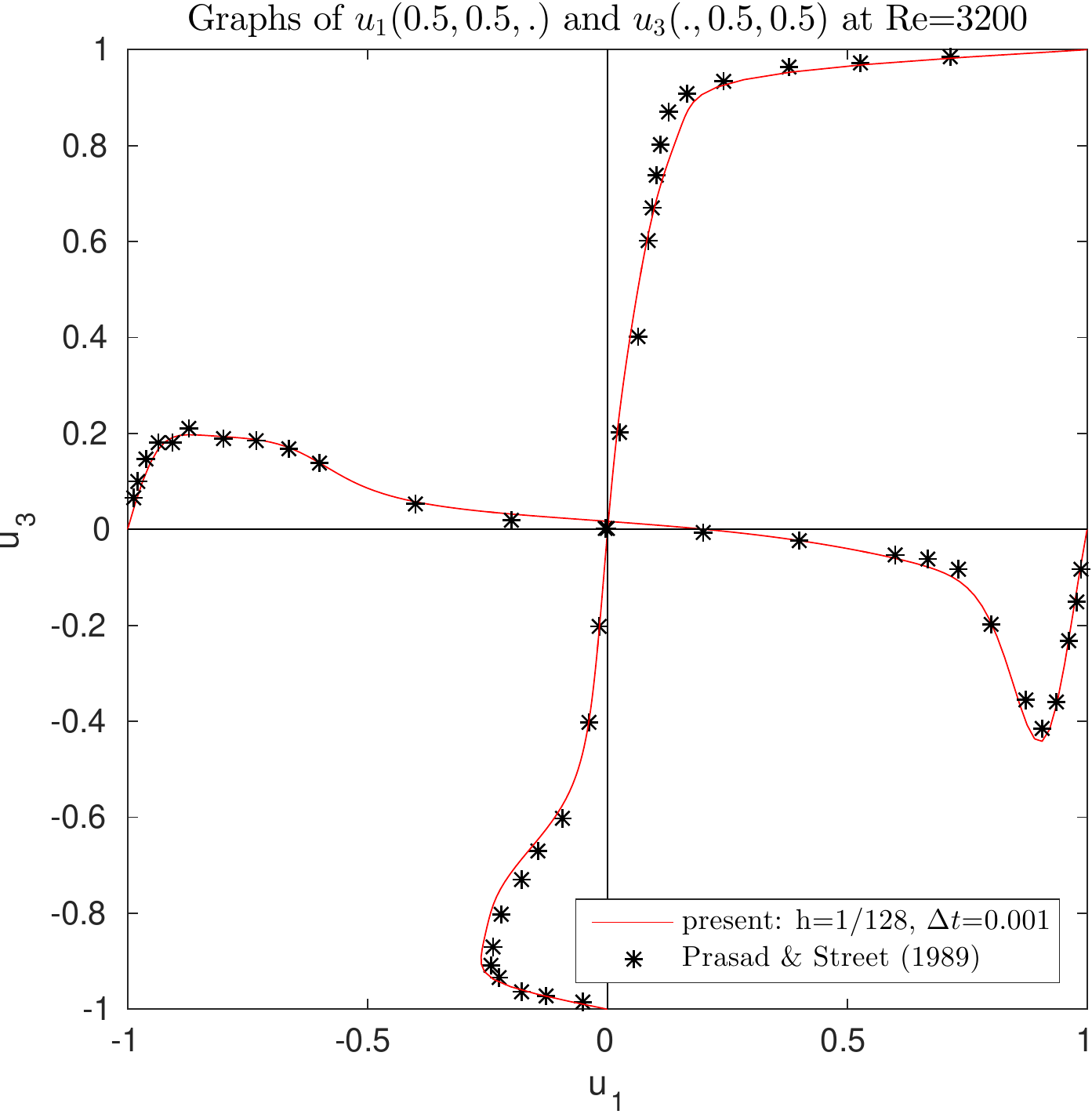}  
\end{center}
\caption{Comparisons of the numerical results at Re$=$3200  obtained for $h=1/80$  and $220 \le t \le 400$ 
(left) and for $h=1/128$ and $158 \le t \le 278$ (right).}\label{fig4}
\end{figure}

\begin{figure}[!t]
\begin{center}
\leavevmode 
\includegraphics[width=0.4true in]{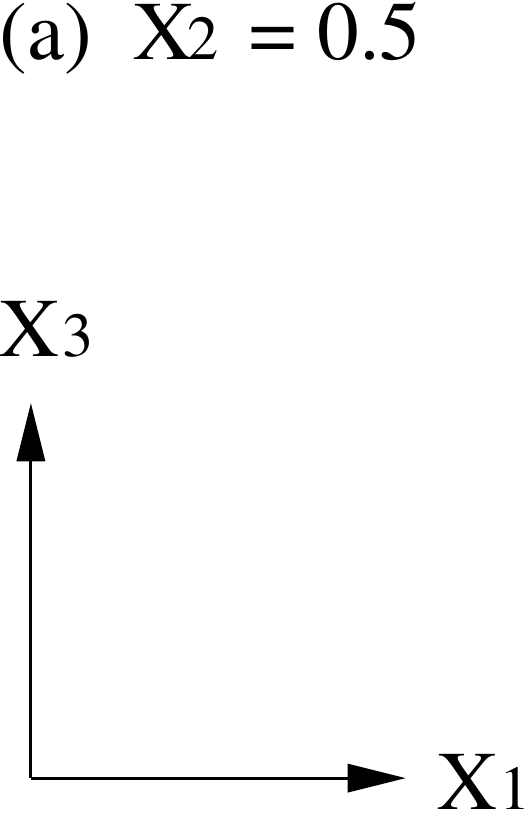}  
\includegraphics[width=1.8true in]{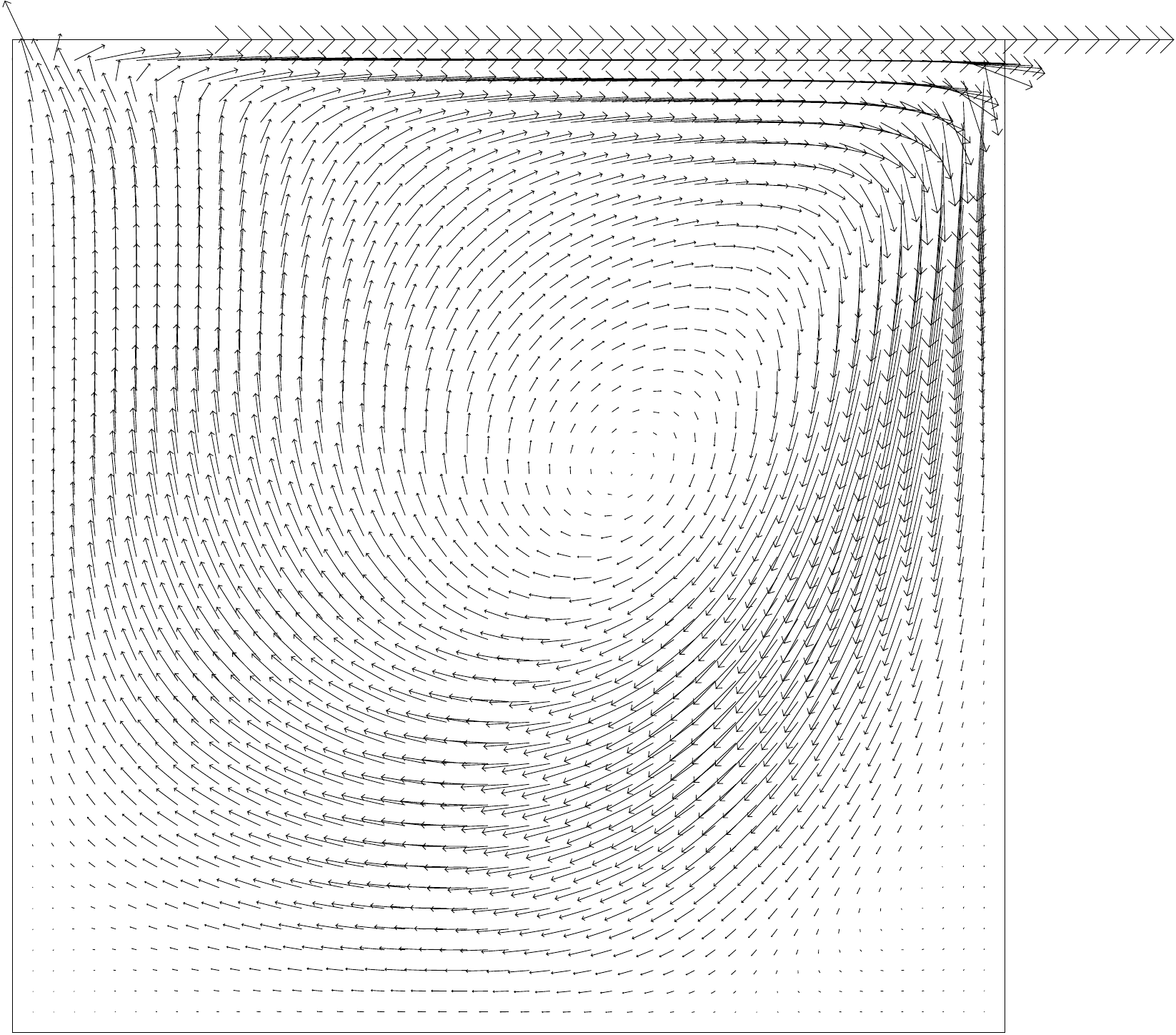}  \hskip -18pt
{\includegraphics[width=0.4in]{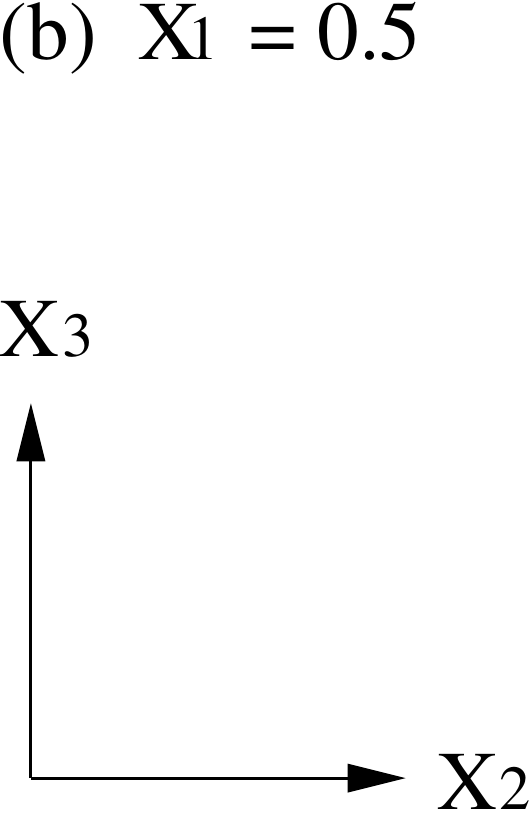} }
\includegraphics[width=1.5in]{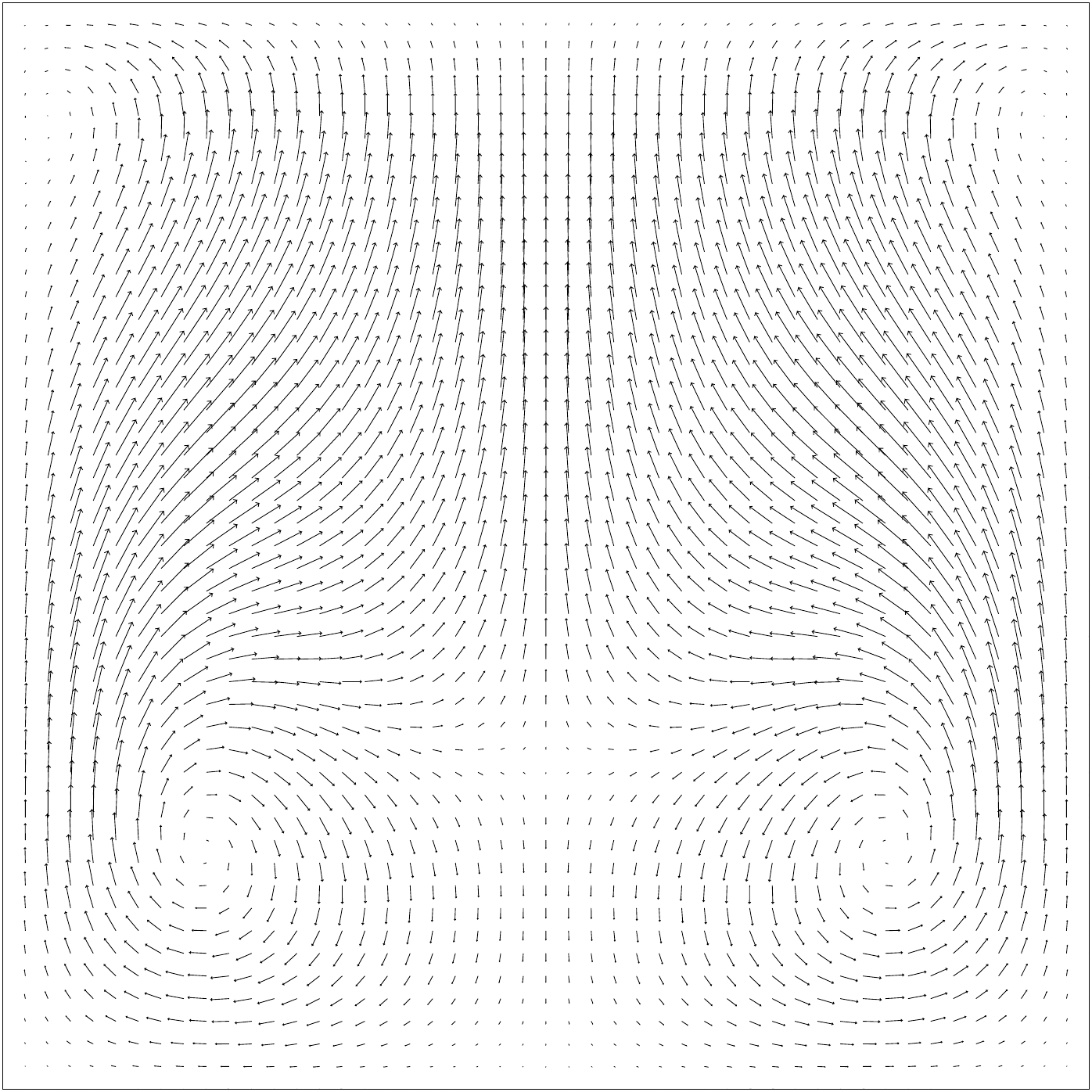} 
\includegraphics[width=0.4in]{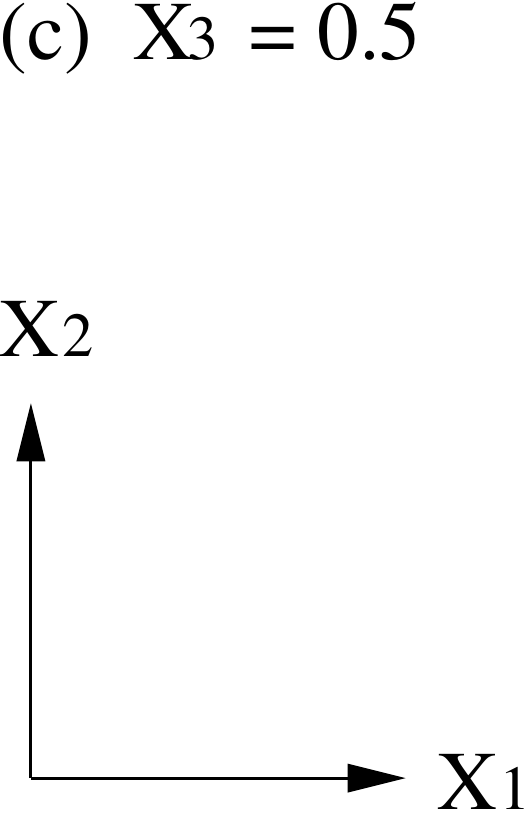}  
\includegraphics[width=1.5in]{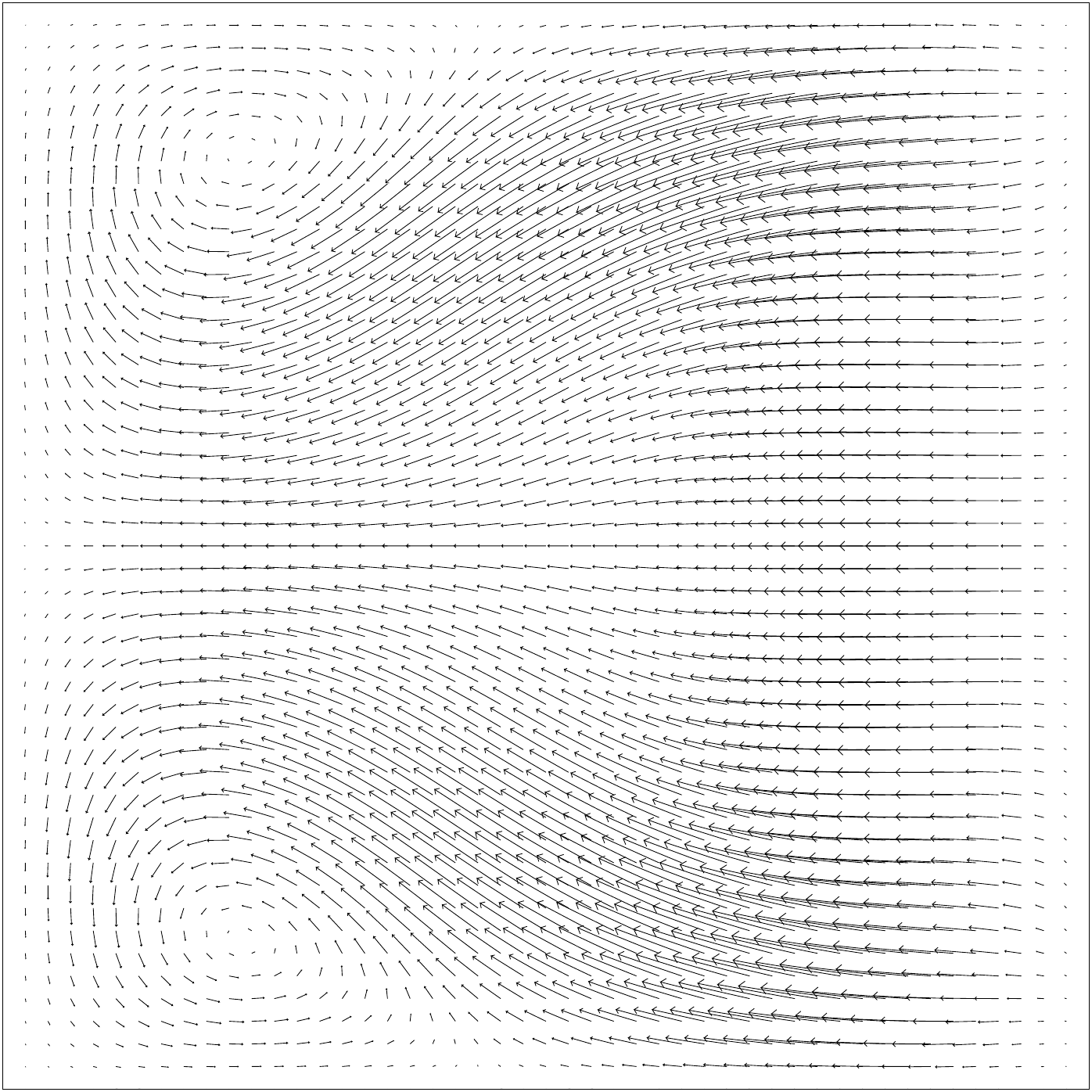} 
\end{center}
\begin{center}
\leavevmode 
\includegraphics[width=0.4true in]{cord-a.pdf}
\includegraphics[width=1.8true in]{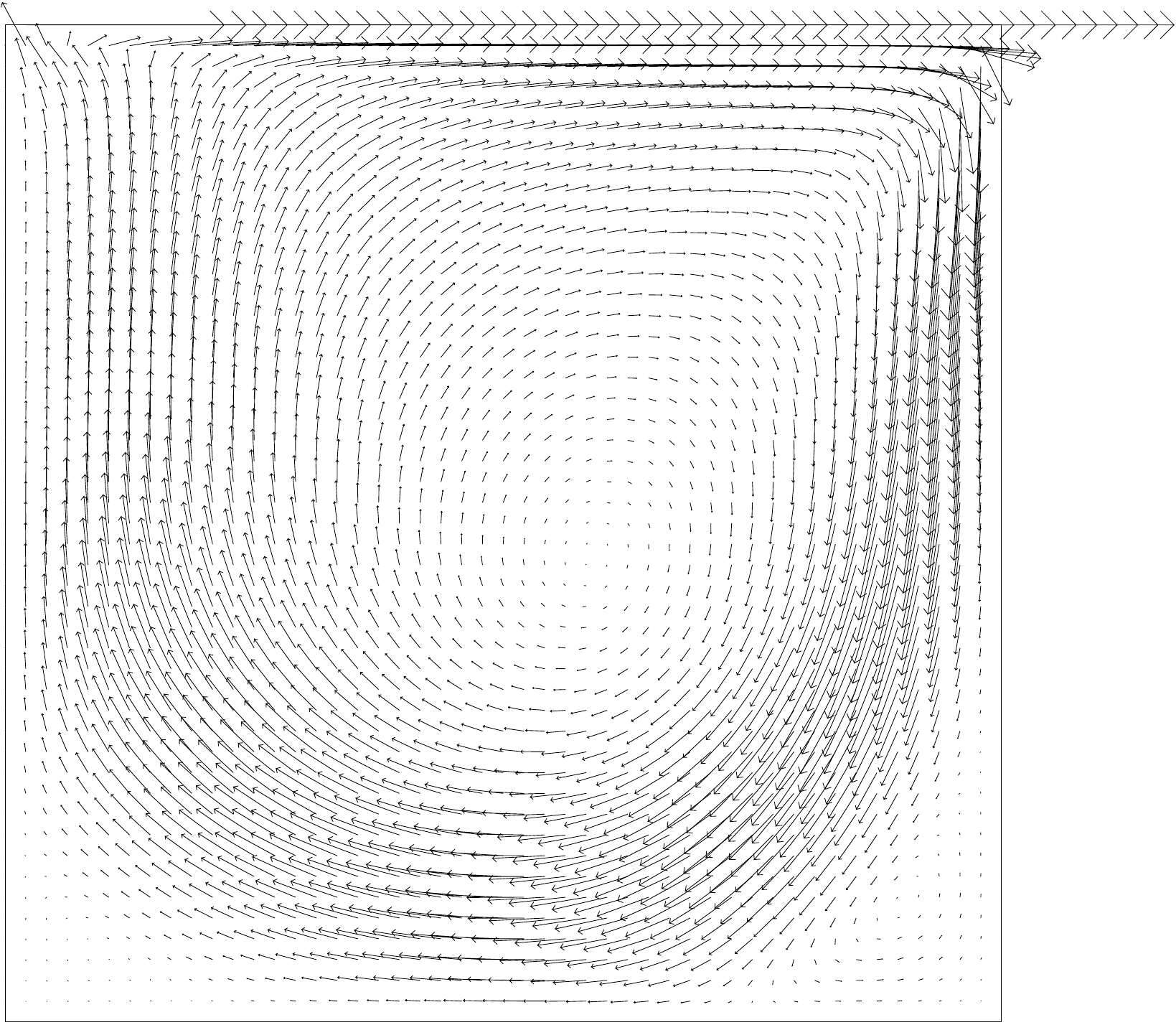}  \hskip -18pt
{\includegraphics[width=0.4in]{cord-b.pdf} }
\includegraphics[width=1.5in]{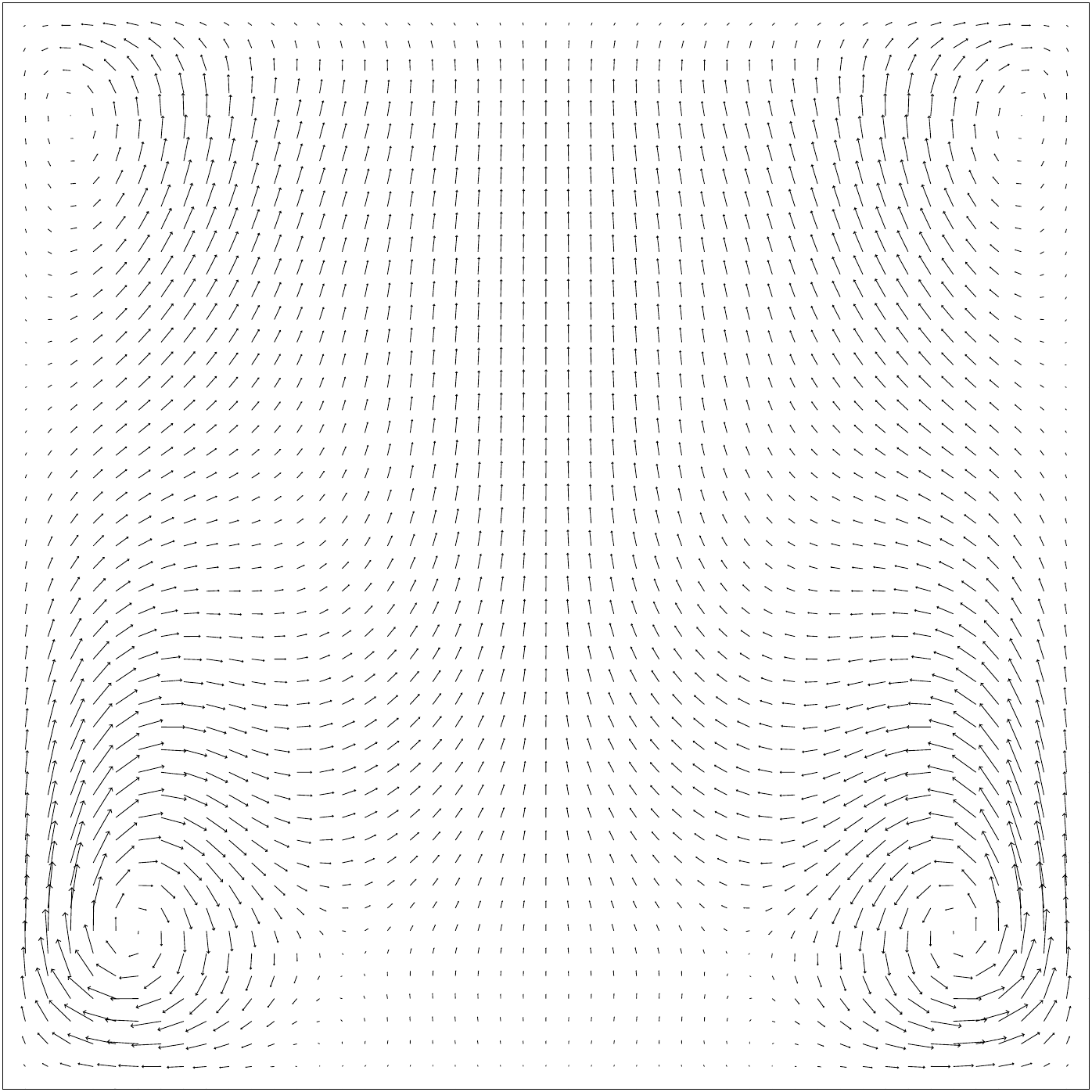} 
\includegraphics[width=0.4in]{cord-c.pdf}  
\includegraphics[width=1.5in]{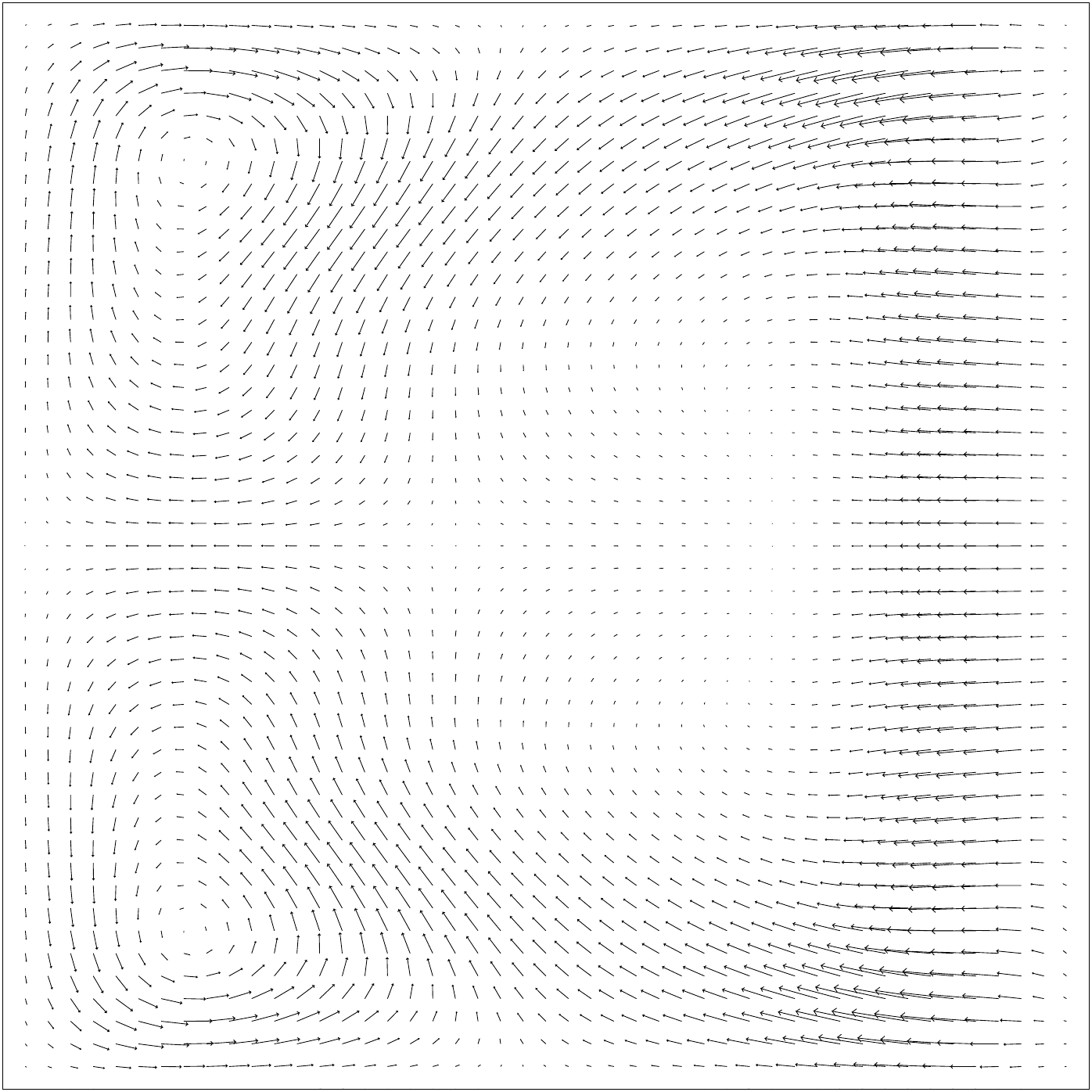}
\end{center}
\caption{Velocity vector of steady flow for Re=400 (top) and 1000 (bottom) projected on 
the planes  $x_2=0.5$ (left),  $x_1=0.5$ (middle), and $x_3=0.5$ (right) for  $h=1/96$ 
and   $\triangle t$=0.001. The vector scale in the middle and right plots 
is twice that  of the actual size.}\label{fig3}
\end{figure}
\begin{figure}[!t]
\begin{center}
\includegraphics[width=0.45in]{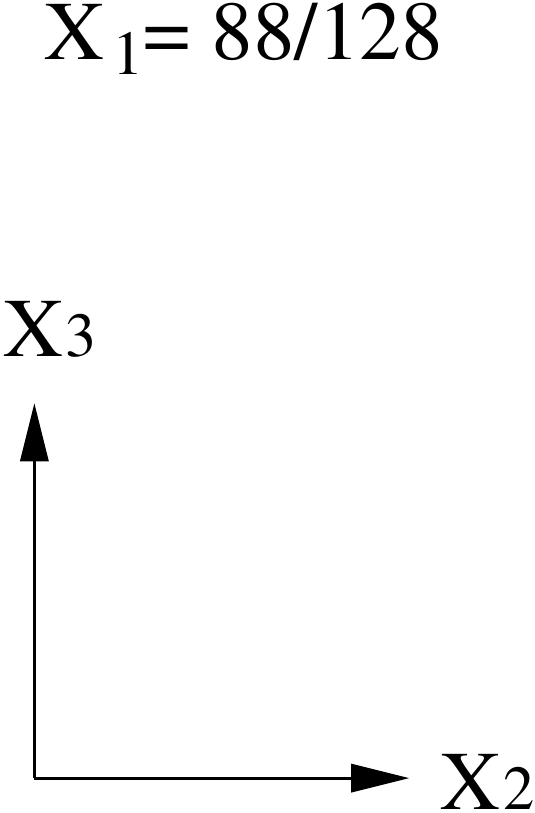} 
\includegraphics[width=2.8in]{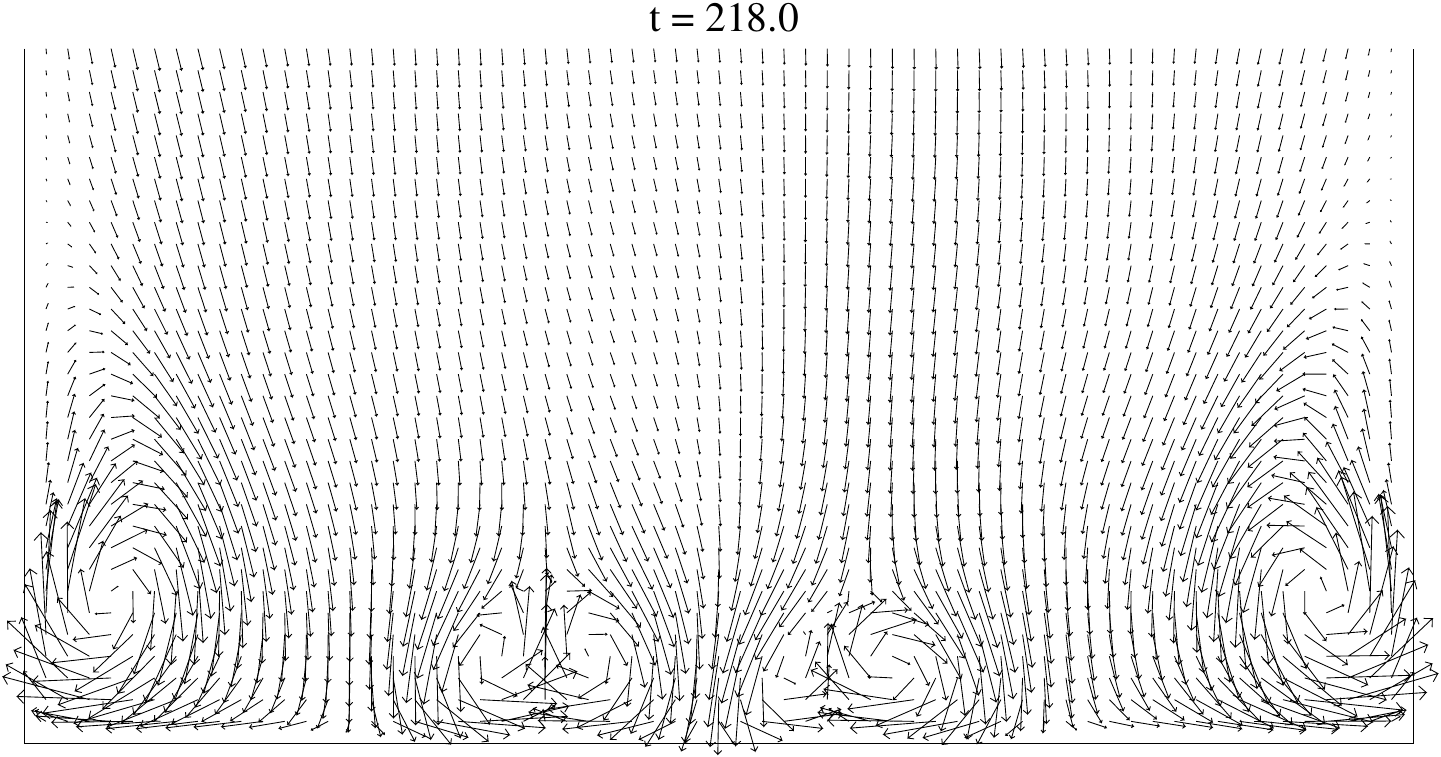}  \hskip 10pt 
\includegraphics[width=2.8in]{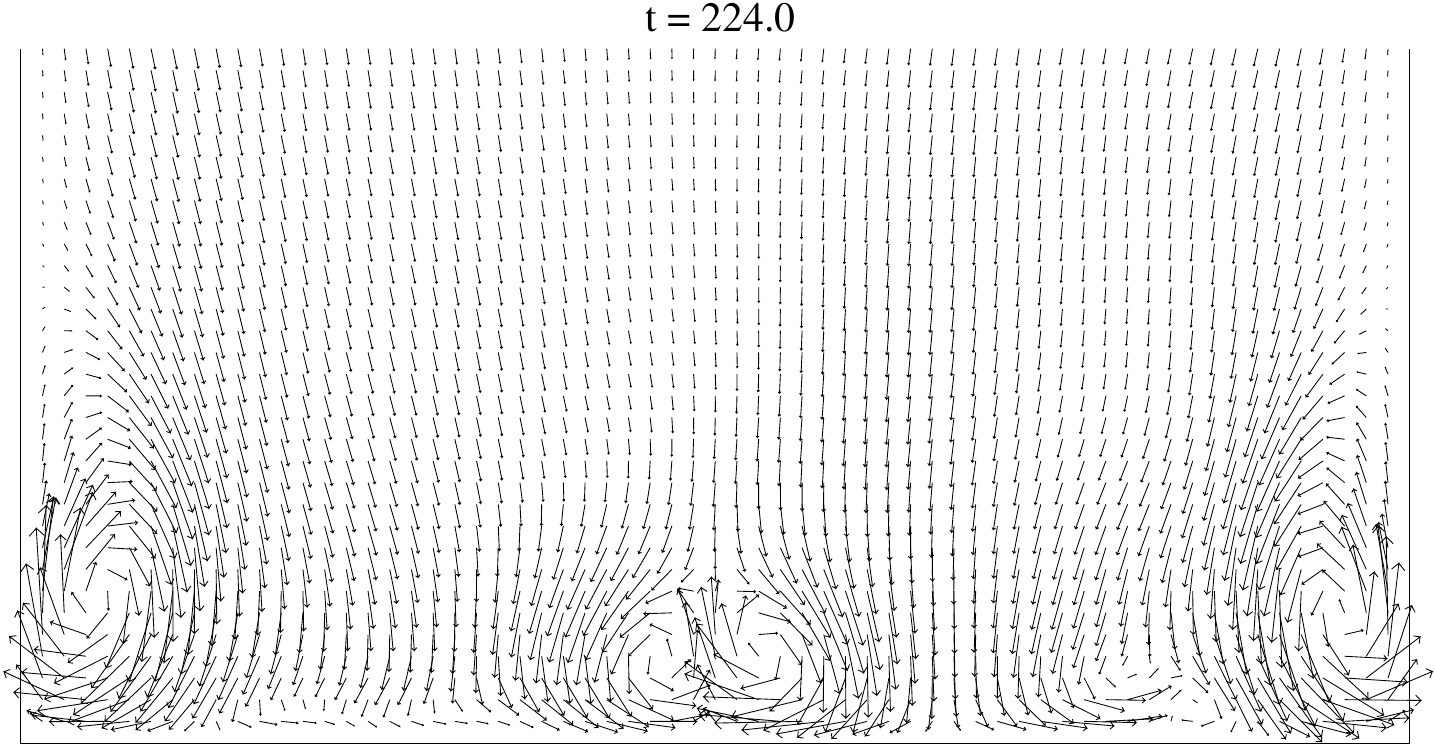}  \\
\hskip 0.46in         
\includegraphics[width=2.8in]{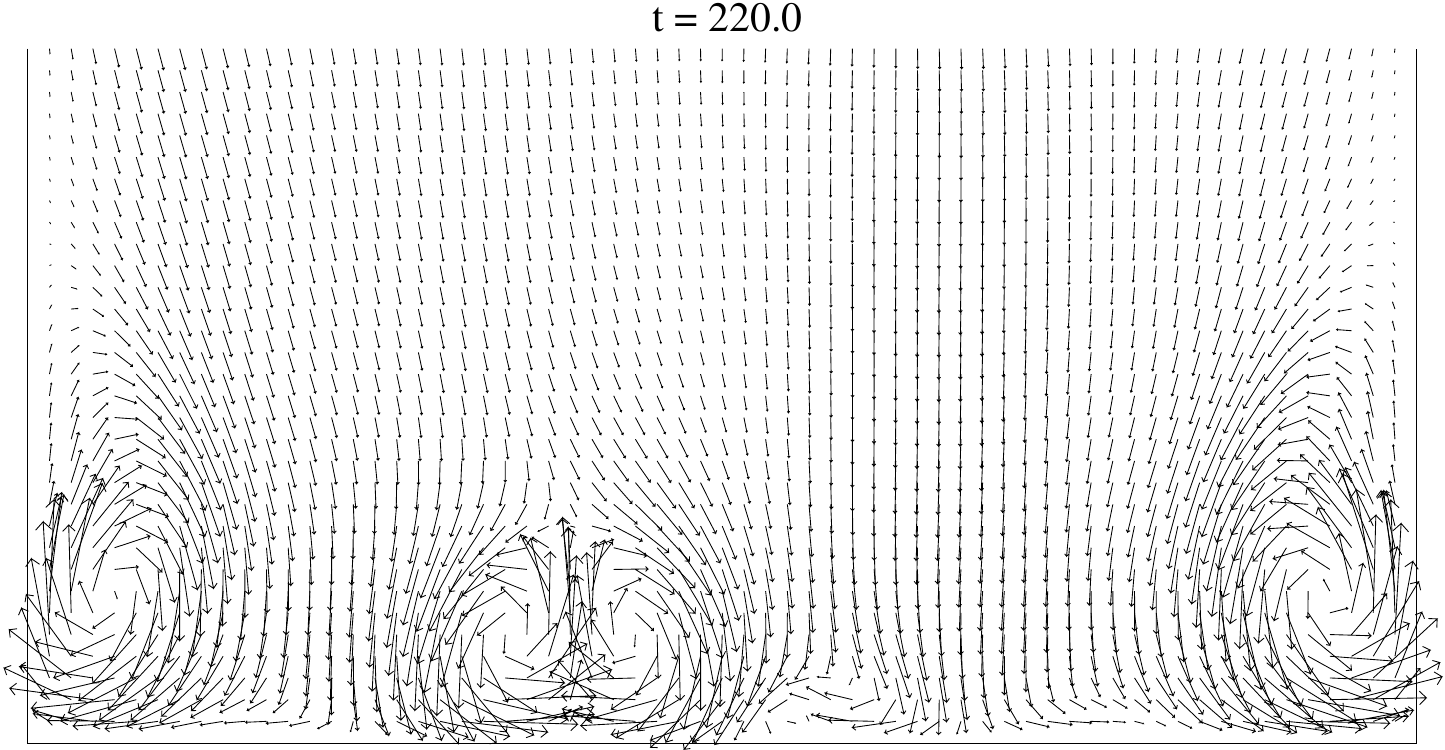}   \hskip 10pt \includegraphics[width=2.8in]{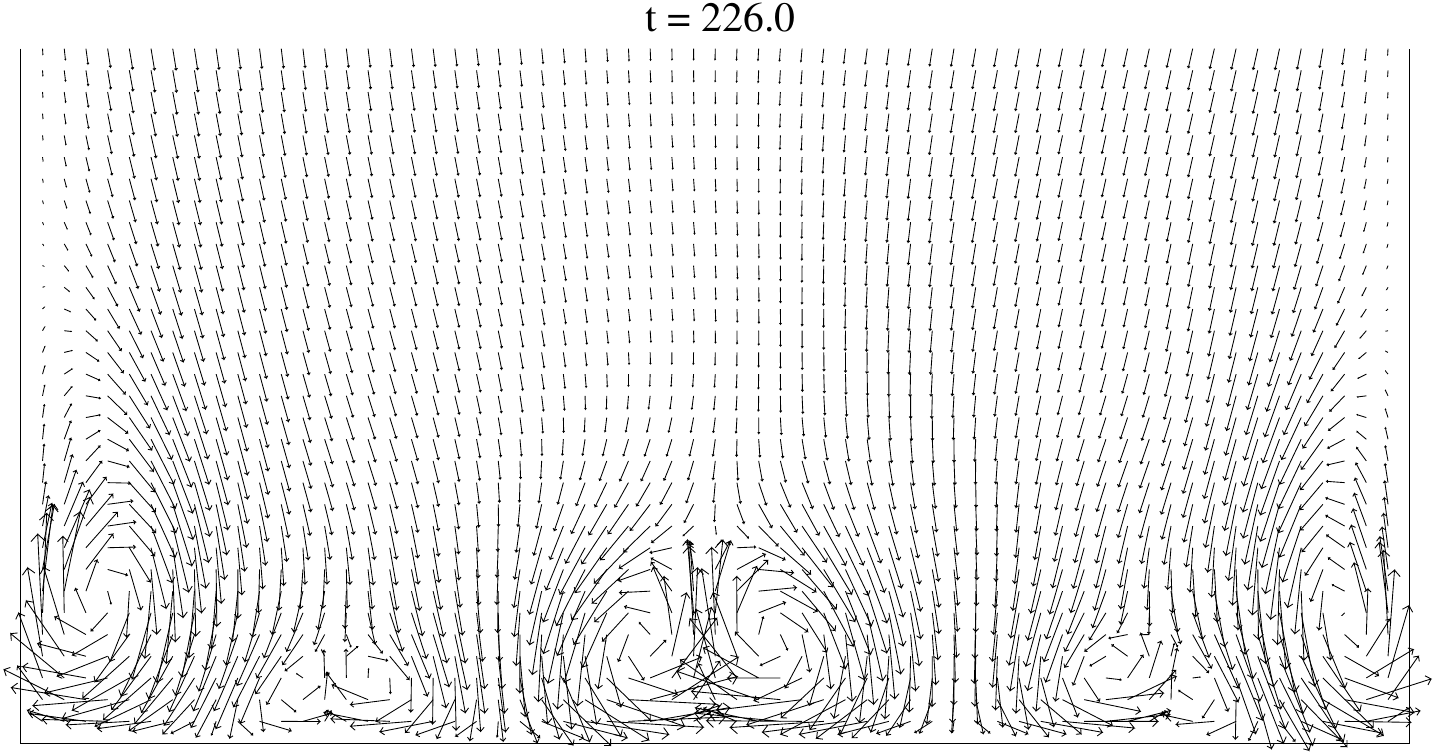} \\
\hskip 0.46in         
\includegraphics[width=2.8in]{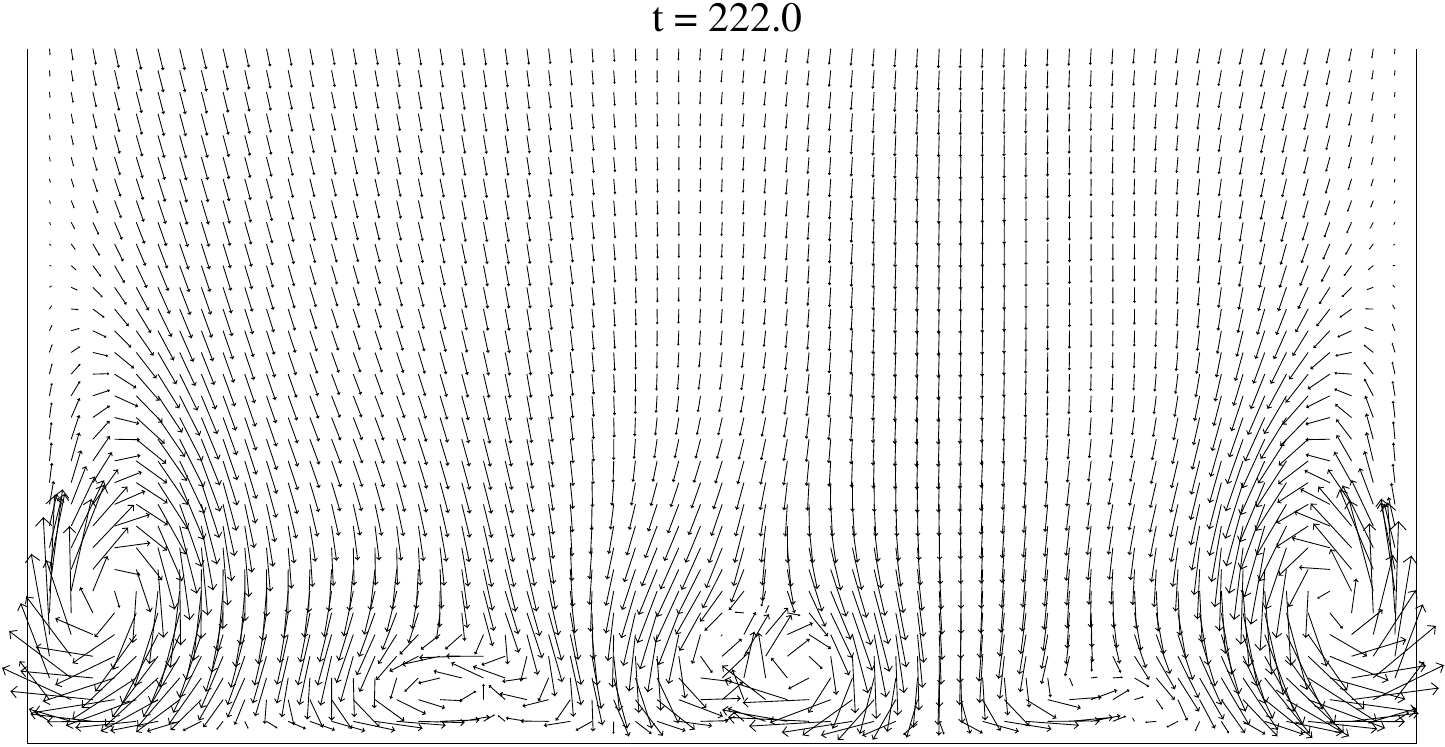}   \hskip 10pt \includegraphics[width=2.8in]{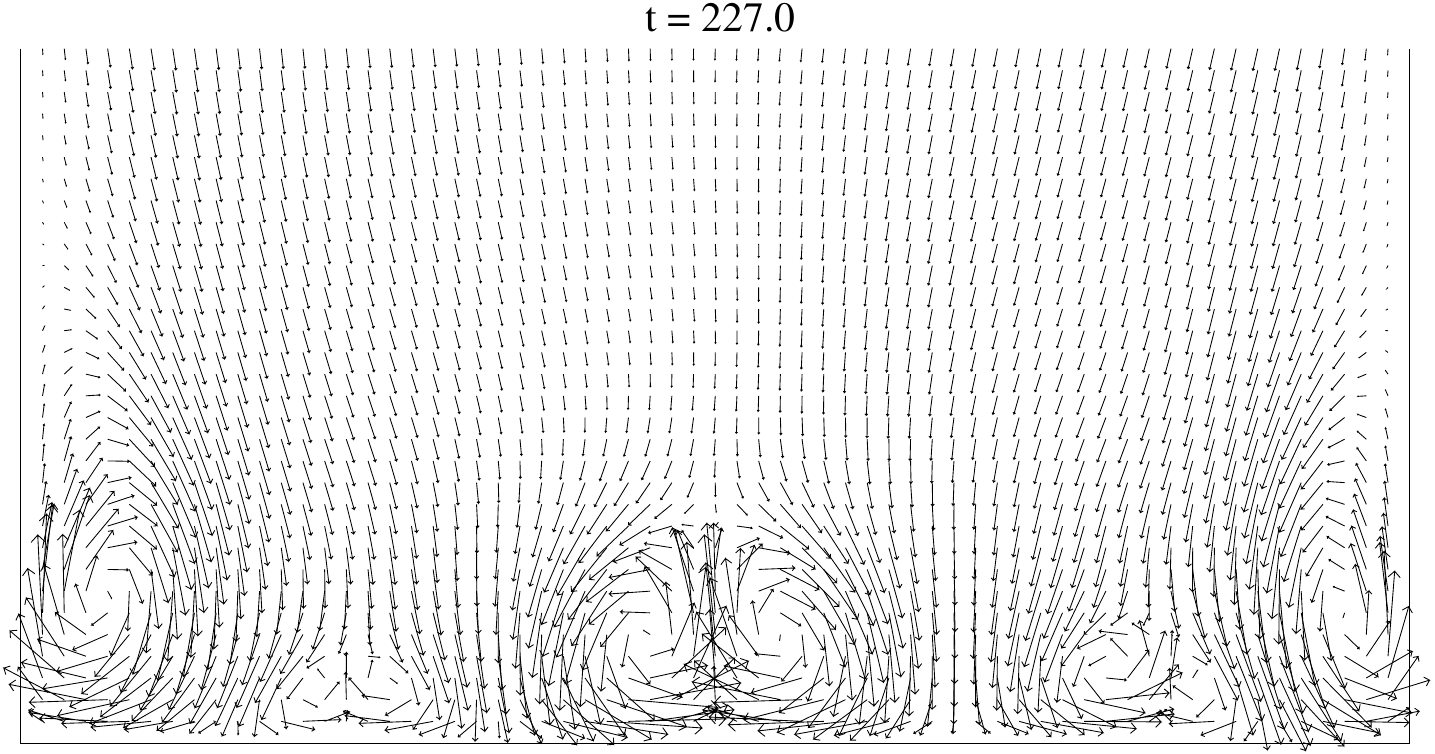}
\end{center} 
\caption{Projected velocity  vectors  on $x_1=88/128$ for Re$=3200$ at different instants of time shows
interaction between TGL vortices and corner vortices. The scale of vectors is twice that 
of the actual one.}\label{fig4a}
\end{figure}

\subsection{Validation}

To validate the numerical methodology, we have considered for the velocity mesh size the values $h=1/60$ 
and 1/96  associated with the time step $\triangle t$=0.001. For Re=400 and 1000, 
the results reported in Figure \ref{fig2} show a very good agreement  with 
those obtained in  \cite{1136}, \cite{1137} and \cite{1138}.
Velocity vectors of steady flows obtained for Re=400 and 1000 are shown in Figure \ref{fig3}. 
Those velocity field vectors are projected orthogonally to the three planes, $x_2 = 0.5$, 
$x_1 = 0.5$, and $x_3 = 0.5$, and the length of the vectors has been doubled in the 
two later planes to observe the flow more clearly.  The plots show that the center of the primary vortex 
moves down as the Reynolds increases from 400 to 1000 and secondary vortices appear in two lower corners, 
which is similar, in some sense, to what happens for the  two-dimensional wall-driven cavity flow. At 
$x_1 = 0.5$, a pair of secondary vortices moves  toward the lower corners when the Reynolds increases. 
Also another pair of vortices appears at the top corners. At $x_3 = 0.5$, there is a pair of secondary 
vortices near the upstream wall.

Another validation has been done at Re=3200 for   $h=1/80$ and 1/128 and   $\triangle t$=0.001. 
For Re=3200 in a cubic cavity, experiments  \cite{Prasad1989} indicates that there are usually two pairs of TGL 
vortices. Moreover, these vortices are not stationary; they meander to and fro over the bottom wall closer to the end wall
in the spanwise direction. In \cite{Perng1989} and \cite{Teixeira1997}, the number of pairs of TGL vortices obtained 
numerically varies between two and three.  Figure \ref{fig4} shows  the  averaged speed  profiles 
$u_1(0.5, 0.5, \cdot)$ and $u_1(\cdot,0.5, 0.5)$ are in a good agreement with the experimental values obtained 
in \cite{Prasad1989}. Our simulation results show two to three pairs of TGL vortices at Re=3200  in Figure \ref{fig4a}.
For $218 \le t \le 220$, two pairs  of TGL vortices interacts with each other and corner vortices and one pair 
of TGL vortices is diminished later. Then another two pairs of TGL vortices gradually show up for $222\le t \le 227$.  

\begin{figure}[tp]
\begin{center}
\leavevmode 
\includegraphics[width=2.75true in]{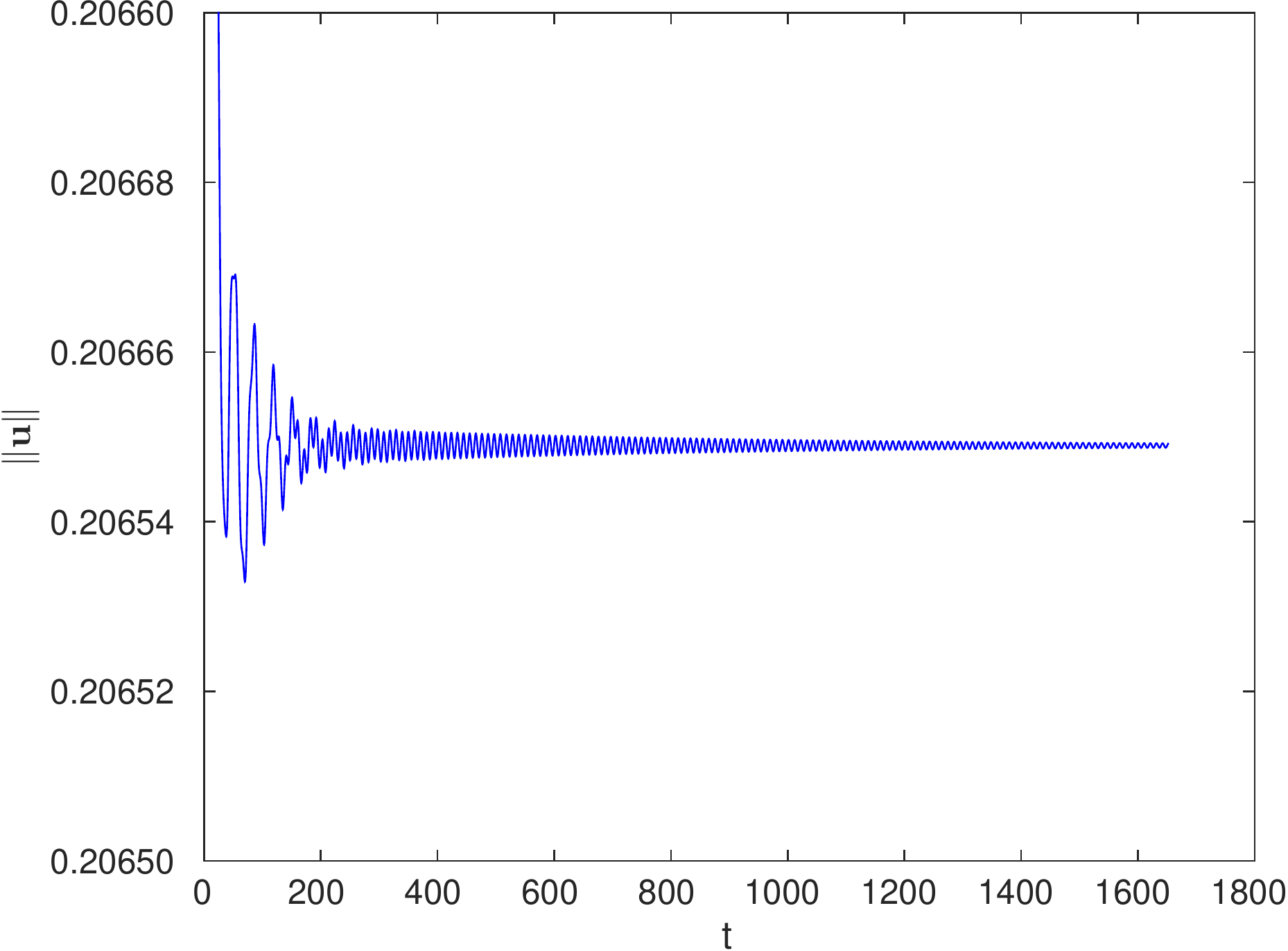} \hskip 10pt
\includegraphics[width=2.75true in]{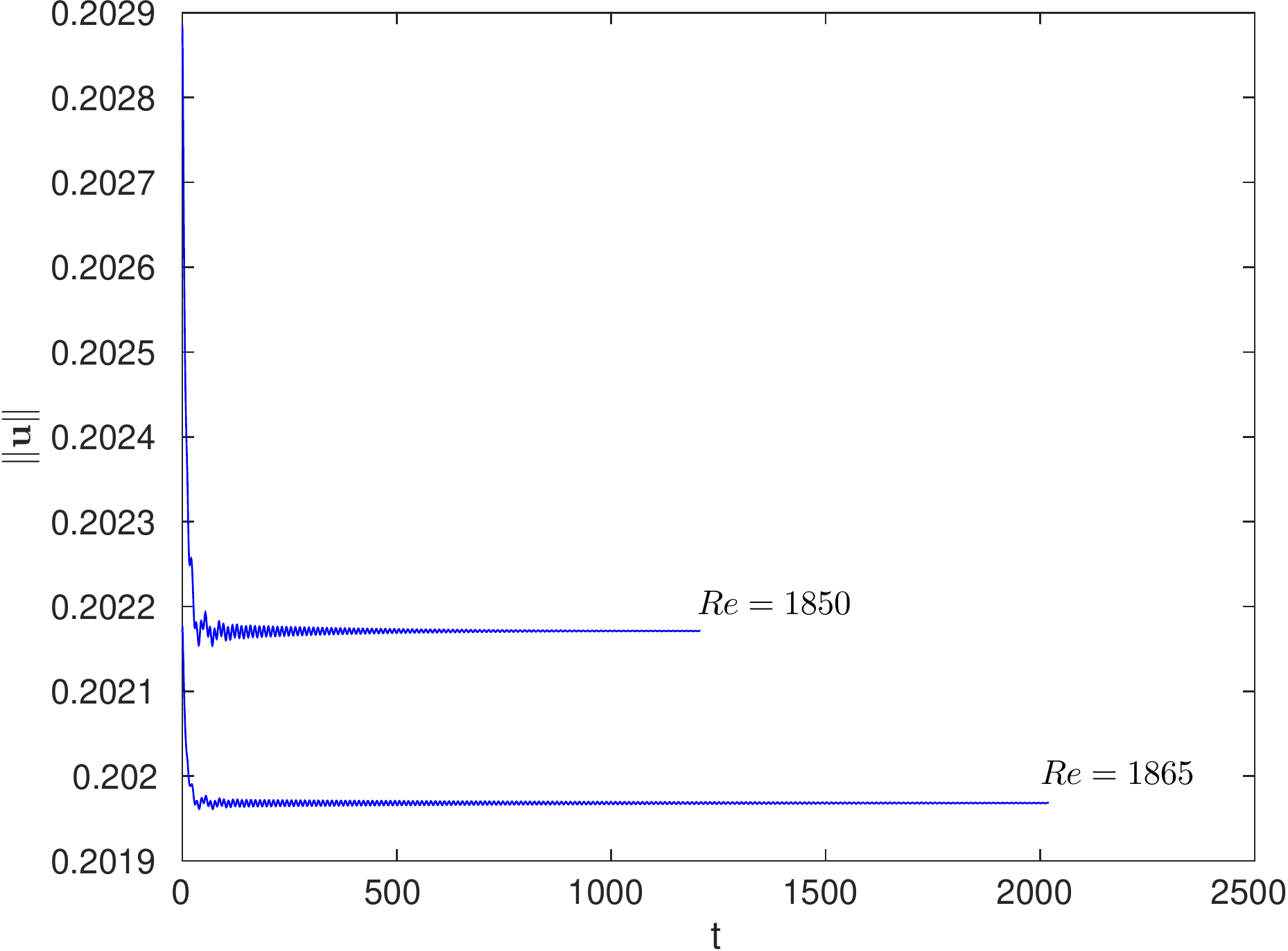} \\
\includegraphics[width=2.75true in]{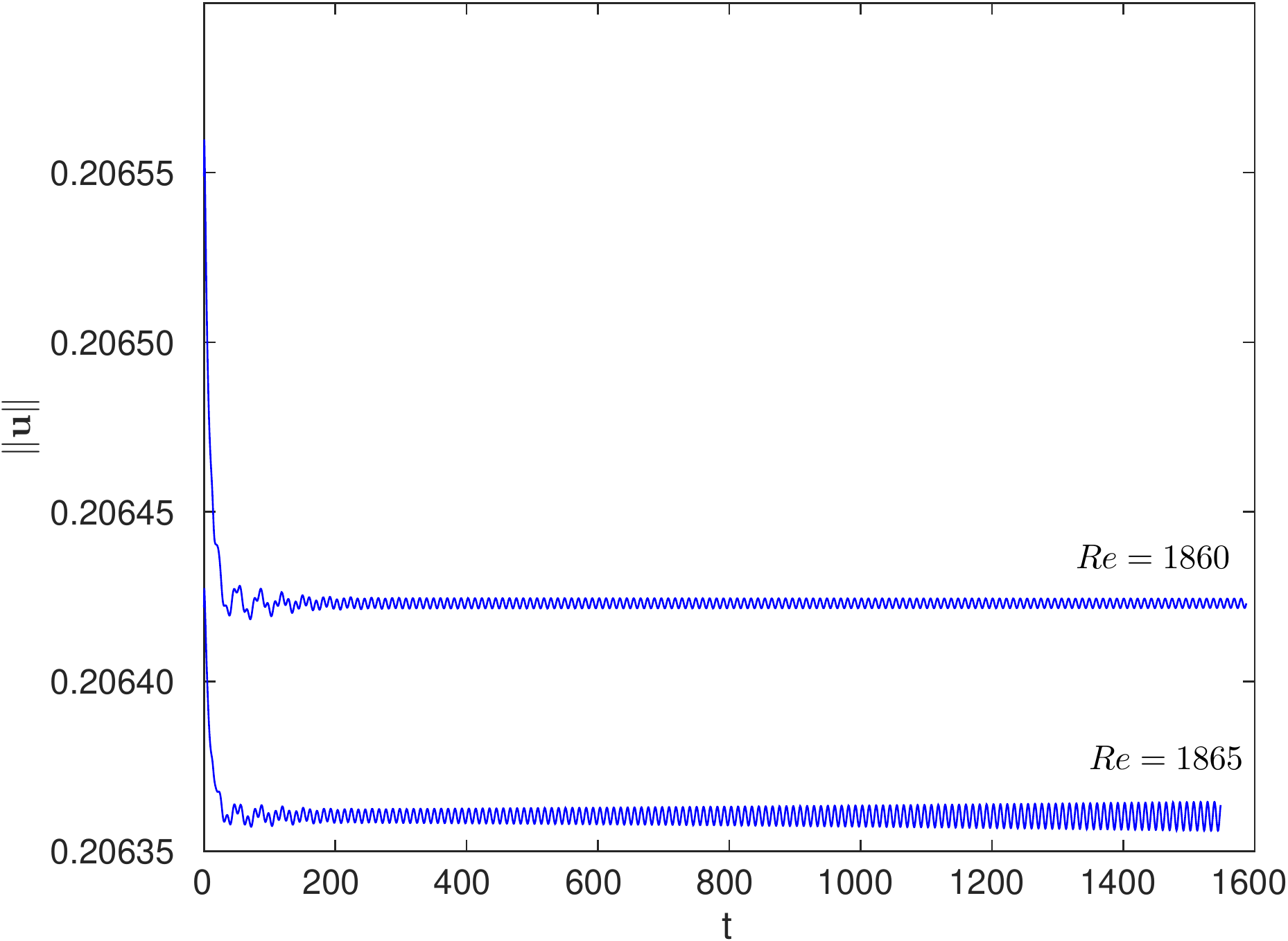} \hskip 10pt
\includegraphics[width=2.75true in]{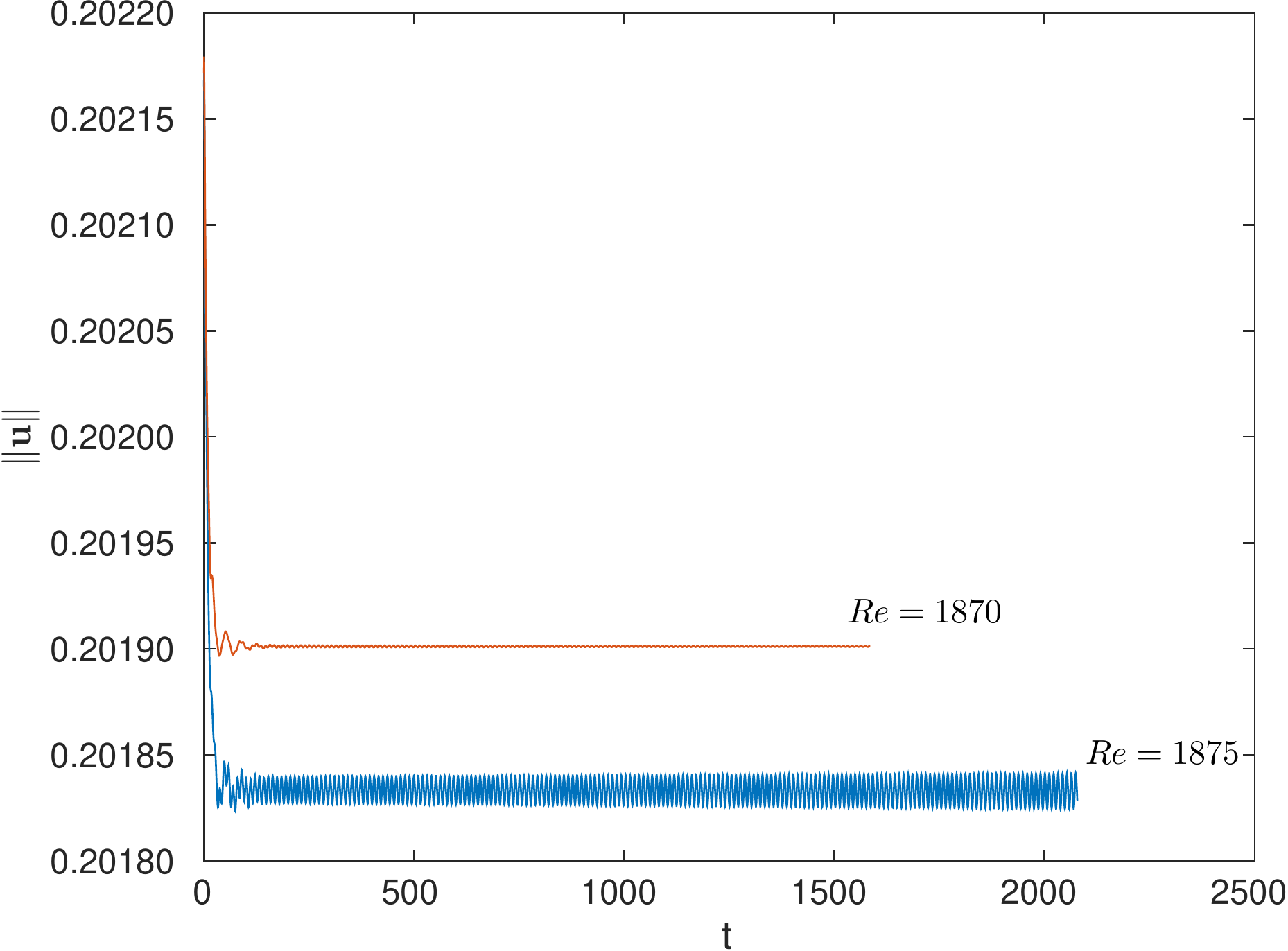} \\
\hskip -10pt\includegraphics[width=2.75true in]{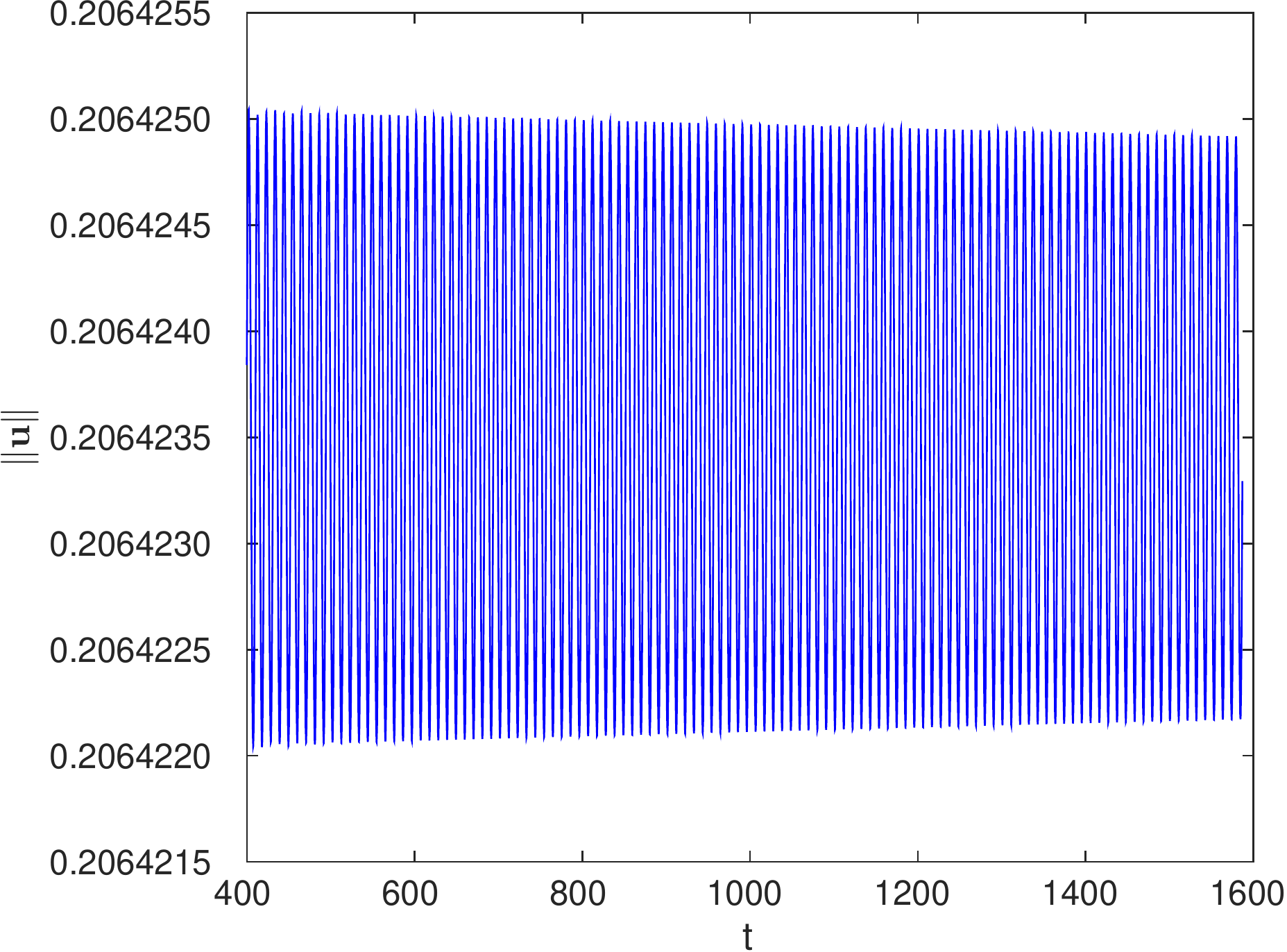} \hskip 11pt
\includegraphics[width=2.75true in]{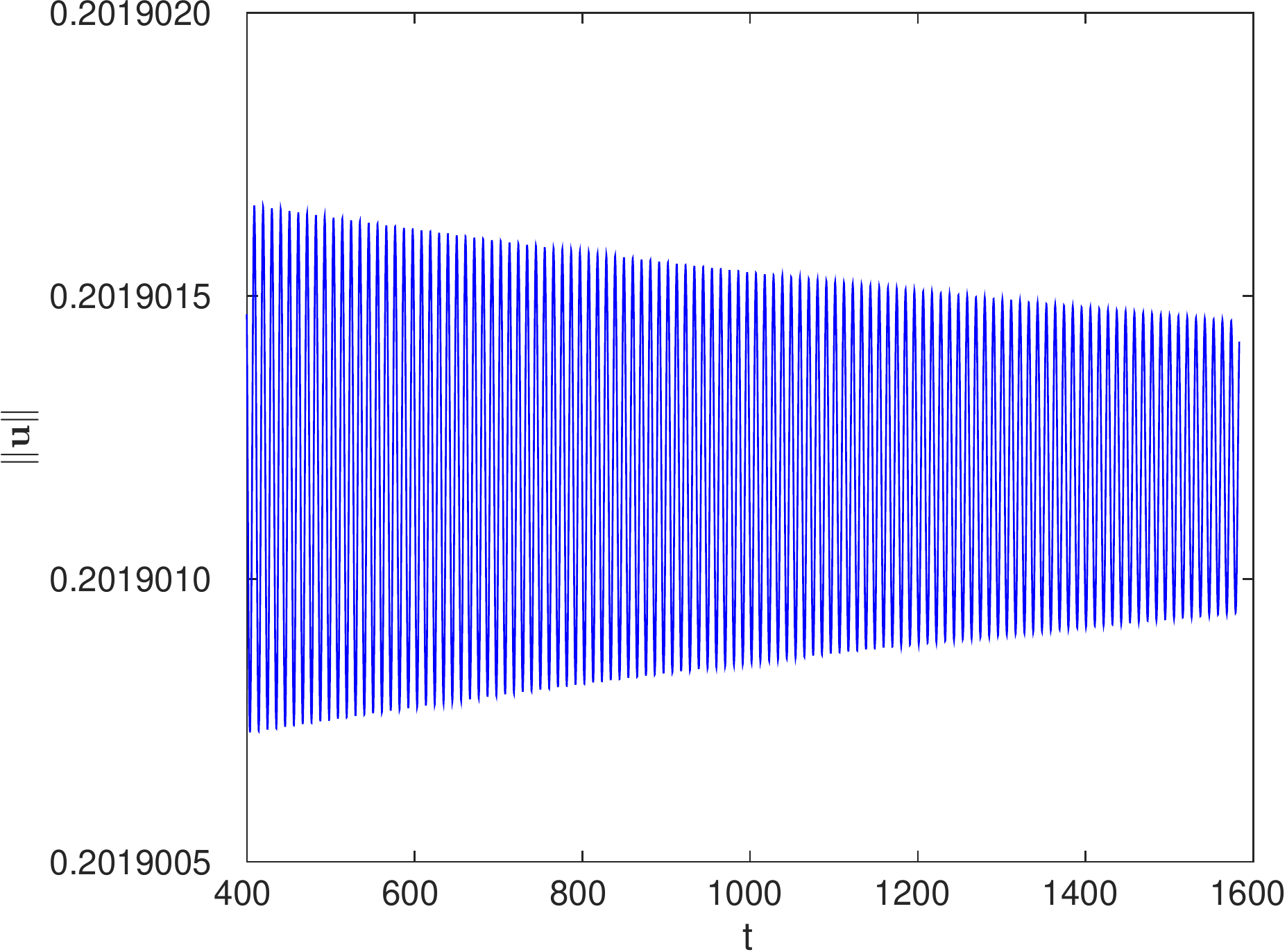}
\end{center}
\caption{Histories of the flow field $L^2$-norm for $h=1/60$ (left) and 1/96 (right): 
(a) Re$=1850$  (top left),  Re$=1860$ and 1865  (middle left) and an enlargement of the case 
of Re$=1860$ (bottom left); (b) Re$=1850$ and 1865 (top right), Re$=1870$ and 1875  (middle right) and 
an enlargement of the case of Re$=1870$ (bottom right).}\label{fig5}
\end{figure}

\subsection{Transition from steady flow to oscillatory flow}

A documented feature of three-dimensional lid-driven cavity flows is that they may exhibit Taylor-Görtler-like (TGL) 
vortices if Re is sufficiently large. In this section, we focus on  the transition from steady flow to oscillatory flow 
and TGL vortices for the Reynolds number $Re$ just below 2000.  Indeed,   Iwatsu {\it et al.}  \cite{Iwatsu1990}  
obtained a pairs of Taylor-G\"ortler-like (TGL) vortices at Re$=2000$. Also  as predicted in \cite{Feldman2010} 
and \cite{Liberzon2011}, a transition from steady flow to oscillatory flow occurs at Re$_{cr} < 2000$. On the other 
hand, using a global linear stability analysis, Gianetti {\it et al.}  found (ref. \cite{Giannetti2009}) that cubic 
lid-driven cavity flow becomes unstable for Re just  above 2000.  All these results (ours in particular) lead us to 
suspect that the Hopf bifurcation is connected to the TGL vortices at Re slightly below 2000.

To study the transition from steady flow to oscillatory flow, we have analyzed the history of the  $L^2$-norm
of the flow field, $\|\bu^n_h\|$, for different values of Re and of the mesh size $h$. For $h=1/60$,  the flow 
field evolves to steady state for $Re \le 1860$ and the amplitude of the oscillation of the flow field  $L^2$-norm 
decreases also in time. At Re$=1865$,  the steady state criterion is not satisfied and the amplitude of the 
oscillation increases in time as in Figure  \ref{fig5}. Thus we conclude that the critical Reynolds number Re$_{cr}$ for 
the occurrence of the transition is somewhere between 1860 and 1865. The frequencies of the flow field  $L^2$-norm 
oscillation are 0.09449 and 0.09456 for $h=1/60$ at Re=1860 and 1865, respectively. Applying a similar analysis, 
Re$_{cr}$  is in $(1870, 1875)$ for $h=1/96$ and the associated histories of the flow field  $L^2$-norm being shown 
in Figure  \ref{fig5}. The frequencies of the  $L^2$-norm oscillation are 0.095147 and 0.095057 for $h=1/96$ at Re=1870 and 1875, respectively.

\begin{figure}[t]
\begin{center}
\leavevmode 
\includegraphics[width=0.4in]{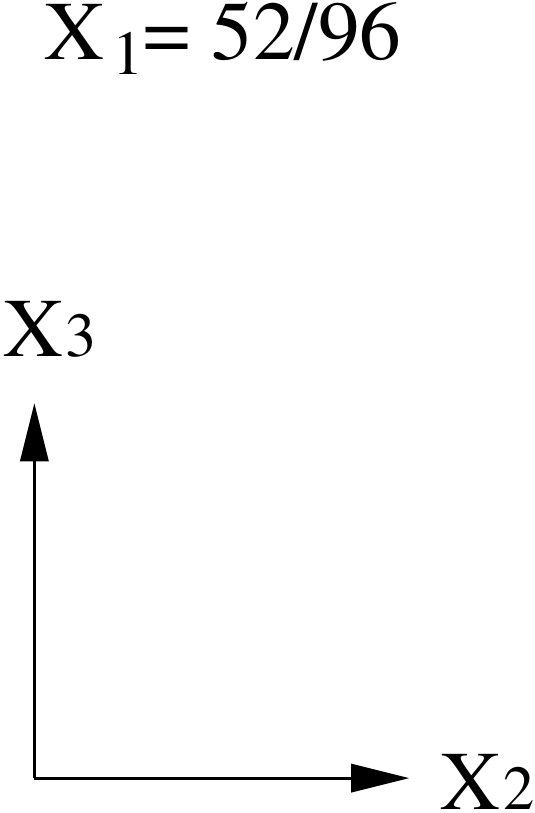} \hskip 2pt
\includegraphics[width=1.8in]{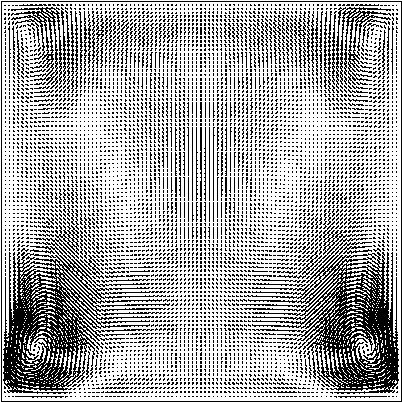} \hskip 2pt
\includegraphics[width=1.8in]{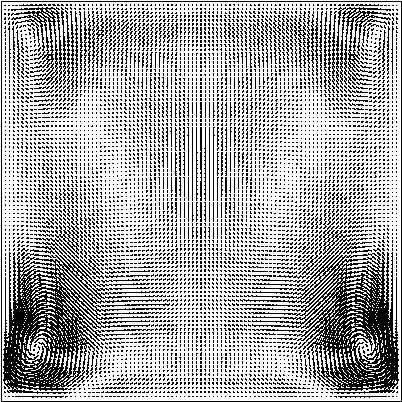}  \hskip 2pt
\includegraphics[width=1.8in]{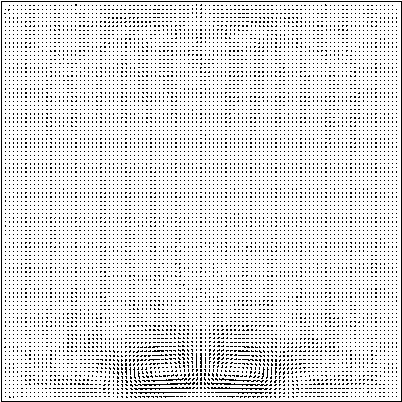}
\end{center}
\begin{center}
\leavevmode 
\includegraphics[width=0.4in]{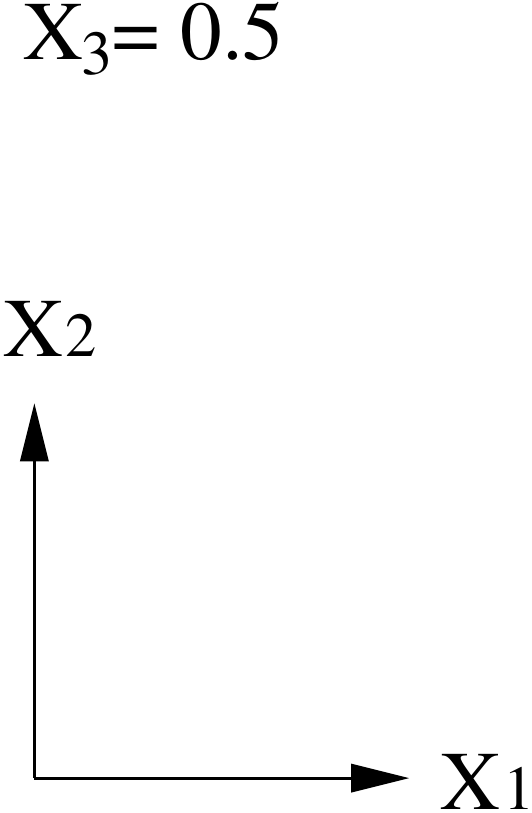} \hskip 2pt
\includegraphics[width=1.8in]{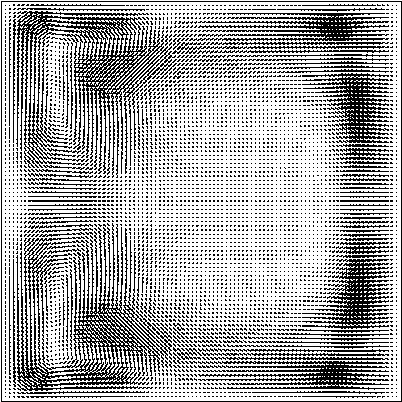} \hskip 2pt
\includegraphics[width=1.8in]{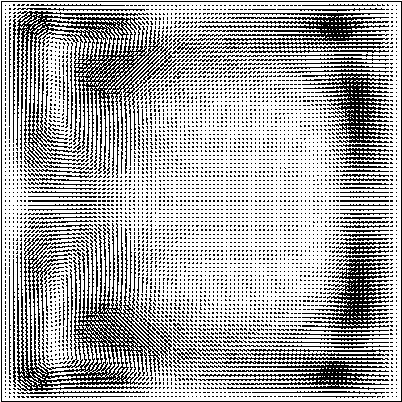} \hskip 2pt
\includegraphics[width=1.8in]{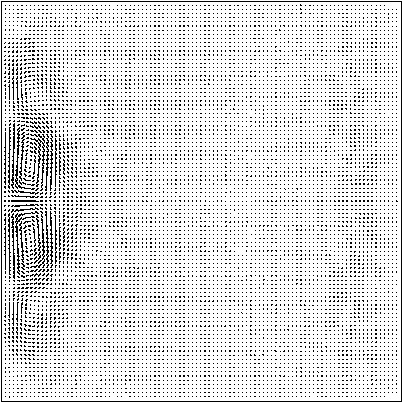}
\end{center}
\caption{Left and middle: Projections (at Re = 1870) of the cavity flow velocity vector fields associated with the peak (left) 
and bottom (middle) of the velocity $L^2$-norm during an oscillation. Right: Projections (at Re = 1870) of the vector 
field obtained by difference of the velocity vector fields associated with the peak and bottom of the velocity $L^2$-norm. 
All the vector fields are projected on the planes $x_1=52/96$ (top) and $x_3=0.5$ (bottom). The vector scale for the field 
obtained by difference (right) is 500 times that of the actual one, while the scale for the two other fields (left
and middle) is twice that of the actual one.}\label{fig6}
\end{figure}

\begin{figure}[t]
\begin{center}
\includegraphics[width=0.4in]{cord-b1.pdf} \hskip 2pt
\includegraphics[width=1.8in]{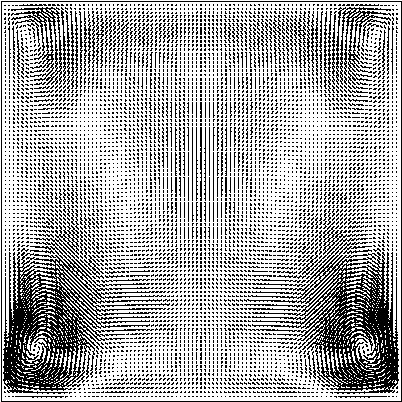} \hskip 2pt
\includegraphics[width=1.8in]{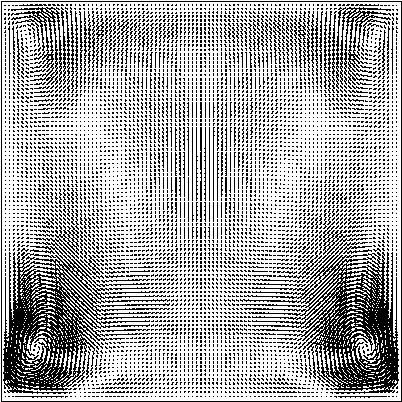} \hskip 2pt
\includegraphics[width=1.8in]{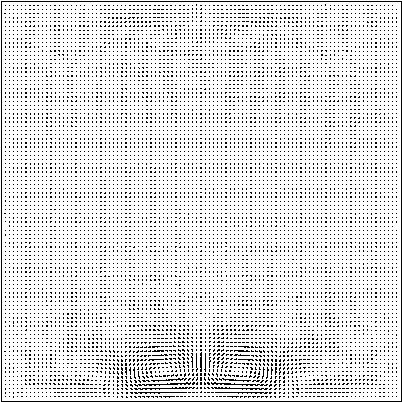}
\end{center}
\begin{center}
\leavevmode 
\includegraphics[width=0.4in]{cord-c1.pdf} \hskip 2pt
\includegraphics[width=1.8in]{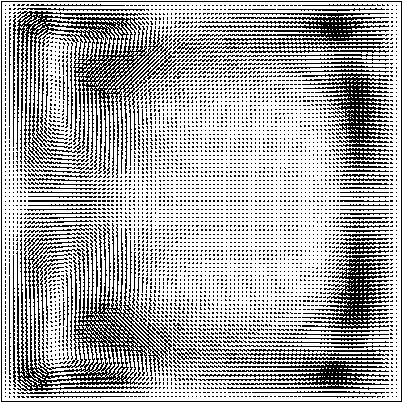} \hskip 2pt
\includegraphics[width=1.8in]{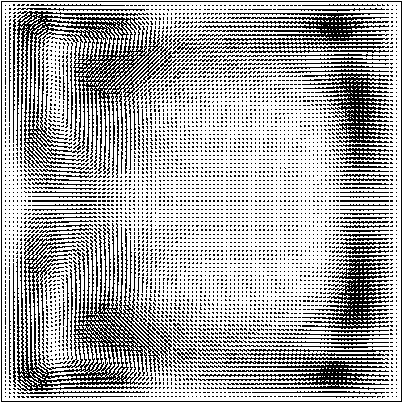} \hskip 2pt
\includegraphics[width=1.8in]{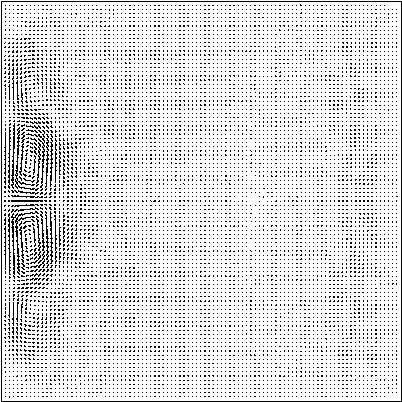}
\end{center} 
\caption{Left and middle: Projections (at Re = 1875) of the cavity flow velocity vector fields associated with the peak (left) 
and bottom (middle) of the velocity $L^2$-norm during an oscillation. Right: Projections (at Re = 1875) of the vector 
field obtained by difference of the velocity vector fields associated with the peak and bottom of the velocity $L^2$-norm. 
All the vector fields are projected on the planes $x_1=52/96$ (top) and $x_3=0.5$ (bottom). The vector scale for the field 
obtained by difference (right) is 15 times that of the actual one, while the scale for the two other fields (left
and middle) is twice that of the actual one.}\label{fig7}
\end{figure}

\begin{figure}[!thp]
\begin{center}
\includegraphics[width=3.2in]{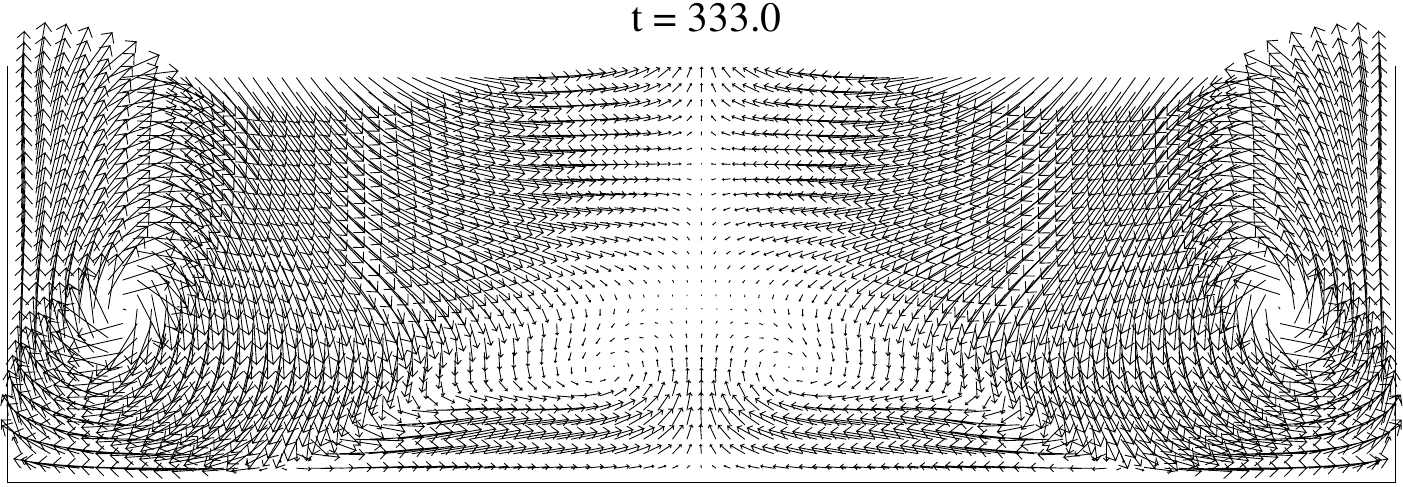}  
\includegraphics[width=3.2in]{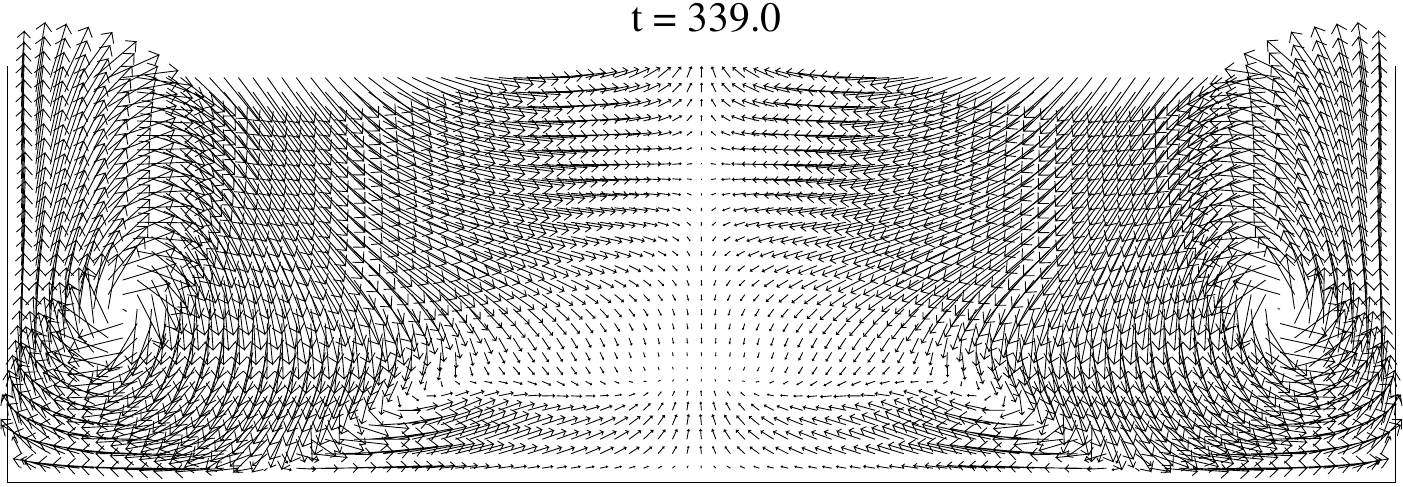}  \\
\includegraphics[width=3.2in]                              
{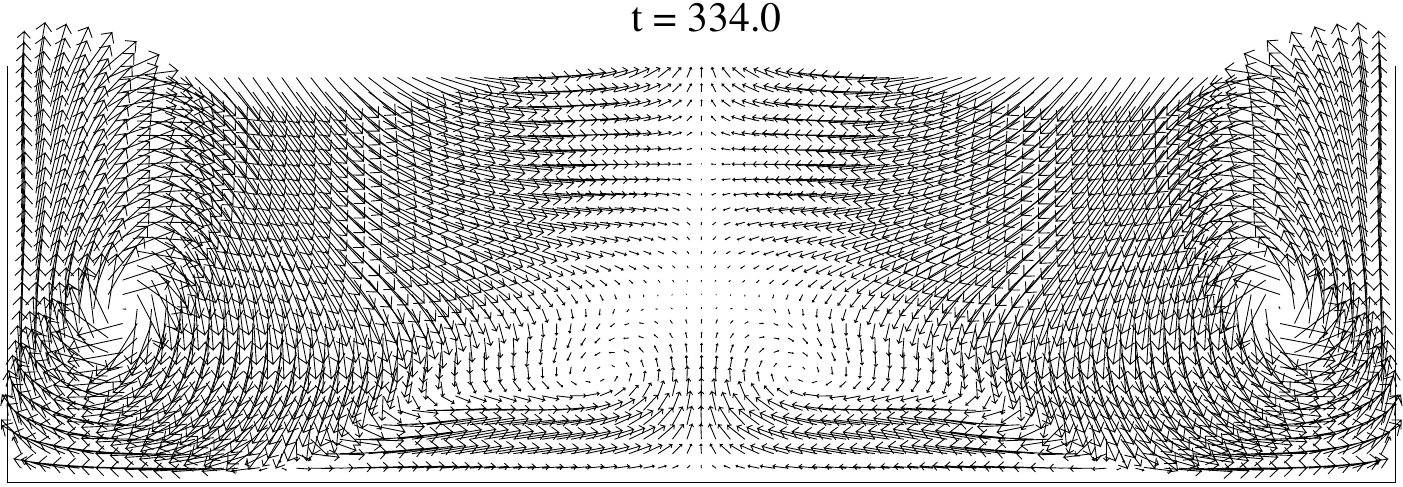}  
\includegraphics[width=3.2in] {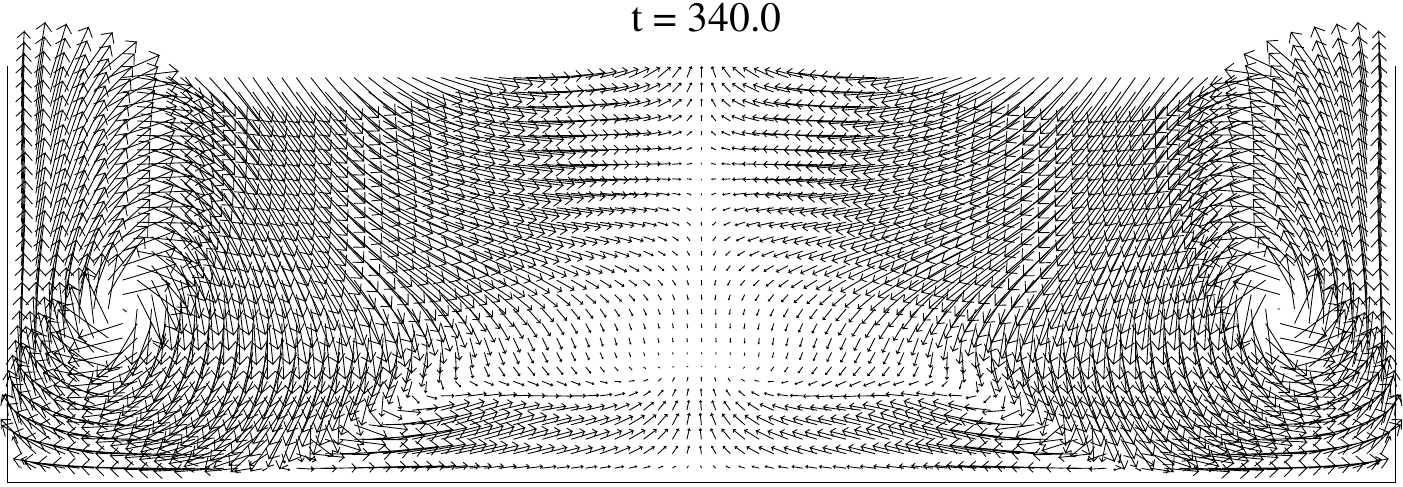} \\
\includegraphics[width=3.2in]                                                  
{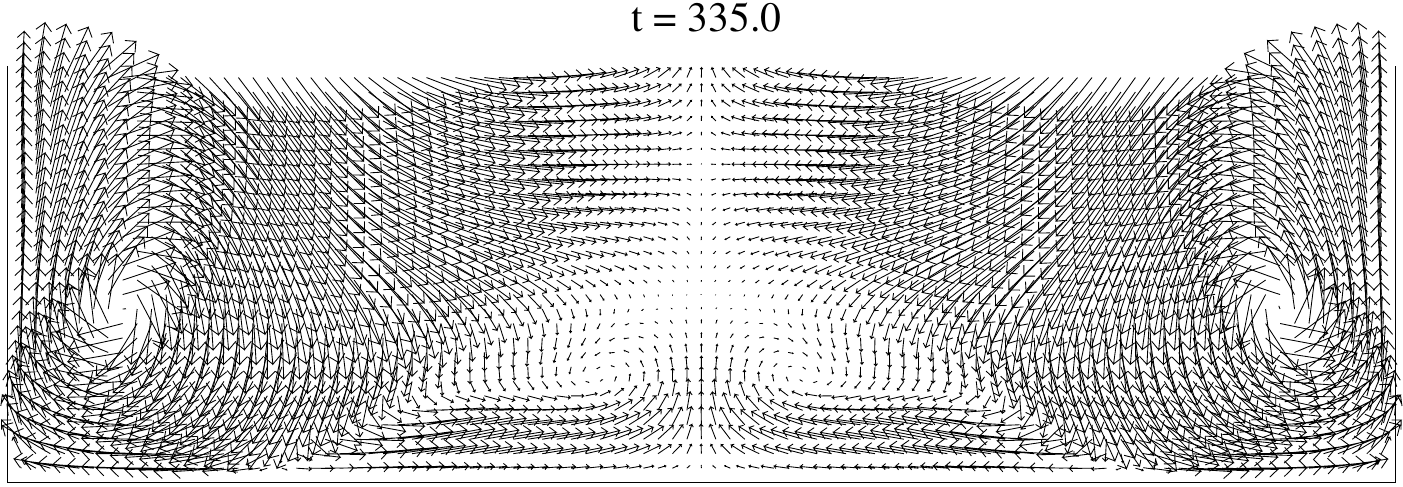}  
\includegraphics[width=3.2in] {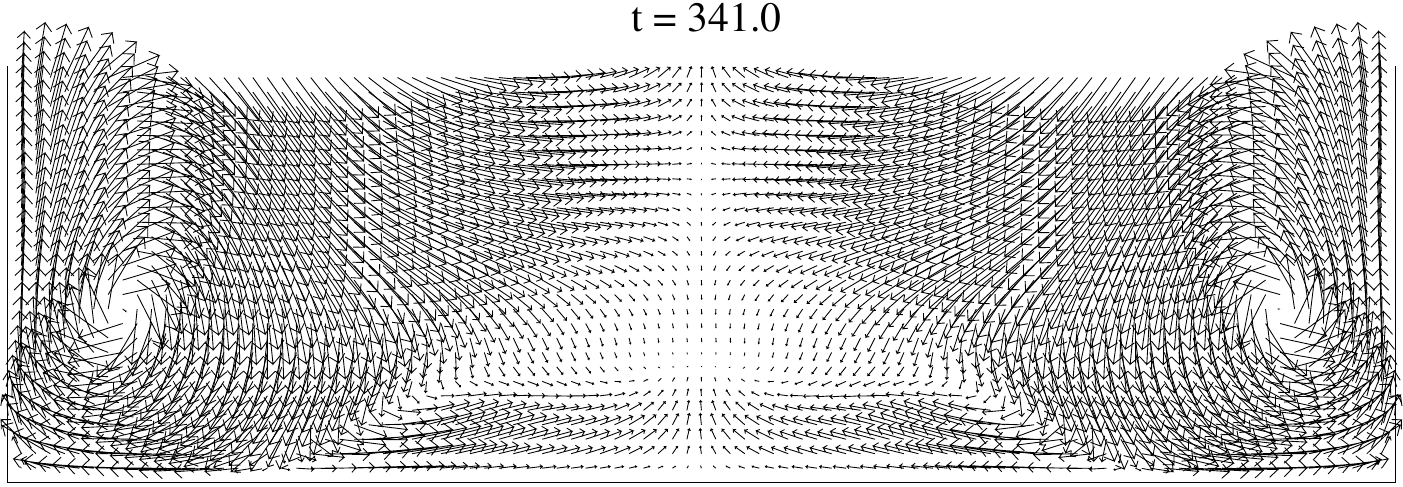} \\
\includegraphics[width=3.2in]                                                  
{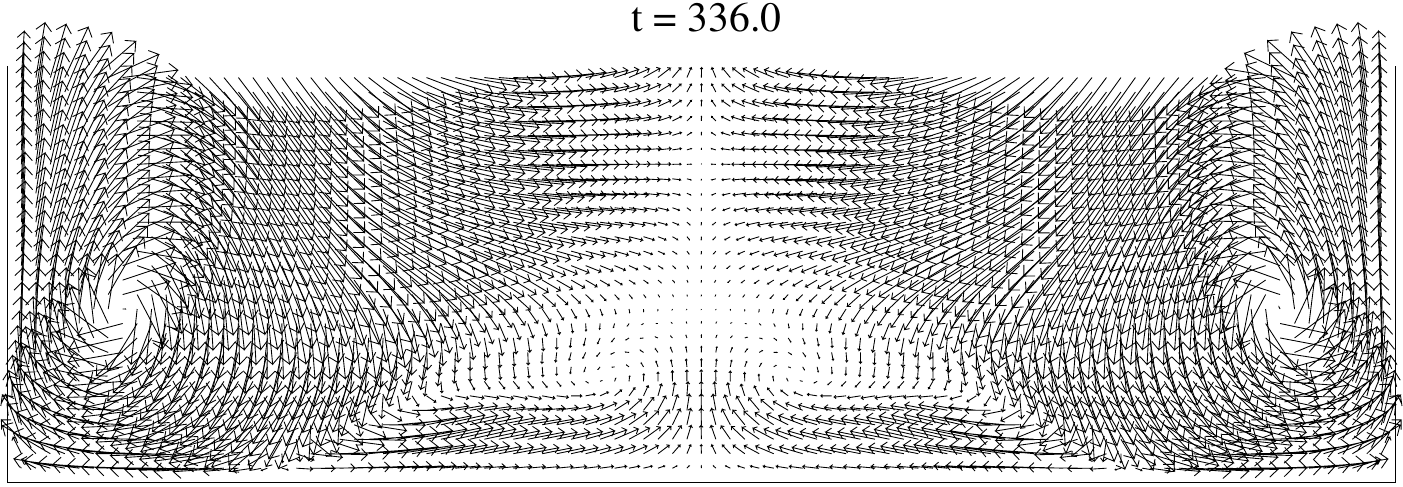}  
\includegraphics[width=3.2in] {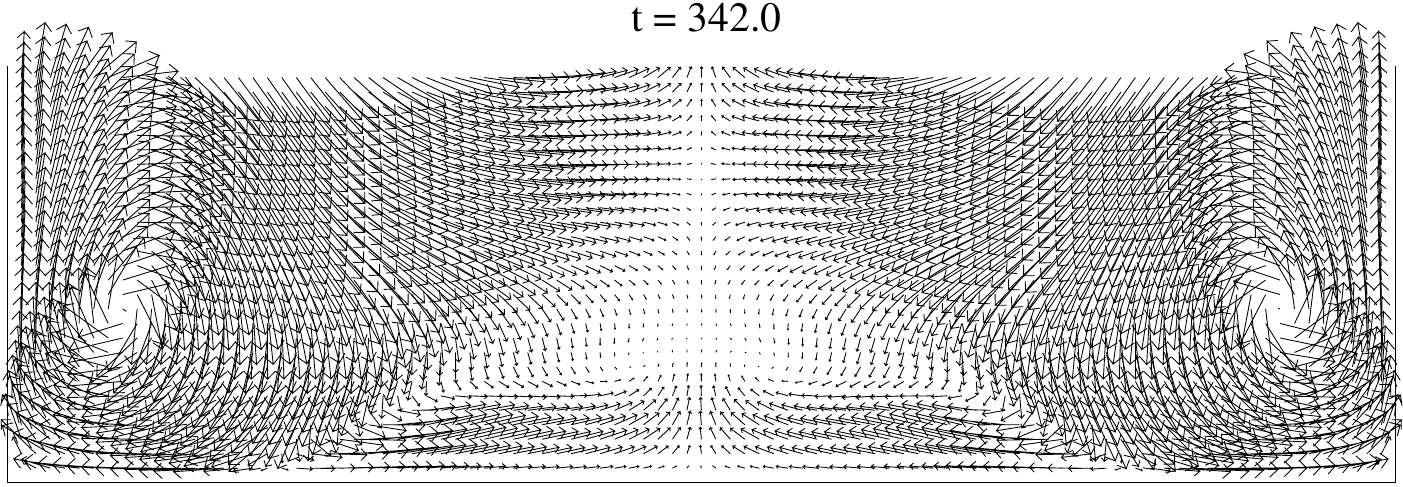} \\
\includegraphics[width=3.2in]                                                   
{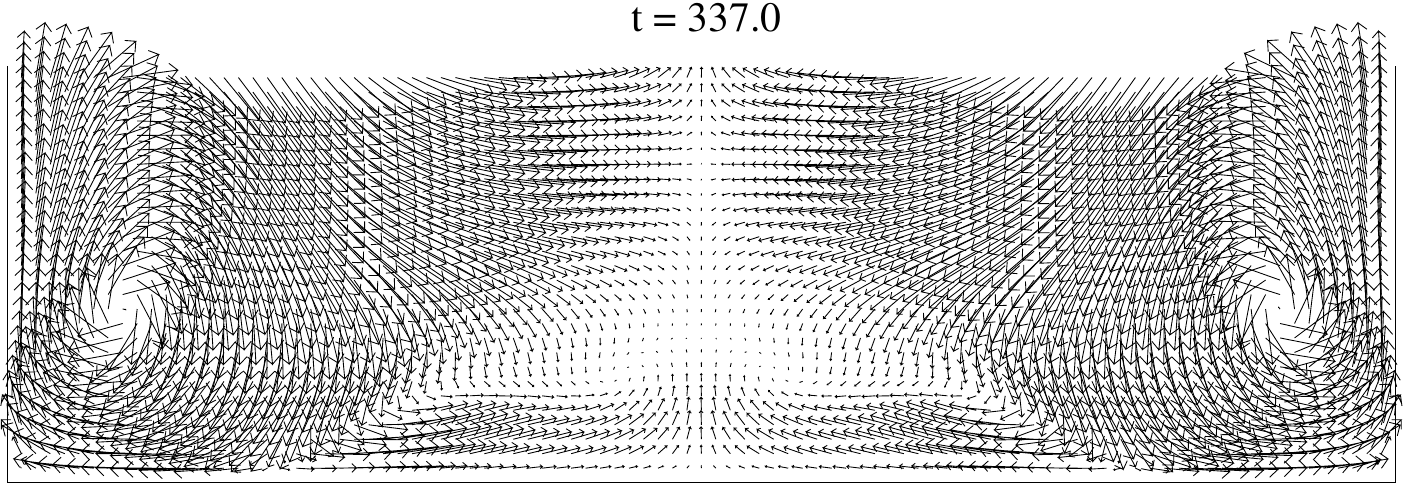}  
\includegraphics[width=3.2in] {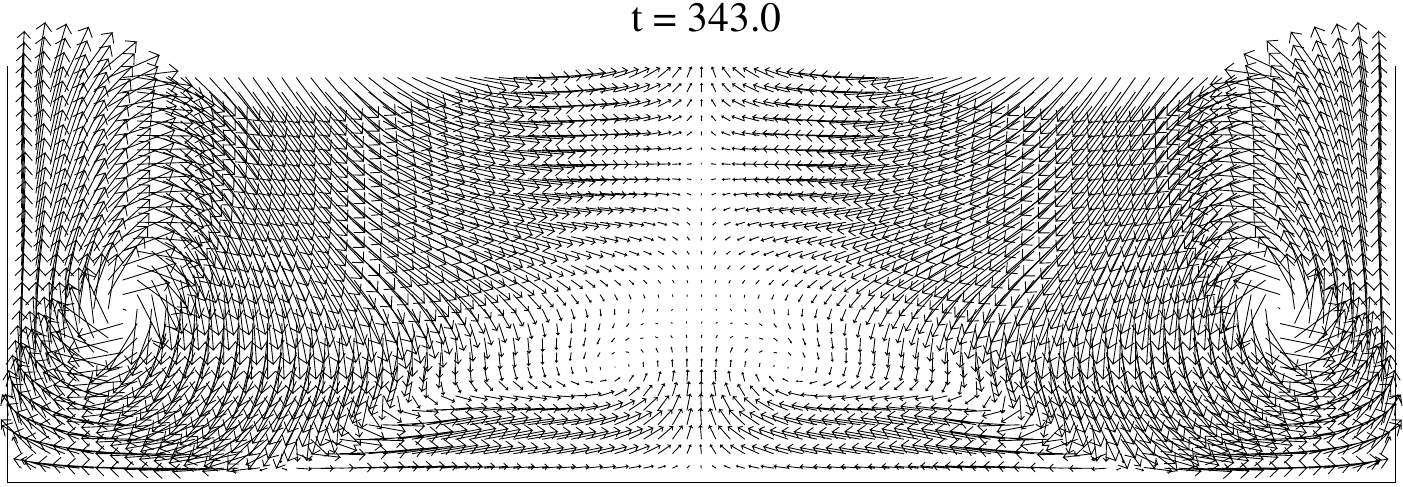}\\
\includegraphics[width=3.2in]                                                   
{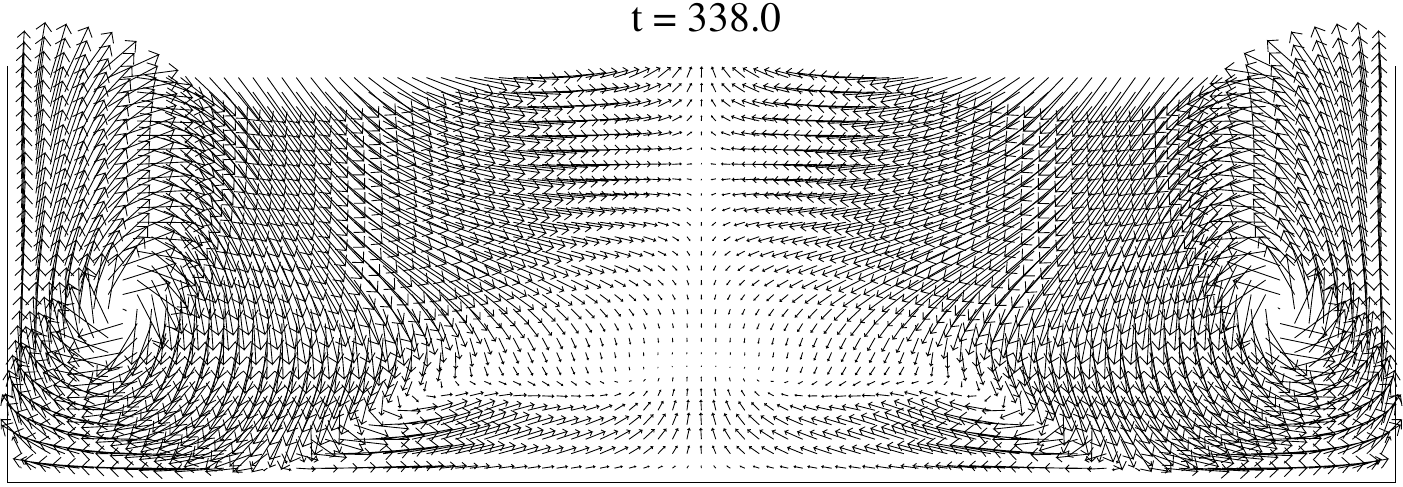}  
\includegraphics[width=3.2in] {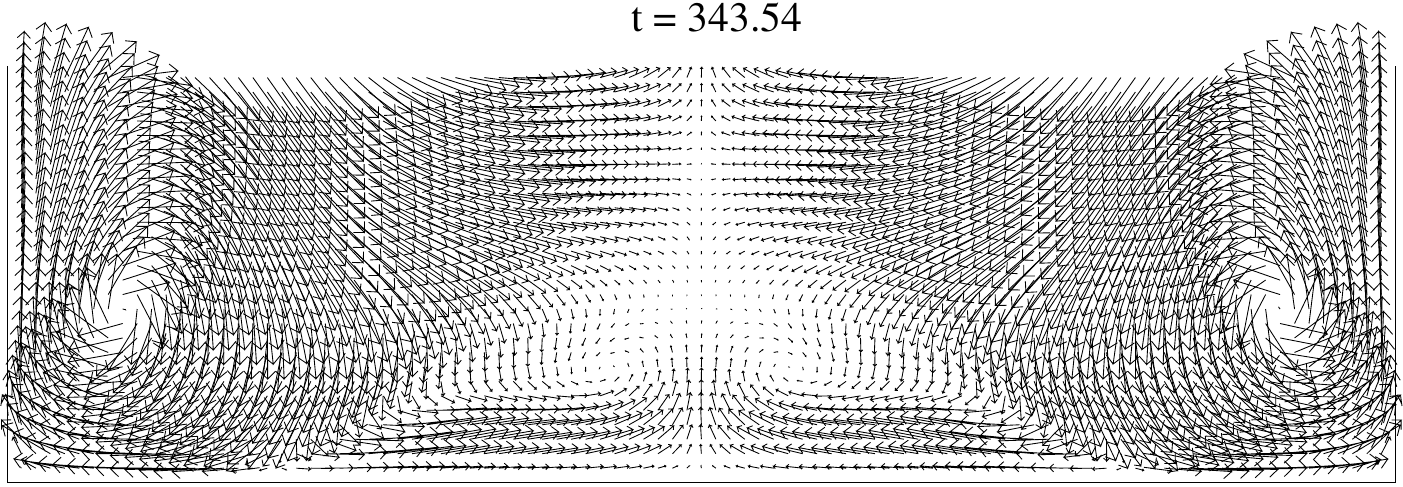}
\end{center} 
\caption{Projected velocity vector field of the cavity flow at Re=1885 on the plane  $x_1=54/96$
at different instants of time during one oscillation of the flow field $L^2$-norm from $t =333$ to
343.54 (for enhancing the visibility of the TGL vortices we proceeded as follows: (i) for those
projected vectors of length $\le 0.02$ the vector scale is 10 times that of the actual one and (ii) for
those projected vectors of length $> 0.02$, the length is reduced to 0.02 first and then plotted as in (i)).}\label{fig8}
\end{figure}

\begin{figure}[!thp]
\begin{center}
\includegraphics[width=3.2in]
{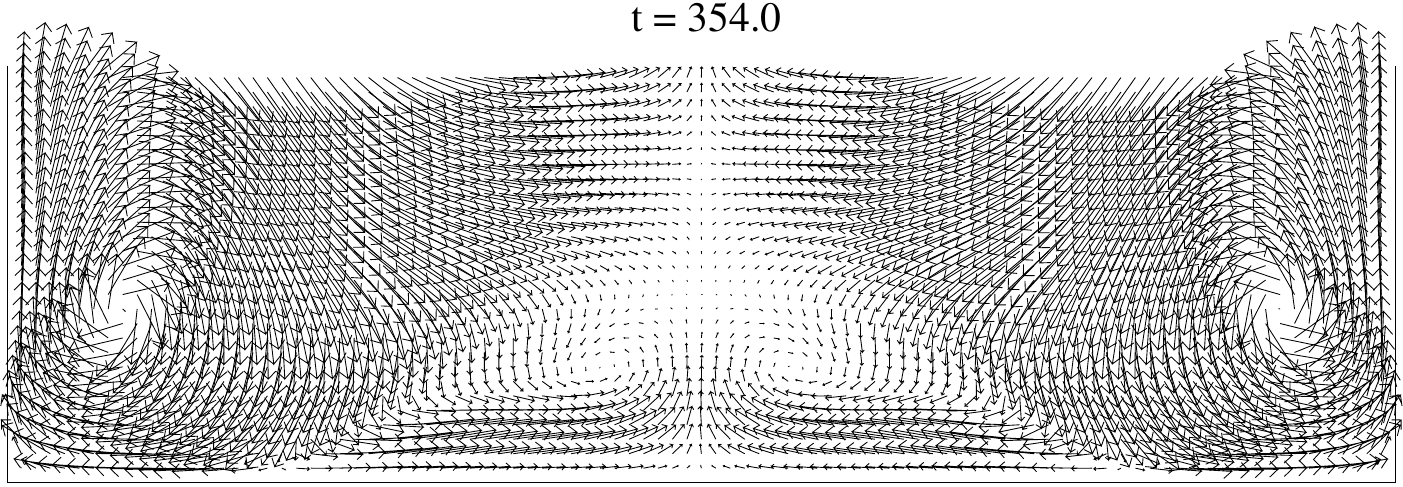}  \includegraphics[width=3.2in] {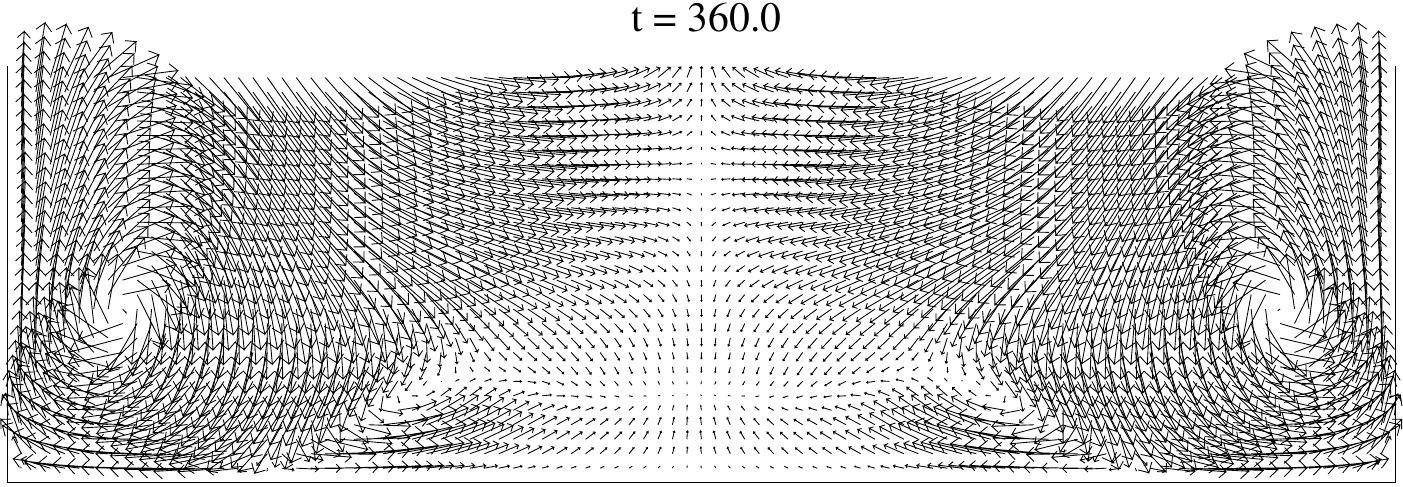}  \\
\includegraphics[width=3.2in]                                                    
{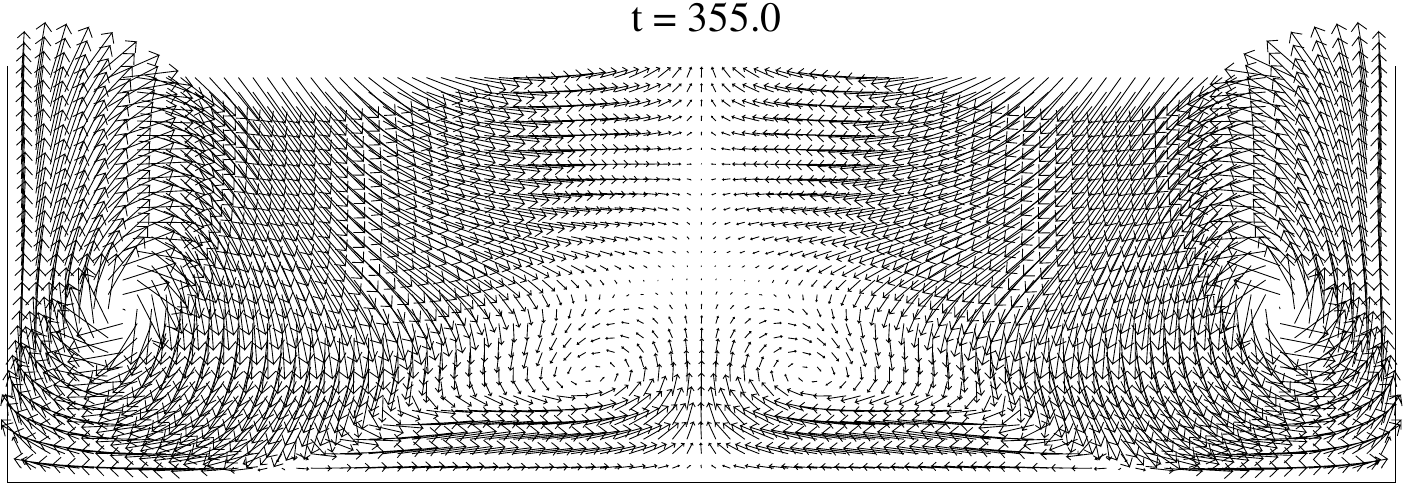}  \includegraphics[width=3.2in] {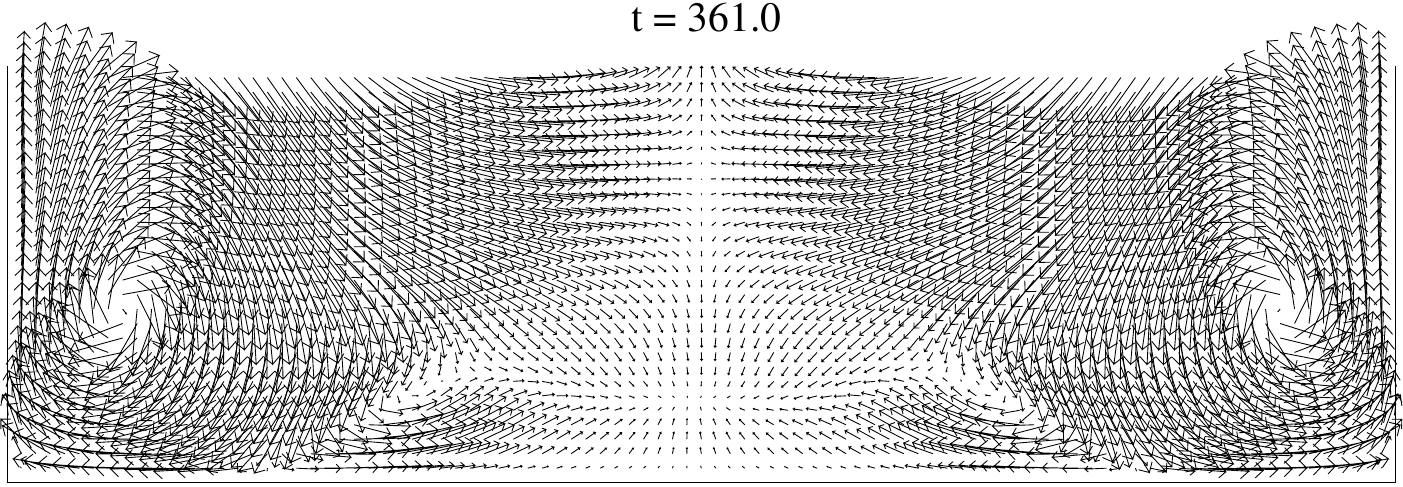} \\
\includegraphics[width=3.2in]                                                    
{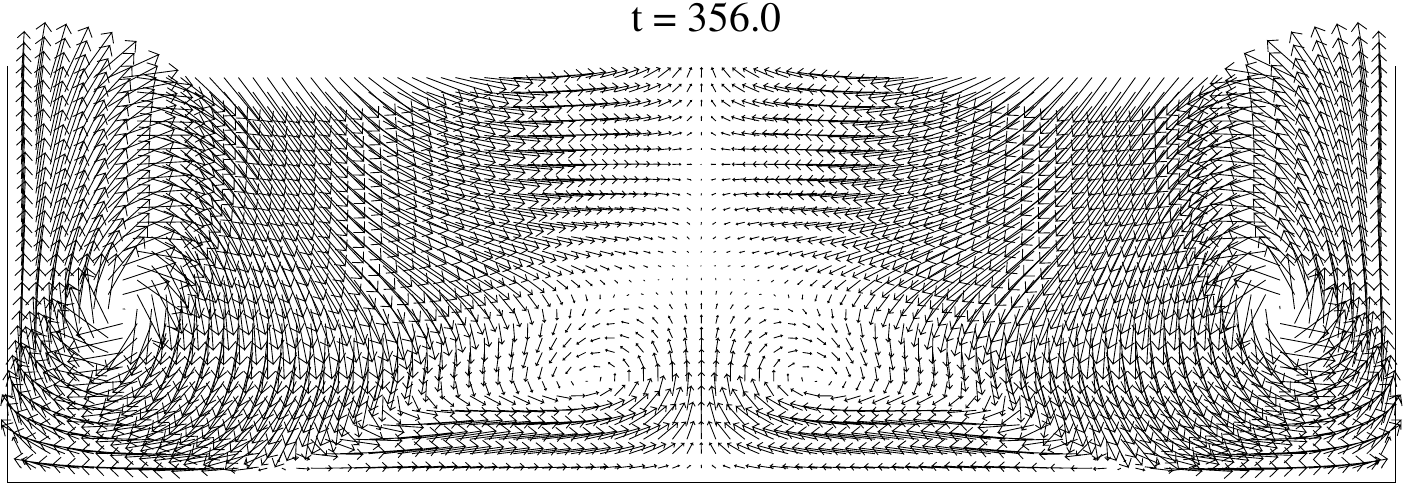}  \includegraphics[width=3.2in] {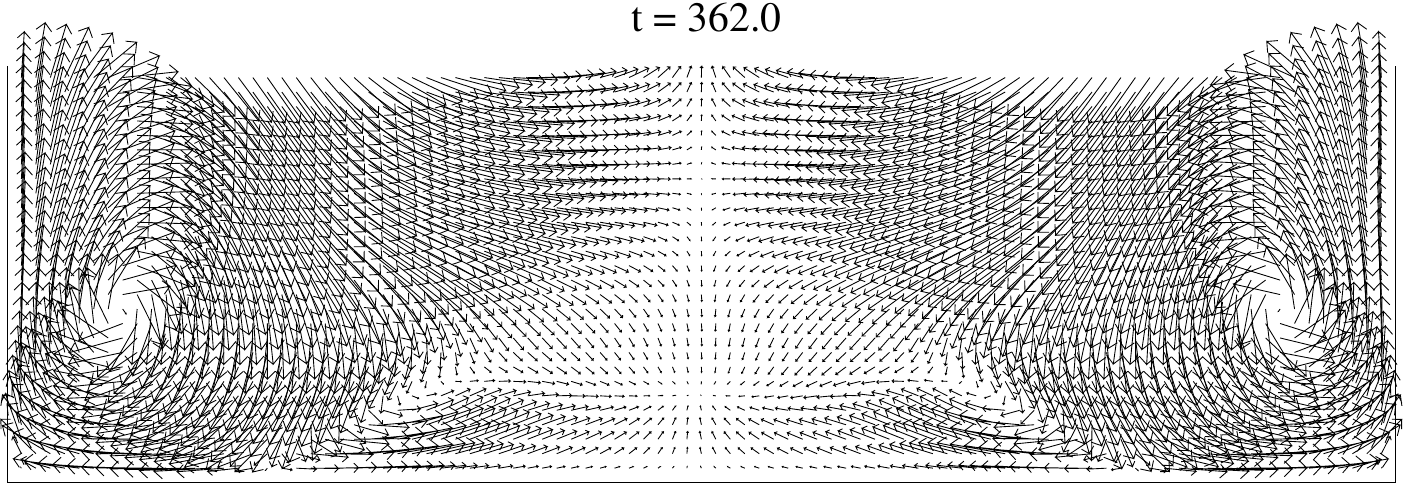} \\
\includegraphics[width=3.2in]                                                    
{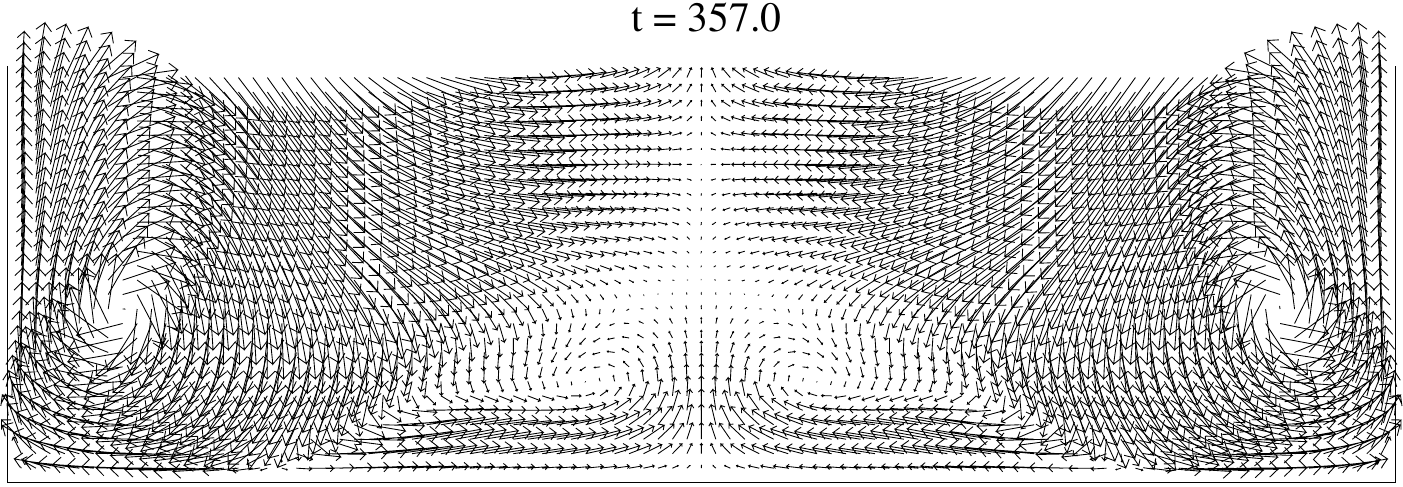}  \includegraphics[width=3.2in] {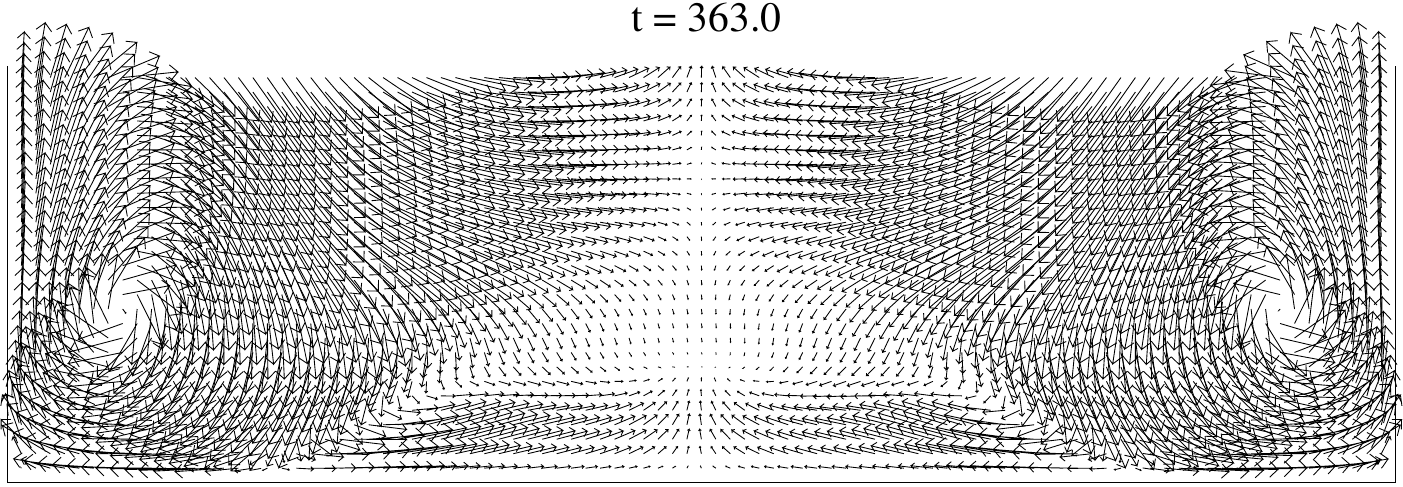} \\
\includegraphics[width=3.2in]                                                    
{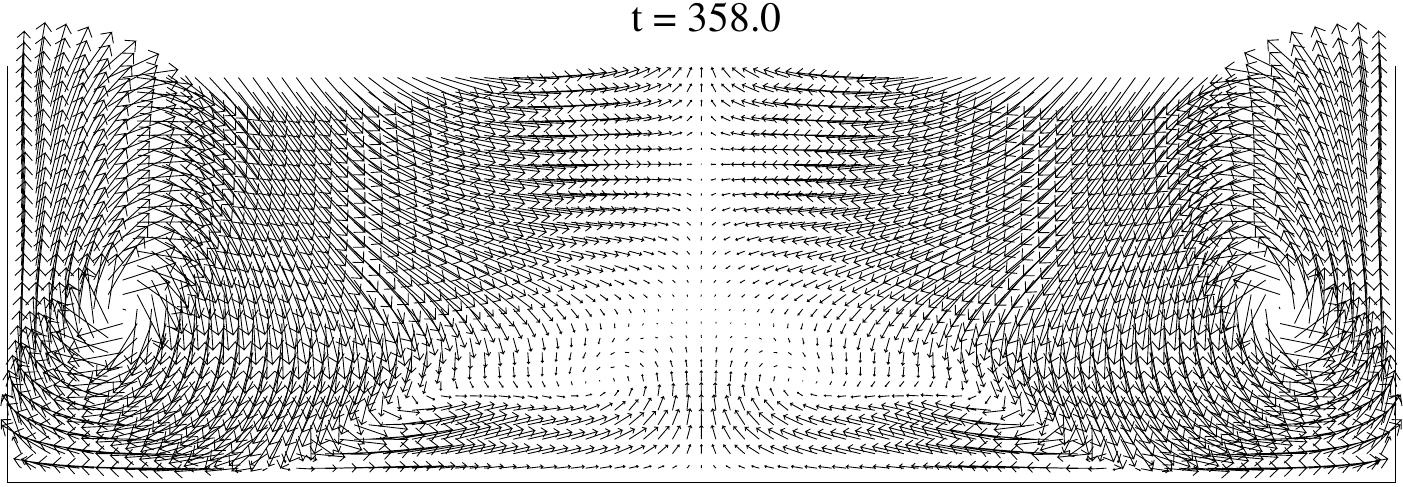}  \includegraphics[width=3.2in] {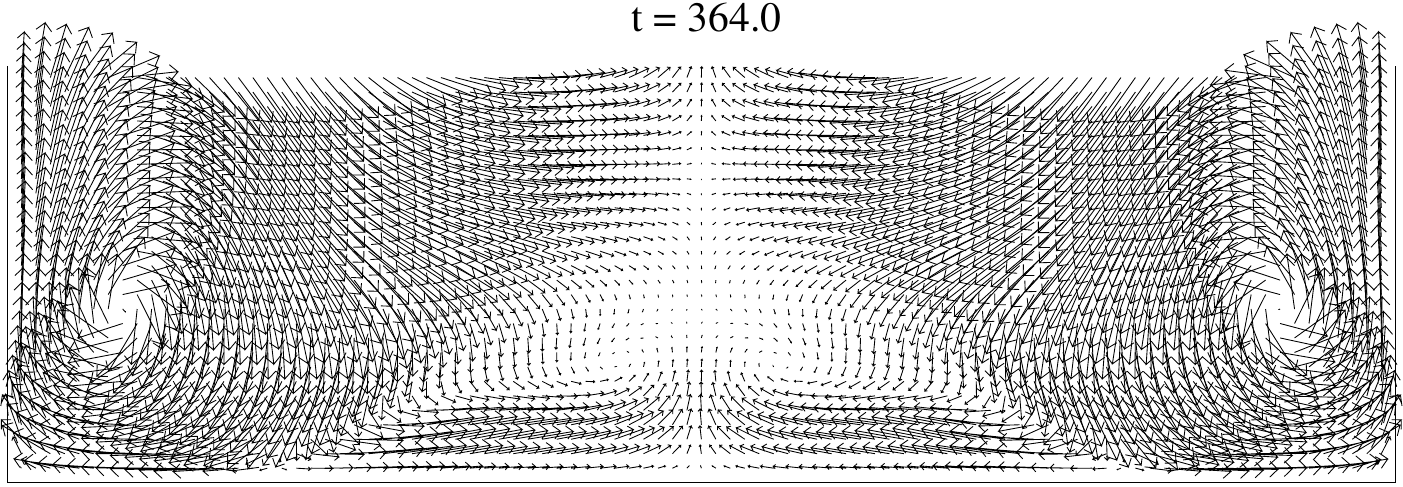}\\
\includegraphics[width=3.2in]                                                    
{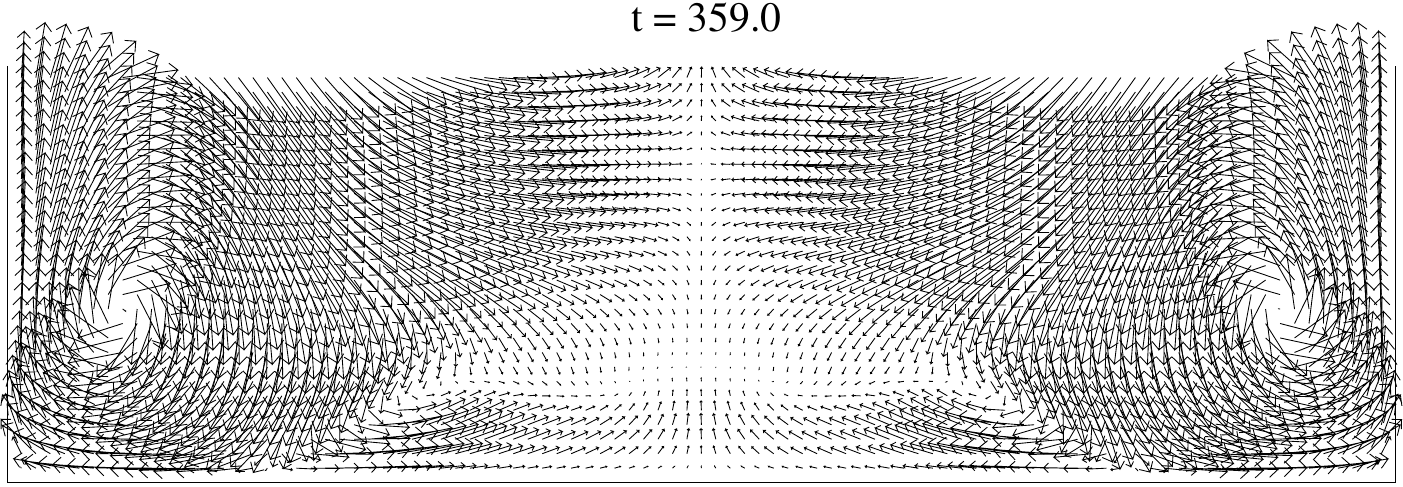}  \includegraphics[width=3.2in] {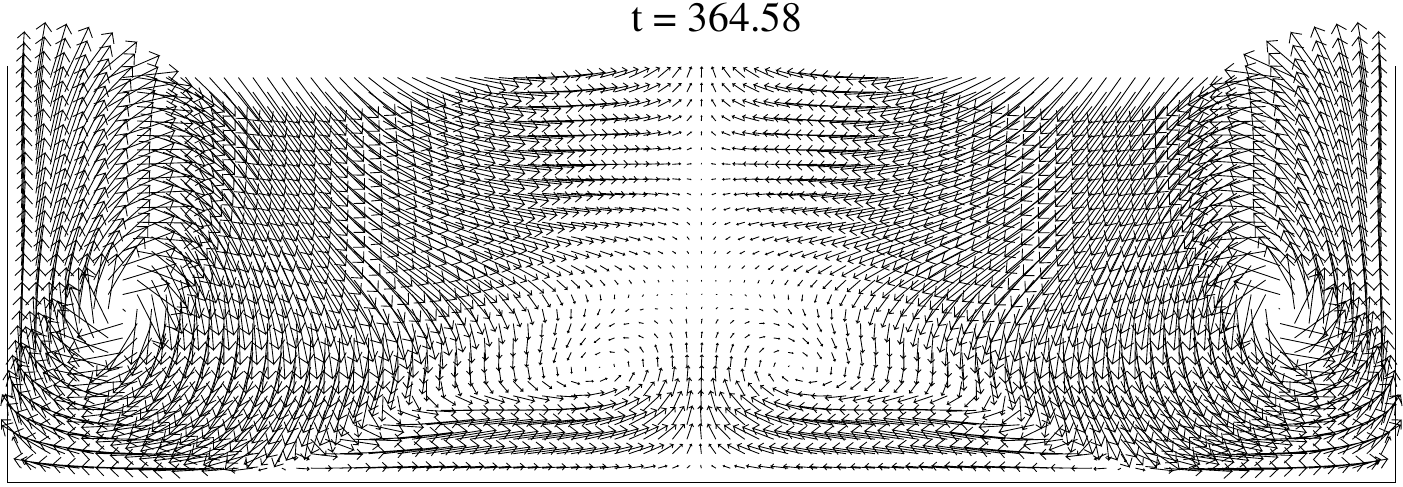}
\end{center} 
\caption{Projected velocity vector field of the cavity flow at Re=1900 on the plane  $x_1=54/96$
at different instants of time during one oscillation of the flow field $L^2$-norm from $t =354$ to
364.58 (for enhancing the visibility of the TGL vortices we proceeded as follows: (i) for those
projected vectors of length $\le 0.02$ the vector scale is 10 times that of the actual one and (ii) for
those projected vectors of length $> 0.02$, the length is reduced to 0.02 first and then plotted as in (i)).}\label{fig10}
\end{figure}

\begin{figure}[!tp]
\begin{center}
\includegraphics[width=0.45in]
{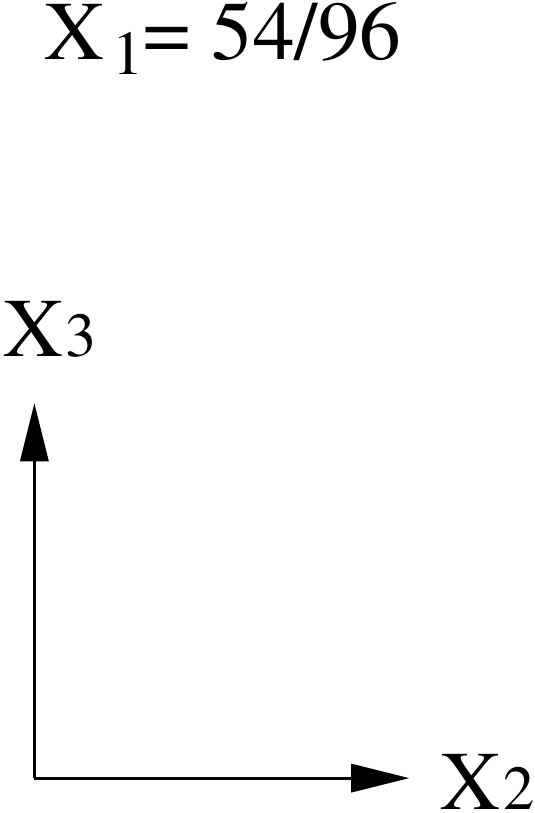} \hskip 2pt
\includegraphics[width=2.0in]
{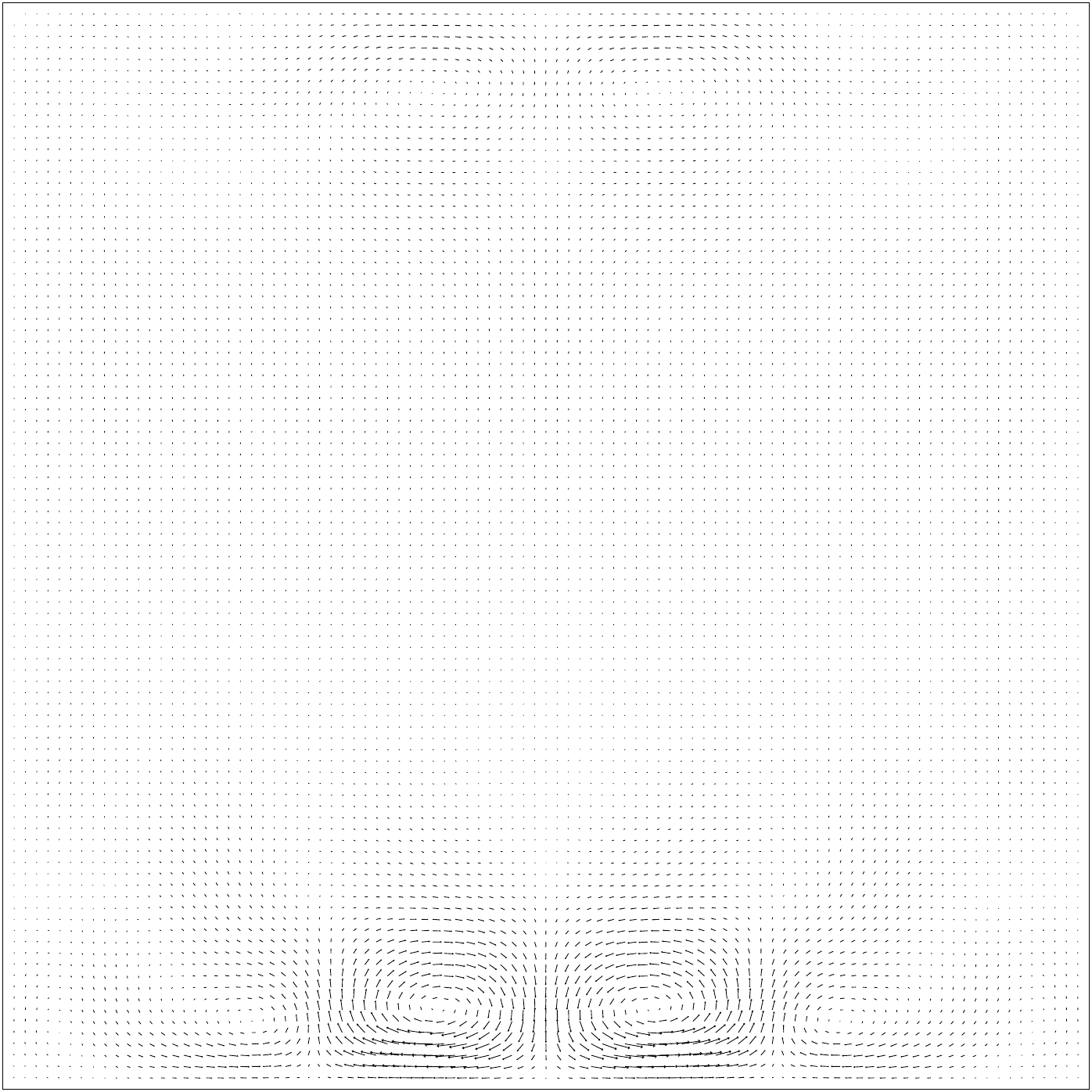}
\includegraphics[width=2.0in]
{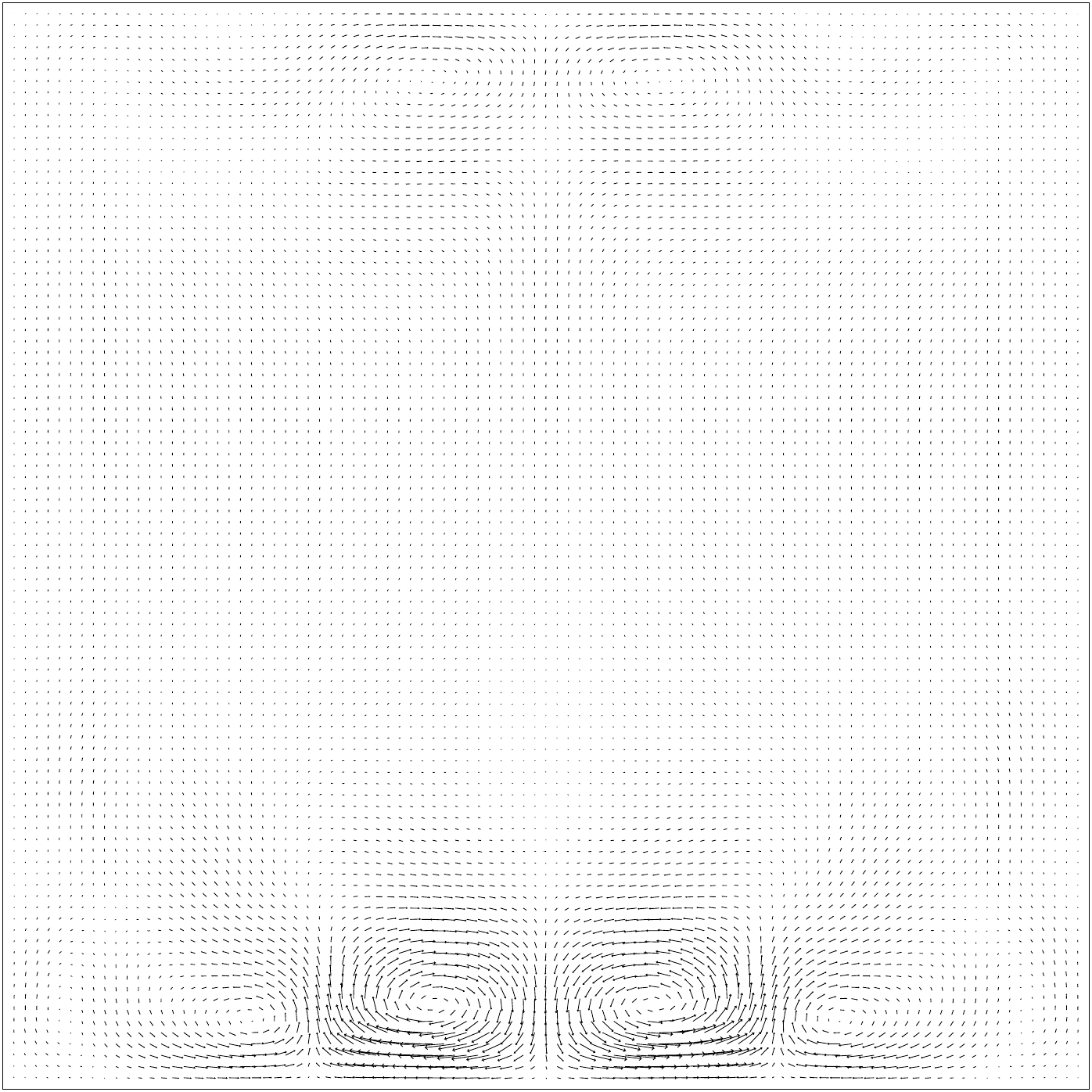}
\end{center}
\caption{Projected velocity vector field of  the difference  of the velocity fields at $t=340$ and $t=342$ on the plane
 $x_1=54/96$ for Re=1885 (left) and that  of  the difference  of the velocity fields at $t=361$ and $t=363$ on the plane
 $x_1=54/96$ for Re=1900 (right). The vector scale  is 25 times that of the actual one for both.}\label{fig9}
\begin{center}
\includegraphics[width=2.3in]
{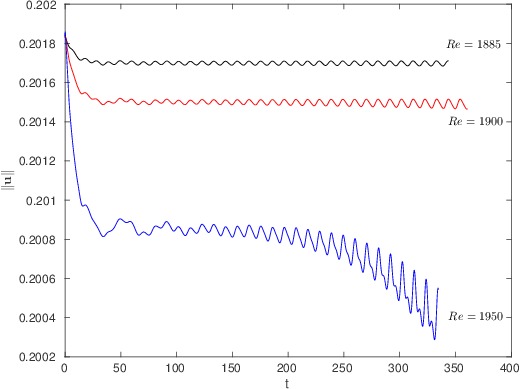}
\includegraphics[width=2.3in]
{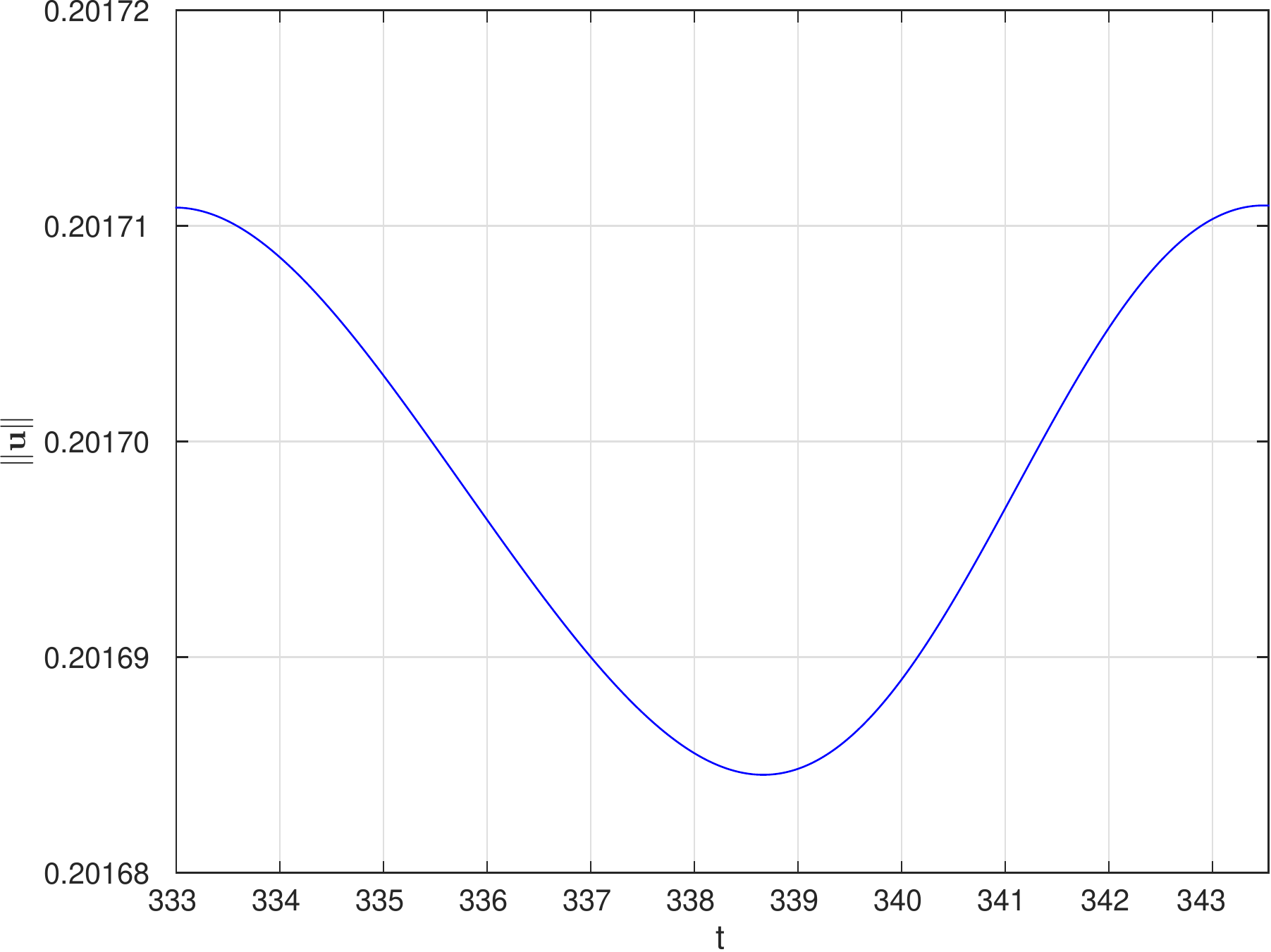} \\
\includegraphics[width=2.3in]
{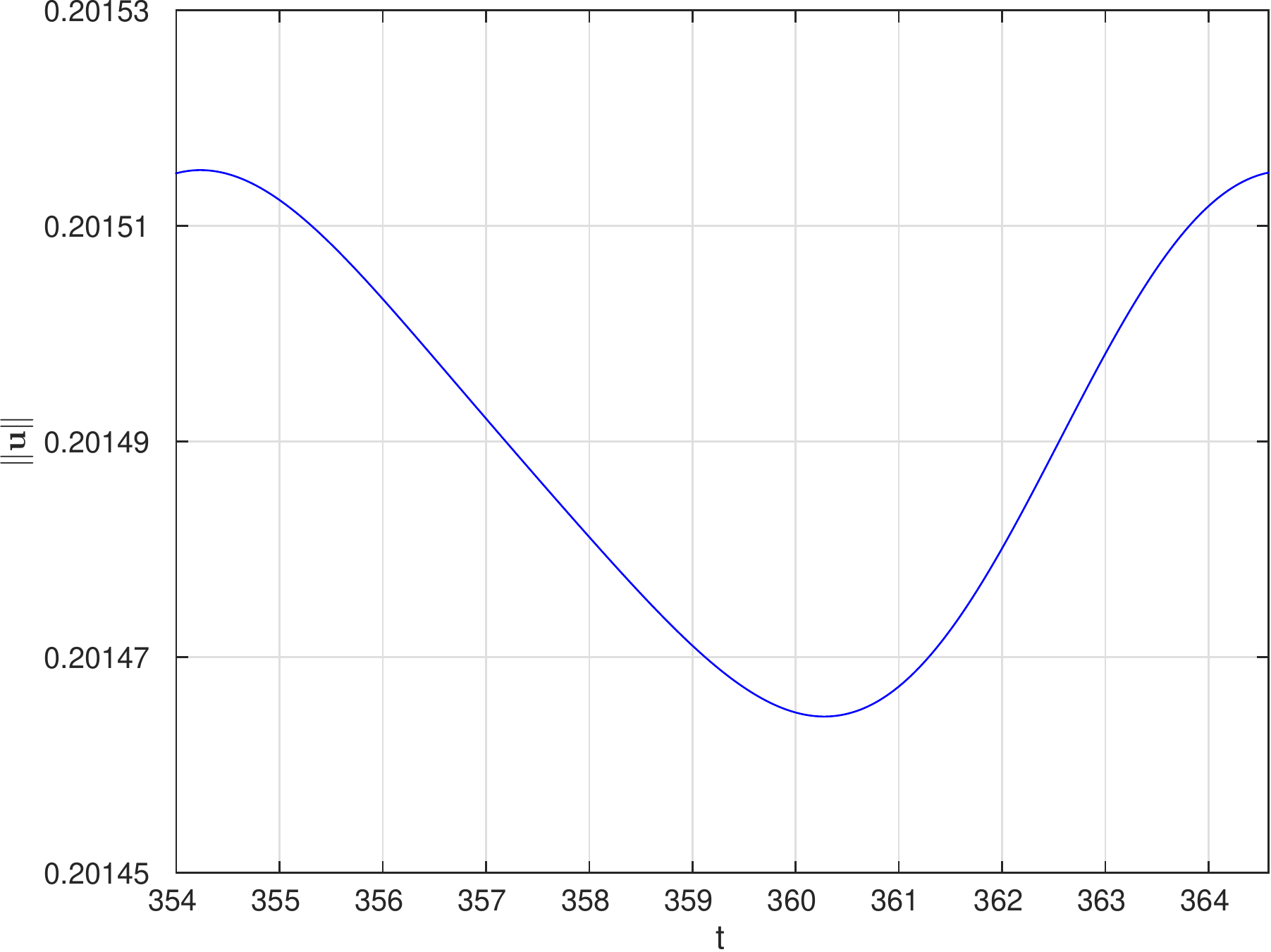}
\includegraphics[width=2.3in]
{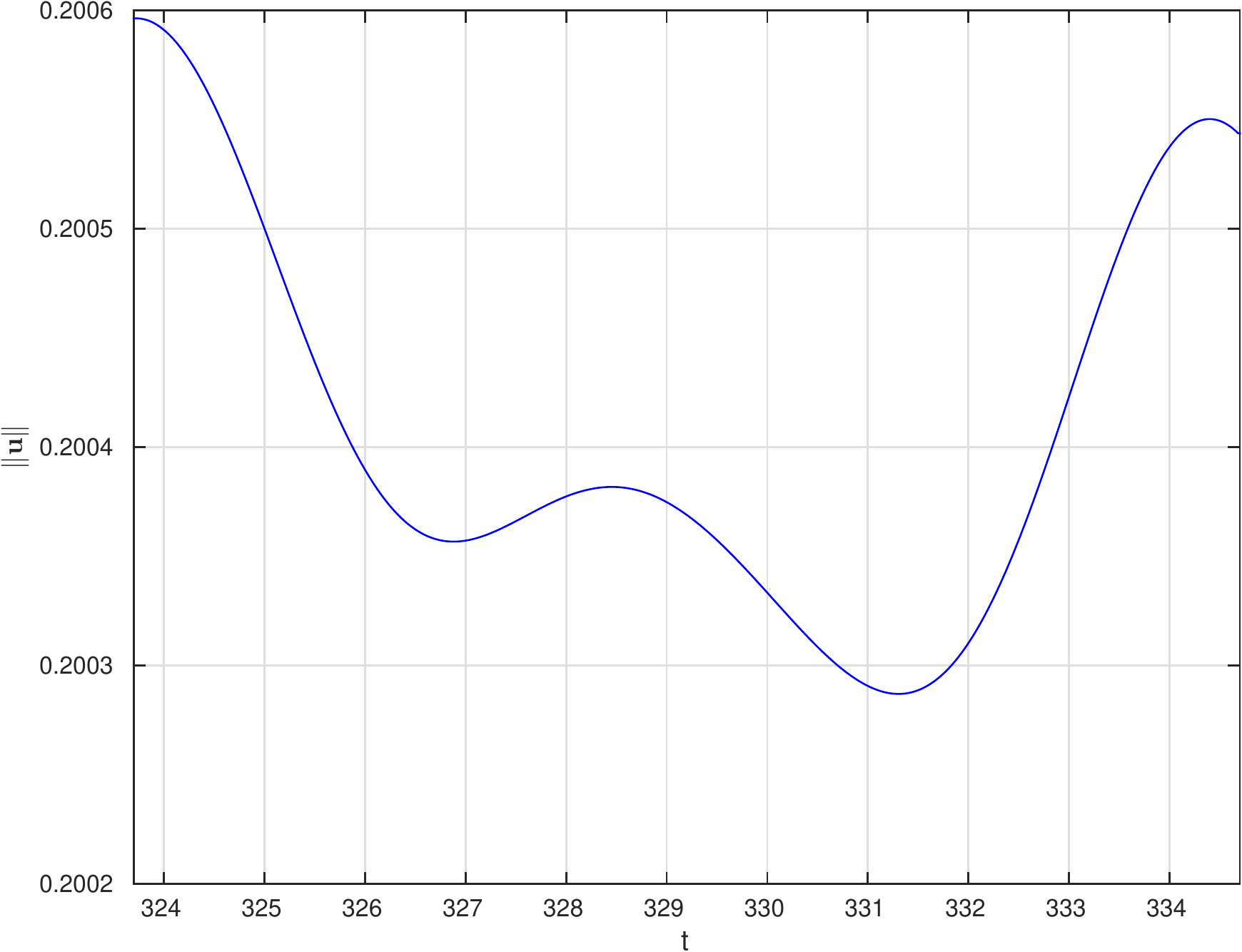}
\end{center}
\caption{Histories of the flow field $L^2$-norm for $h=1/96$  at Re=1885, 1900, and 1950 (top left) and one oscillation of the flow field $L^2$-norm for Re=1885 (top right), Re=1900(bottom left), and  Re=1950 (bottom right). }\label{fig11}
\end{figure}
\begin{figure}[!tp]
\begin{center}
\includegraphics[width=3.2in]
{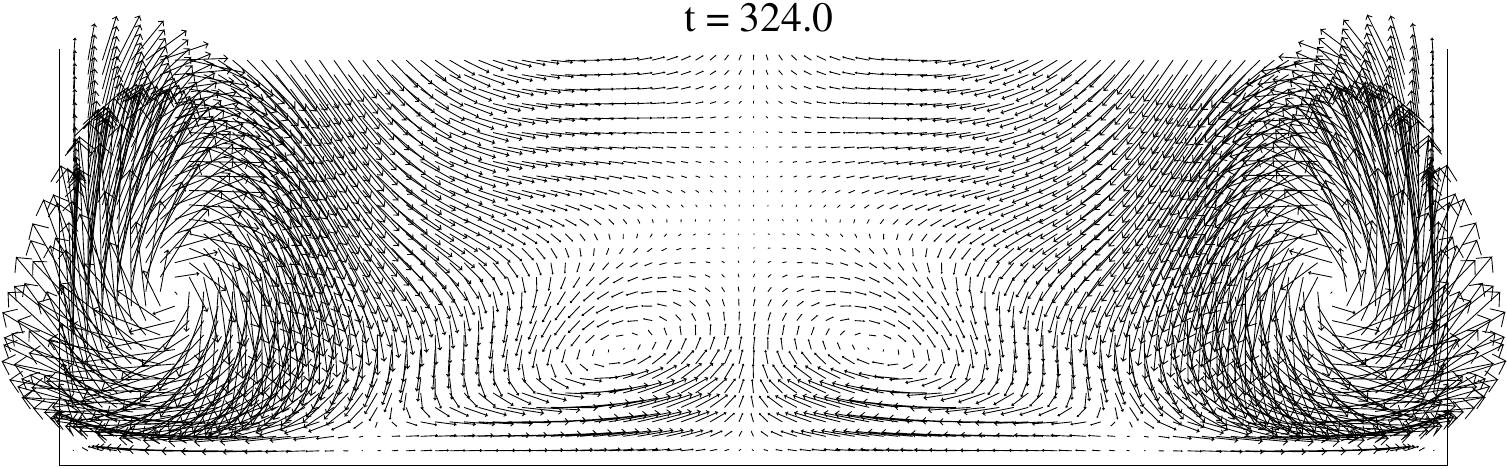}  \includegraphics[width=3.2in] {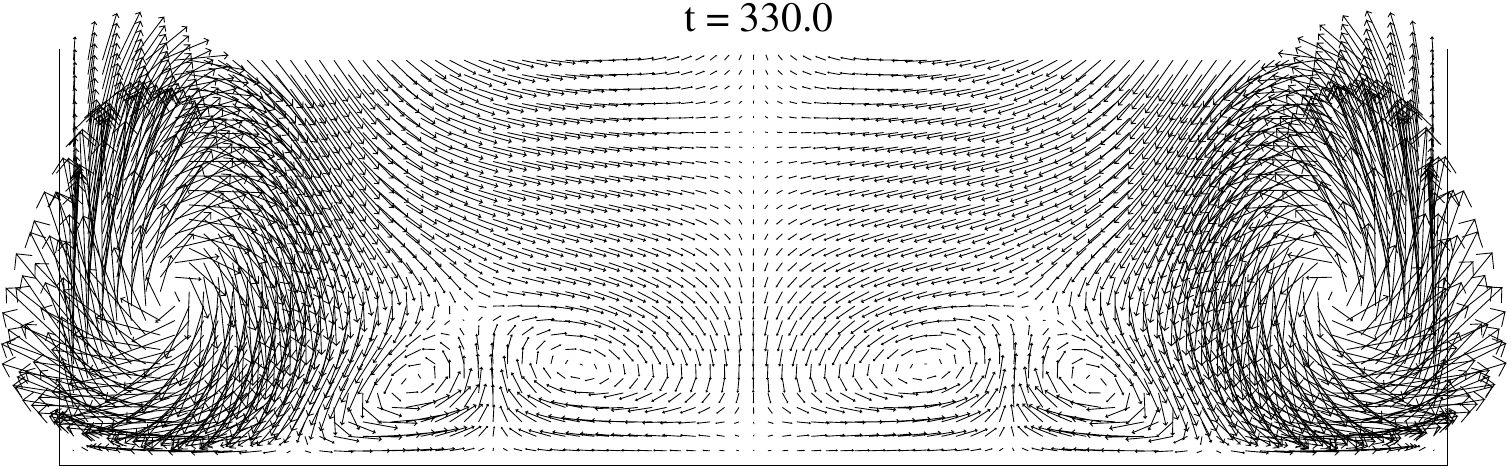}  \\
\includegraphics[width=3.2in]                                                    
{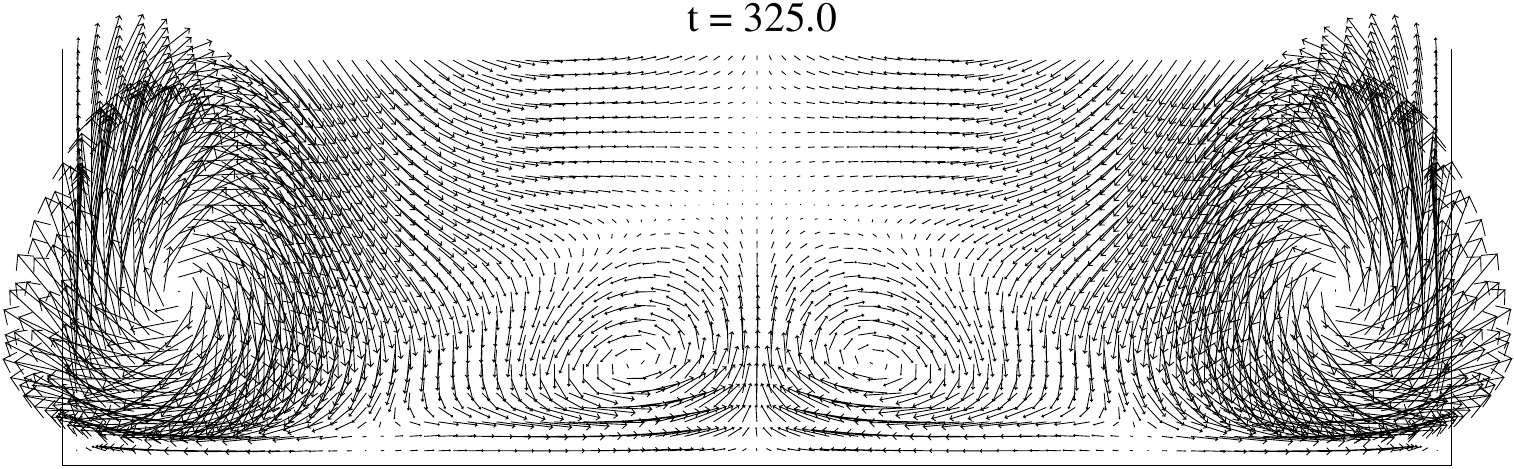}  \includegraphics[width=3.2in] {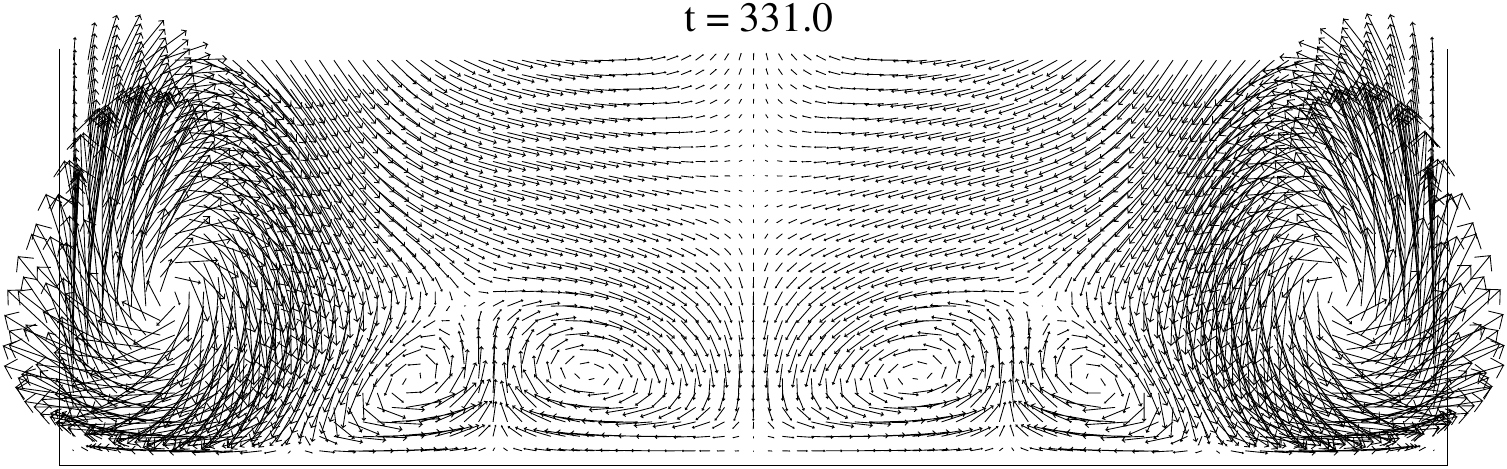} \\
\includegraphics[width=3.2in]                                                    
{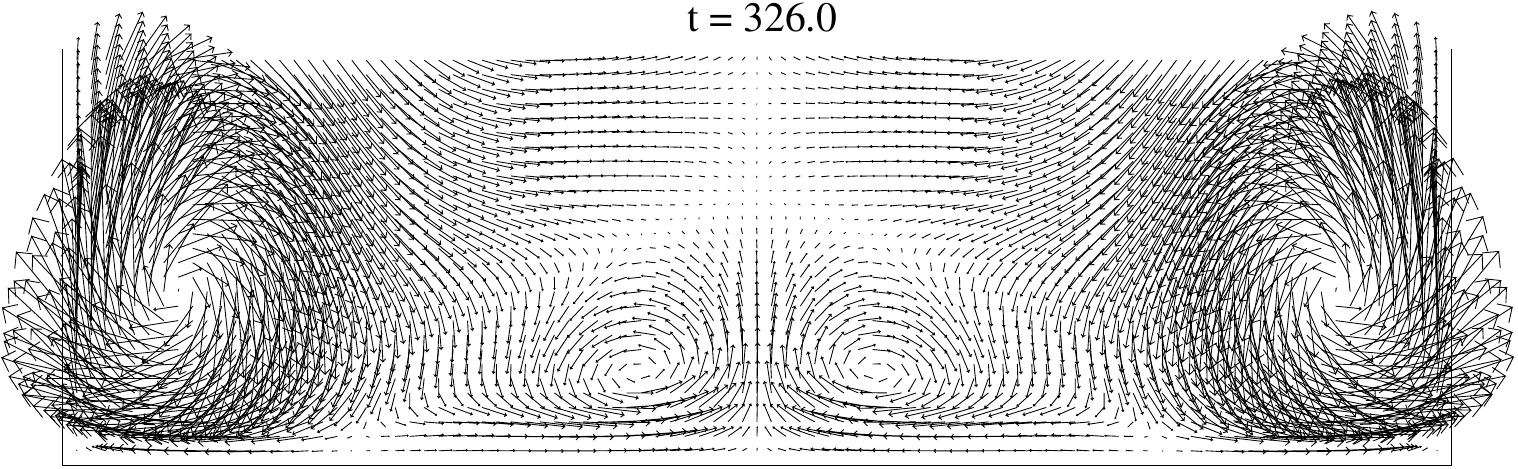}  \includegraphics[width=3.2in] {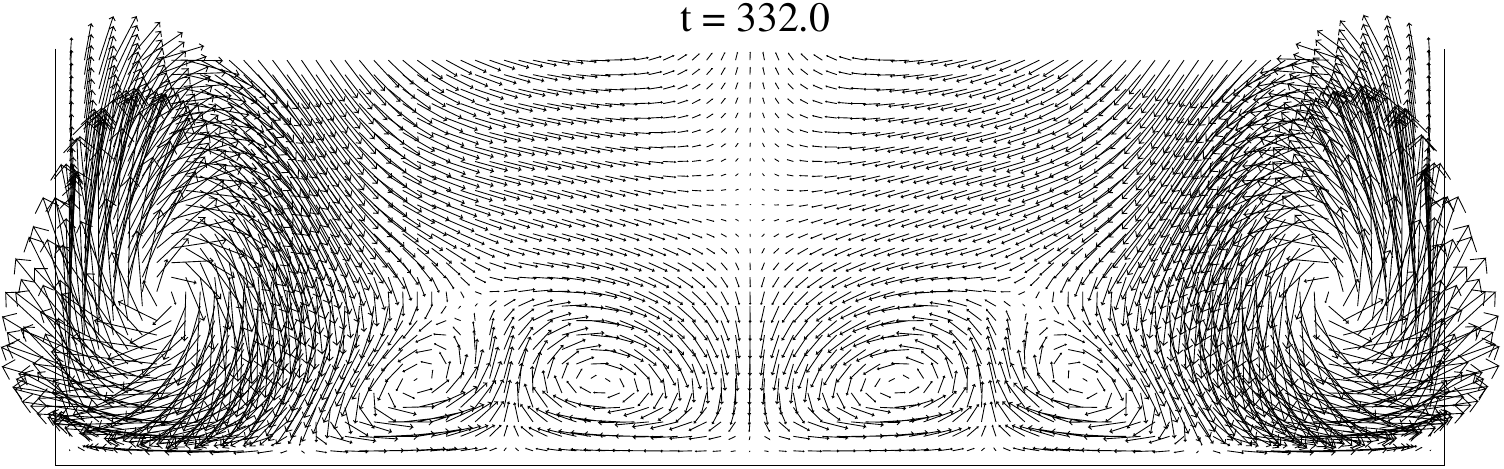} \\
\includegraphics[width=3.2in]                                                    
{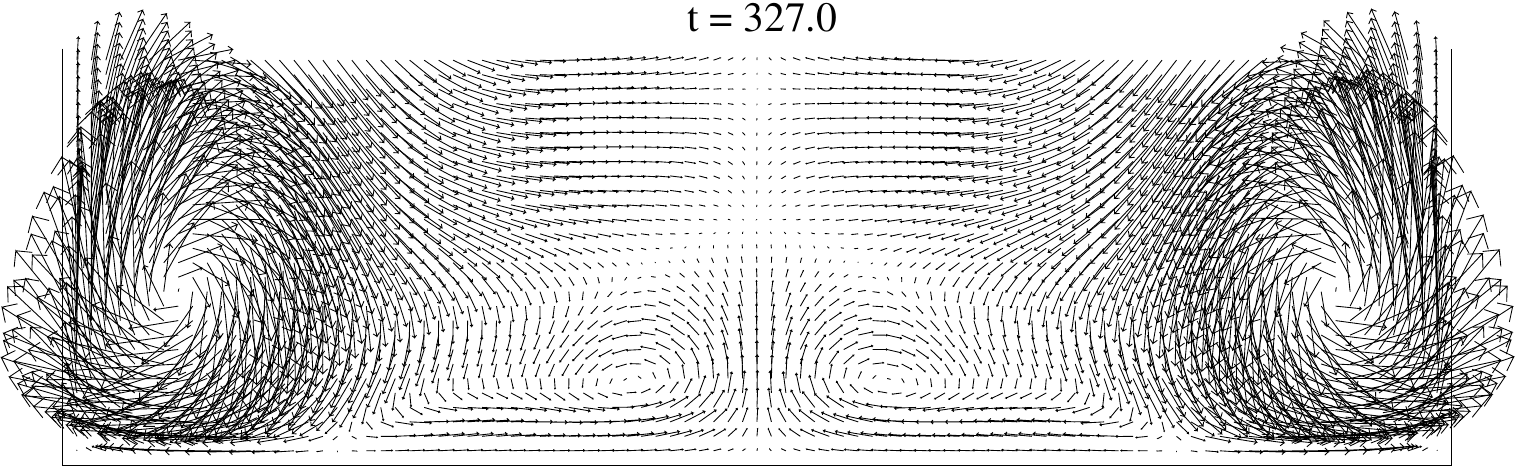}  \includegraphics[width=3.2in] {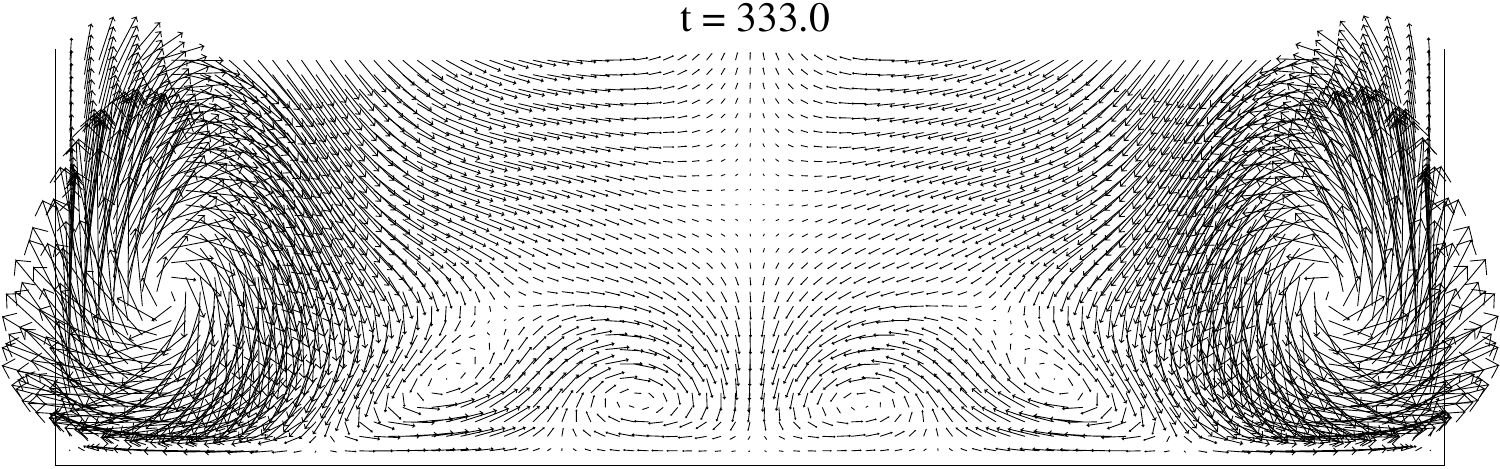} \\
\includegraphics[width=3.2in]                                                    
{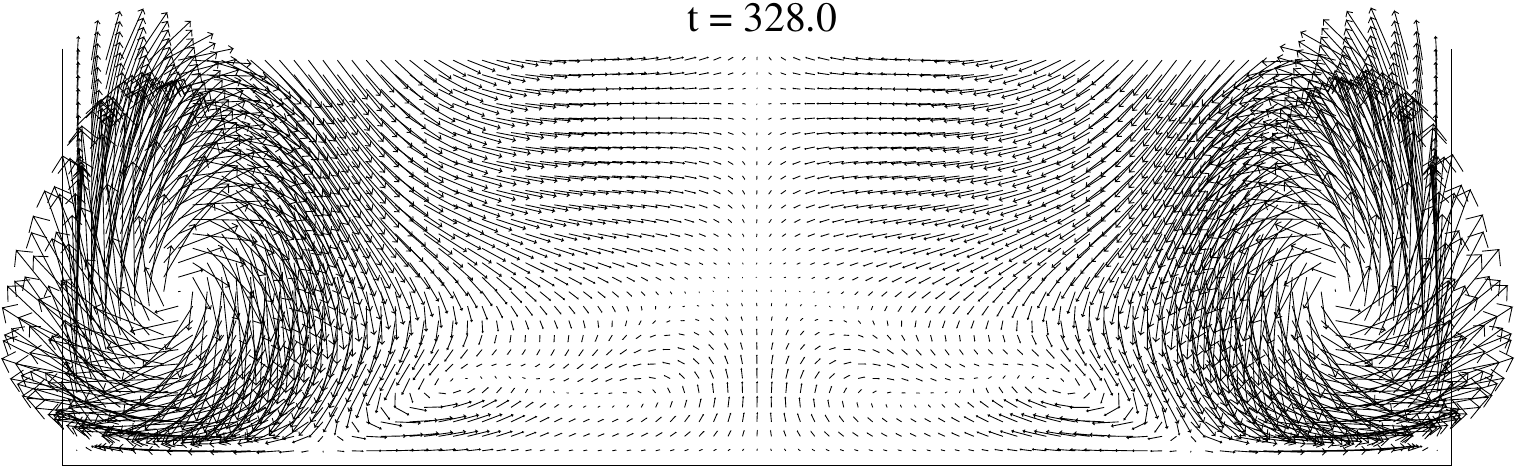}  \includegraphics[width=3.2in] {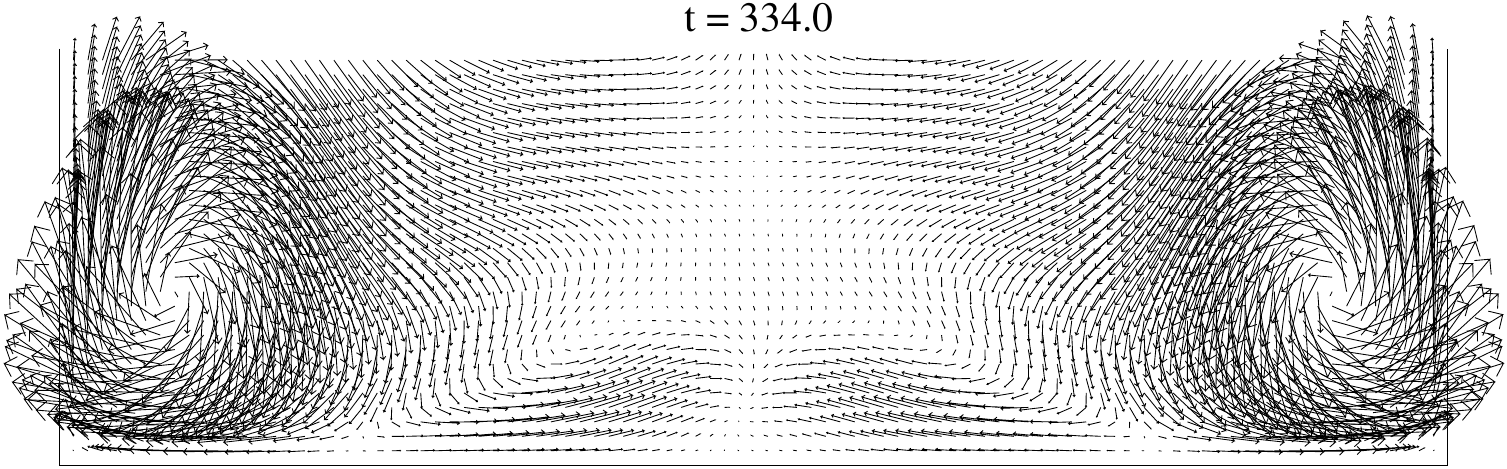}\\
\includegraphics[width=3.2in]                                                    
{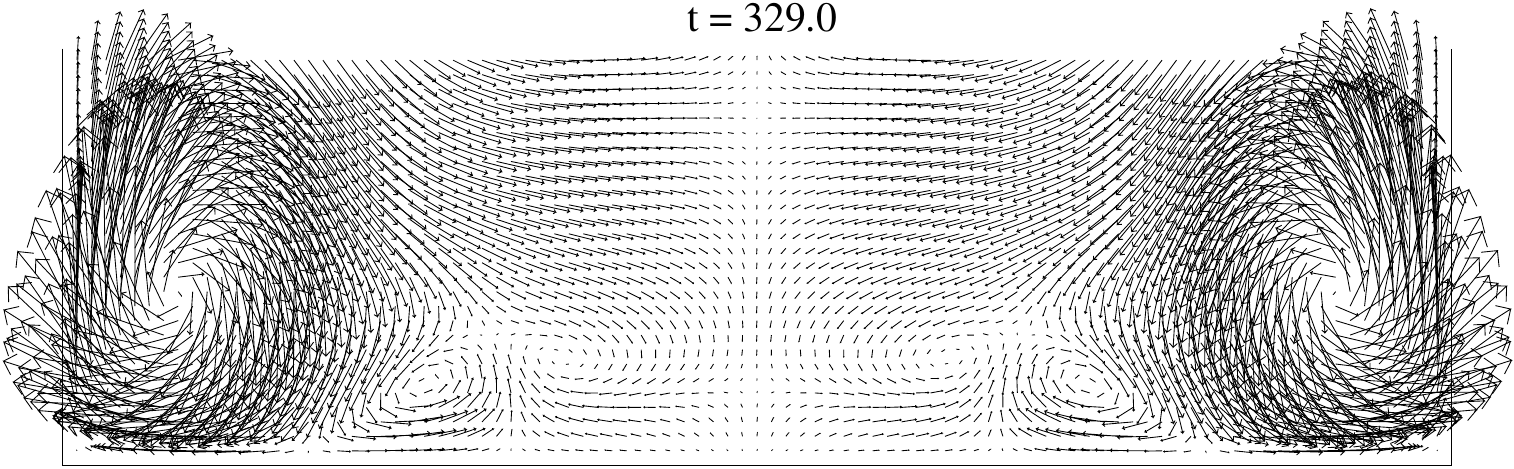}  \includegraphics[width=3.2in] {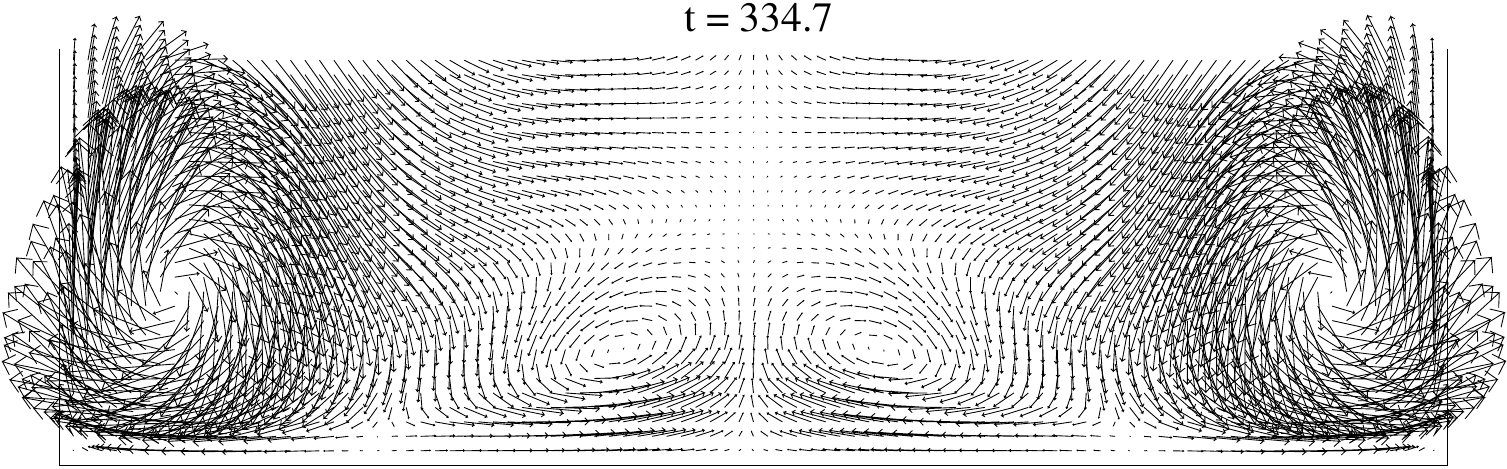}
\end{center} 
\caption{Projected velocity field vector  of the cavity flow for Re=1950 on the plane $x_1=54/96$ at different instants of
time during one oscillation of the flow field $L^2$-norm  from $t=324$ to 334.7. The vector scale  is 5 times that of the 
actual one for both.}\label{fig12}
\end{figure}

The  pictures of Figure  \ref{fig5}  show oscillations of the flow field $L^2$-norm and the oscillation amplitude either 
decreases or increases in time depending on Re. These oscillations indicate that  there is a periodic flow  field distortion.
In order to investigate the computed flow distortion, we have visualized in Figures  \ref{fig6} (for Re=1870) and  \ref{fig7} 
(for Re=1875) the velocity fields for $h=1/96$ associated with the peak and bottom of the velocity field $L^2$-norm, and the 
vector field associated with the difference of the above two velocity fields. The top (resp., bottom) pictures have been obtained 
by projecting the vector fields on the plane $x_1 = 52/96$ (resp., $x_3 = 1/2$). Figures \ref{fig6} and \ref{fig7}
show no evidence of TGL vortices for the velocity fields computed with $h = 1/96$ at Re=1870 and 1875; however, 
the pictures on the right of Figures \ref{fig6} and \ref{fig7}, obtained by the vector field difference detailed 
above, show a pair of vortices reminiscent of the GTL ones with much smaller magnitude. Those vector 
fields have been amplified by a factor of 500 (resp., 15) for Re=1870 (resp., 1875) in order to make them visible. 
For $h=1/60$, we have obtained similar results for Re=1860 and 1865.  This vortex pair keeps hiding there and becomes 
stronger as Re increases. These results suggest that the TGL vortices observed for Re slightly below 2000 are related 
to the  onset of an oscillatory flow. Thus we have further studied the oscillatory flow at Re=1885, 1900, and 1950. 
At Re=1885, a pair of TGL vortices becomes visible. In Figure \ref{fig8},  we have 
visualized (using a nonlinear scaling to enhance visibility) several snap-shots of the velocity field during an 
oscillation time period. A pair of TGL vortices is being formed in the time 
interval [333, 336] and then this pairs of TGL vortices disappear after $t = 337$; finally they reappear during 
the next time-period (at $t=343.54$). This pair of TGL vortices is stationary and remains symmetric with 
respect to the mid-plane $x_2 = 1/2$.   Actually this pair of TGL vortices  for $333 \le t \le 336$ does occur
from the peak of the flow field $L^2$-norm to the middle between the peak and the bottom as shown in Figure  \ref{fig11}.
Then  there is no significant sign of TGL vortices  at the bottom of  the flow field $L^2$-norm for $337 \le t \le 342$. 
But in Figure \ref{fig9},  there is a vortex pair in the flow field distortion obtained by difference of the velocity 
flow fields at $t=340$ and $t=342$,  which is reminiscent of those visualized on the right of Figures \ref{fig6} 
and \ref{fig7}. At Re$=1900$, we have also obtained numerical results  similar to those for Re=1885 (see Figures \ref{fig9},
\ref{fig10} and \ref{fig11}). In the histories of the flow field $L^2$-norm for Re=1885 and 1900 shown
in Figure  \ref{fig11}, there is a dominated single mode  which corresponds to the appearance of TGL vortices 
discussed above.  When increasing Re to 1950, there is a second mode gradually showing up in time. In the 
snapshots of velocity field projected on the plane $x_1=54/96$ for one oscillation of the flow field $L^2$-norm 
(see Figures \ref{fig11} and \ref{fig12}), the pair of TGL vortices (like those the dominated modes for Re=1885 and 1900) 
again occurs from the peak to the first bottom (for $324 \le t \le 327$). But  for $329 \le t \le 333$, 
a tertiary vortex next to the corner vortex appears, interacts with the symmetric one, and disappear; 
finally the symmetric pair show up again to start another period. These results  support that the TGL vortices observed 
for Re slightly below 2000 are related  to the oscillatory flows.

\section{Conclusion} 
In this article, we have studied numerically the transition from steady flow to oscillatory flow and  
occurrence of the TGL vortices via a three-stage Lie's scheme  The numerical results obtained for 
Re=400, 1000 and 3200 show a good agreement with available numerical and experimental  results in literature. 
Our simulation results predict that the critical Re$_{cr}$ for the transition from steady flow to oscillatory 
(a Hopf bifurcation) is somewhere in (1870, 1875) (resp., (1860, 1865)) for $h=1/96$ (resp., $h=1/60$).
For the connection between the occurrence of TGL vortices and the transition from steady flow to oscillatory flow, 
we have investigated the flow field at $Re$ close to $Re_{cr}$. The difference of flow fields at different instants of time
shows  a pair of vortices reminiscent of the GTL ones, but with much smaller magnitude  for Re slightly smaller than
Re$_{cr}$. For Re slightly larger than Re$_{cr}$, this flow field distortion is  visible and then invisible  periodically 
as a pair of TGL vortices. When increasing Re to 1950, two tertiary vortices next to the corner vortices 
(one from each corner) start interacting with this pair periodically and the number of TGL pairs varies between one and two.
All these pairs are symmetric. Thus our computational results suggest that the first pair of symmetric TGL vortices 
for Re < 2000 is actually the first  oscillating mode flow, which exists all the way through the transition.

\section*{Acknowledgments.} 
 
We acknowledge also the support of NSF (grant DMS-1418308).

\end{document}